\documentclass[twoside,12pt]{article}
\usepackage{epsfig}

\def\Journal#1#2#3#4{{#1} {#2} (#4) #3 }
\def\NC{{\em Nuovo Cimento} }			%JML 220819	

\def\NPA{{\em Nucl. Phys.} A}

\def\NPB{{\em Nucl. Phys.} B}

\def\PLB{{\em Phys. Lett.} B}

\def\PRL{\em Phys. Rev. Lett.}

\def\PREP{\em Phys. Rep.}

\def\PRD{{\em Phys. Rev.} D}
\def\PRC{{\em Phys. Rev.} C}

\def\ZPC{{\em Z. Phys.} C}
\def\ZPA{{\em Z. Phys.} A}

\def\RMP{{\em Rev. Mod. Phys.}}

\def\EPJC{{\em Eur. Phys. J.} C}		%JML 250119
\def\EPJA{{\em Eur. Phys. J.} A}		%JML 190319

\def\rpip{$\gamma p\to n \pi^{+}$}

\def\rpim{$\gamma n\to p \pi^{-}$}

\def\beqn{\begin{eqnarray}}
\def\eeqn{\end{eqnarray}}
\def\barr{\begin{array}}
\def\earr{\end{array}}
\def\btab{\begin{tabular}}
\def\etab{\end{tabular}}
\def\bite{\begin{itemize}}
\def\eite{\end{itemize}}
\def\bcen{\begin{center}}
\def\ecen{\end{center}}

\newcommand{\be}{\begin{equation}}
\newcommand{\ee}{\end{equation}}
\newcommand{\bea}{\begin{eqnarray}}
\newcommand{\eea}{\end{eqnarray}}

\begin{document}

\title{ \vspace{1cm} Exclusive Meson Photo- and Electro-production, a Window on the Structure of Hadronic Matter}
\author{J.M.\ Laget,$^{1,2}$ 
\\
$^1$JLab, Newport News, USA\\
$^2$Retired from CEA-Saclay, Gif sur Yvette, France\\
}
\maketitle
\begin{abstract} 
At high energy, exclusive meson photo- and electro-production give access to the structure of hadronic matter. At low momentum transfers, the exchange of a few Regge trajectories leads to a comprehensive account of the cross-sections. Among these trajectories, which are related to the mass spectrum of families of mesons, the Pomeron plays an interesting role as it is related to glue-ball excitations. At high momentum transfers, the exchange of these collective excitations is expected to reduce to the exchange of their simplest (quark or gluon) components.  However, contributions from unitarity rescattering cuts are relevant even at high energies. In the JLab energy range, the asymptotic regime, where the players in the game are current quarks and massless gluons has not been reached yet. One has to rely on more effective degrees of freedom adapted to the scale of the probe. A consistent picture, the “Partonic Non-Perturbative Regime”, is emerging. The properties of its various components (dressed propagators, effective coupling constants, quark wave functions, shape of the Regge trajectories, etc…..) provide us with various links to hadron properties. I will review the status of the field, will put in perspective the  current achievements at JLab, SLAC and Hermes, and will assess future developments that are made possible  by continuous electron beams at higher energies.
\end{abstract}

%\eject
%\tableofcontents

\section{Introduction}

Exclusive photo-production of mesons at high momentum transfer offers us a fantastic tool to investigate the partonic structure of hadrons. The high momentum transfer implies that the impact parameter is small enough to allow for a short distance interaction between, at least, one constituent of the probe  and one constituent of the target. In addition, the exclusive nature of the reaction implies that all the constituents of the probe and the target be in the small interaction volume, in order to be able to recombine into the well defined particles emitted in the final state.

\begin{figure}[tbhp]
\begin{center}
\begin{minipage}[t]{8 cm}
\hspace{-2.5cm}
\epsfysize=8.0cm
\epsfig{file=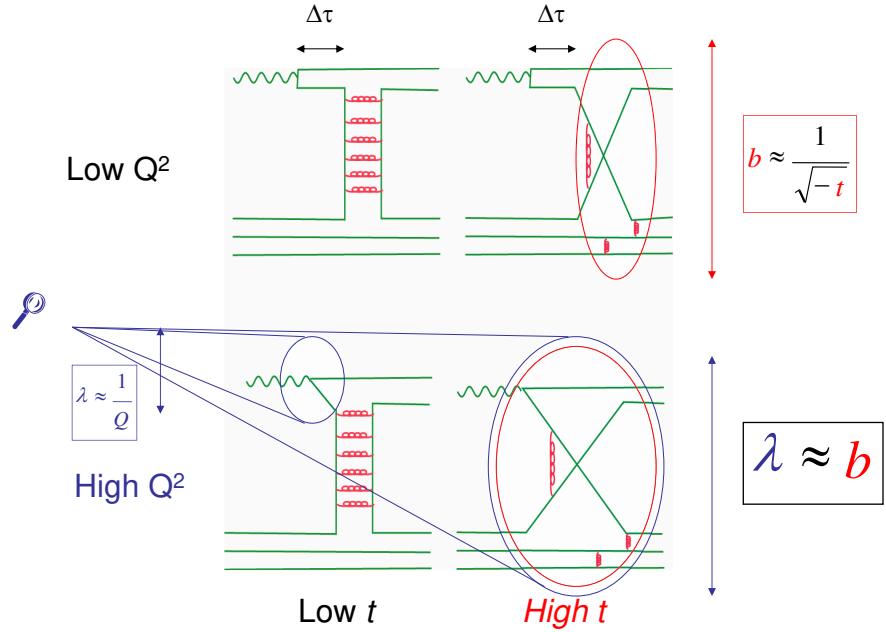,scale=0.5}
\end{minipage}
\begin{minipage}[t]{16.5 cm}
\caption{(Color on line) A schematic view of the evolution of the reaction mechanism toward hard scattering. See the description in the text.
\label{space_time}}
\end{minipage}
\end{center}
\end{figure}

The relevant constituents depend on the scale of observation. At low momentum transfer $-t=-(k-\mu)^2$ (or $-u=-(k-p_f)^2$), being $k$, $\mu$ and $p_f$ respectively the four-momentum of the incoming photon, the emitted meson and the recoiling hadron, a comprehensive description~\cite{gui97,jml} of available data is achieved by the exchange of a few Regge trajectories between the probe and the target. At very high (asymptotic) momentum transfer $-t$ (and $-u$), the interaction should reduce to the exchange of the minimal number of gluons needed to share the momentum transfer between all the current quarks which combine into the initial and final hadrons~\cite{bro}. Here, dimensional counting rules lead to the famous power law behavior of the various cross sections. This is schematically depicted~\cite{lag04} in the top part of Figure~\ref{space_time}.

When the incoming photon becomes virtual (as it is exchanged between the scattered electron and the hadrons) two things happen. On the one hand, the lifetime $\Delta \tau=2\nu/(Q^2+m_V^2)$ of its hadronic component decreases (being $\nu$, $Q^2$ and $m_V$ the energy and the virtuality of the photon and the mass of the vector meson respectively): its coupling becomes more point-like. On the other hand, the transverse wave length ($\lambda \sim 1/Q$) of the photon decreases: it probes processes which occur at shorter and shorter distances. When both the virtuality $Q^2=-q^2=-\nu^2+\vec{k}^2$ of the photon and the momentum transfer $-t$ are small (top left panel), the photon behaves as a beam of vector mesons which passes far away from the nucleon target (large impact parameter $b \sim 1/\sqrt{-t}$): the partons which may be exchanged have enough time to interact which each other and build the various mesons, whose the exchange drives the cross section. At high $-t$ (top right), the small impact parameter $b$ is comparable in size to the hadronization length of the partons which must be absorbed or recombined into the final particles, within the interaction volume of radius $b$, before they hadronize. In other words, the two partons, which are exchanged between the meson and the nucleon, have just the time to exchange a single gluon.  When $Q^2$ increases, the resolving power of the photon increases and allows it to resolve the structure of the exchanged quanta. When $-t$ is small (bottom left), it ``sees'' the partons inside the quantum which is exchanged between the distant meson and nucleon. When $-t$ is large (bottom right), its wave length $\lambda$ becomes comparable to the impact parameter $b$: the virtual photon ``sees'' the partons which are exchanged during the hard scattering.

Varying independently the momentum transfer and the virtuality of the photon  enables us to identify, to determine the role and to access the interactions between each constituent of hadrons.

Pioneering works have been performed at SLAC, Cornell, Daresbury, etc.. At Jefferson Laboratory (JLab), the Continuous Electron Beam Accelerator Facility (CEBAF) at 6 GeV allowed to reach  momentum transfers $-t$ up to 6 GeV$^2$ in {\em exclusive} reactions, at reasonably high energy ($s$ up to 12 GeV$^2$). Also, its high intensity continuous beam made it possible to systematically study the virtual photon sector up to photon virtuality Q$^2 \sim$ 6 GeV$^2$. These ranges are considerably enlarged with the present upgrade of CEBAF at 12 GeV.

The high momentum transfers accessible at JLab correspond to a resolving power of the order of $0.1$ to $0.2$~fm. It is significantly smaller than the size of a nucleon, but comparable to the correlation lengths of partons (the distance beyond which a quark or a gluon cannot propagate and hadronizes).  At this scale, the relevant degrees of freedom are the {\em constituent partons}, whose lifetime is short enough to prevent them to interact and form the mesons which may be exchanged in the $t$-channel, but is too long to allow to treat them as current quarks or gluons. This is the {\em partonic non-perturbative  regime}, where the amplitude can be computed as a set of few dominant Feynman diagrams which involve dressed quarks and gluons, effective coupling constants and quark distributions~\cite{cano}. 

This has been best achieved in the $\phi$ meson production channel. The modeling~\cite{do89} of the Pomeron in terms of two gluons provides a good description~\cite{jml,me95,cano} of the photo-production cross section~\cite{phiprl} at high $-t$. In addition the exchange of meson Regge trajectories are needed to account for the angular distributions in the photo-production of $\rho$ and $\omega$ mesons~\cite{bat01,bat03}.

Those meson exchanges dominate the photo-production of pseudo-scalar mesons. The structure of the amplitudes and the coupling constants  are the same as at lower energies~\cite{physrep}. Simply, the Feynman propagators are replaced by the relevant Regge propagators. Linear Regge trajectories lead to a good description of the angular distributions at low $-t$, without extra parameters. A good description of the cross sections at large $-t$ is achieved when saturating Regge trajectories are used~\cite{gui97}: this is an educated way to extrapolate toward the hard scattering regime. However, unitarity rescattering cuts~\cite{lag07,lag10,lag11} dominate at high $-t$. They involve the propagation of  on-shell hadrons and rely on elementary amplitudes that are well constrained by independent experiments. The cuts involving the propagation of the $\rho$ meson survive even at the highest energies. 
 
At backward angles (small $-u$, highest $-t$), the exchange of the Nucleon and $\Delta$ Regge trajectories is necessary to get a good representation of the cross sections.

Being calibrated at the real photon point such a description has been successfully extended to the virtual photon sector. Here, apart from the Transverse amplitude, also its Longitudinal component contributes to the cross section. At large $-t$, the data recorded at JLab~\cite{mor05,par13} and Hermes ~\cite{air08} are well reproduced when electromagnetic form factors, whose $-t$ dependency is related to the saturation of Regge trajectories, are used~\cite{lag04}. Unitarity rescattering cuts dominate the cross sections of neutral pion electro-production~\cite{lag11} and Virtual Photon Scattering~\cite{lag07} even at low $-t$.

The issue is therefore to identify the channels where the contribution of the unitarity cuts are small and where there may be a chance to access the hard scattering regime.

In the two following sections, the salient features of each channel will be reviewed and put in perspective of an hadronic description. In the fourth section, the players in the game will be presented. Appendix~A lists dedicated referred publications where more ancillary technical details can be found. Appendix~B summarizes the relation between the meson electro-production cross sections and the components of the currents.

\section{The real photon sector}
\subsection{\it Vector meson production channels \label{sec:vector}}

\begin{figure}[tbhp]
\begin{center}
\begin{minipage}[t]{8 cm}
\hspace{-1 cm}
\epsfysize=8.0cm
\epsfig{file=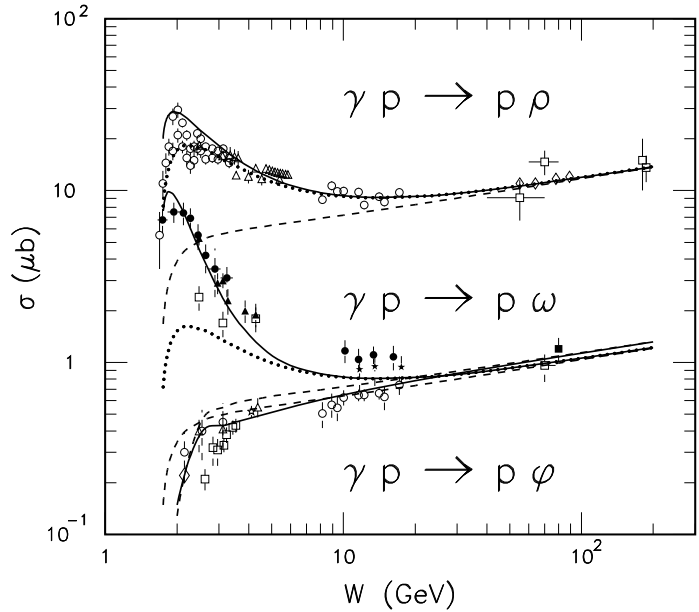,scale=0.5}
\end{minipage}
\begin{minipage}[t]{16.5 cm}
\caption{The total cross section of the photo-production of vector mesons. Dashed curves: Pomeron exchange. Dotted curves: $f_2$ meson exchange included. Solid curves: $\pi$ or $\sigma$ meson exchange included in the $\omega$ or $\rho$ meson production channel respectively; $\pi$ and $f'_2$ meson exchange included in the $\phi$ meson production channel.
\label{vector_cross}}
\end{minipage}
\end{center}
\end{figure}

Figure~\ref{vector_cross} displays the evolution of the cross section of the photo-production of vector mesons with the total c.m. energy $W=\sqrt s$. The details of the model and the references to experiments can be found in~\cite{jml}. At the highest energies (HERA), the cross section rises only moderately with energy and is dominated by the Pomeron exchange in all three channels. At low energies, $f_2$ Regge exchange dominates $\rho$ meson production and contributes significantly to $\omega$ meson production. Pion exchange dominates $\omega$ production, while $\sigma$ meson exchange contributes moderately to $\rho$ meson production. It is remarkable that the available data follow, over two decades of the c.m. energy, the expected evolution of the various dominant Regge exchange amplitudes. 

Due to the dominant $s\overline s$ component of the wave function of the $\phi$ meson, those meson exchange contributions are suppressed in the $\phi$ production channel: $\pi$, $\eta$ and $f'_2$ Regge exchanges contribute a little near threshold. The cross section is therefore dominated by the Pomeron exchange in the entire energy range in Figure~\ref{vector_cross}.

\begin{figure}[tbhp]
\begin{center}
\begin{minipage}[t]{8 cm}
\epsfysize=8.0cm
\epsfig{file=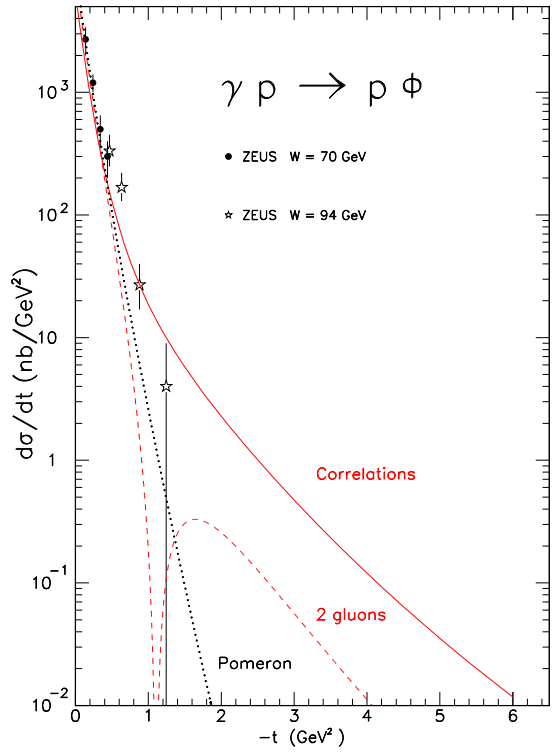,scale=0.3}
\end{minipage}
\begin{minipage}[t]{16.5 cm}
\caption{(Color on line) The differential cross section of the $p(\gamma,\phi)p$ reaction measured at HERA. Dotted line: Pomeron exchange. Dashed line: 2 gluons exchange, without quarks correlations in the nucleon (bottom graphs in Figure~\ref{2gluons_graph}). Solid line: with quark correlations in the nucleon (top graphs in Figure~\ref{2gluons_graph}). References of the experimental data can be found in~\cite{me95,jml}
\label{phi_zeus}}
\end{minipage}
\end{center}
\end{figure}

\subsection{\it The $p(\gamma,\phi)p$ reaction \label{sec:phi}}

\begin{figure}[tbhp]
\begin{center}
\begin{minipage}[t]{8 cm}
\hspace{1 cm}
\epsfysize=8.0cm
\epsfig{file=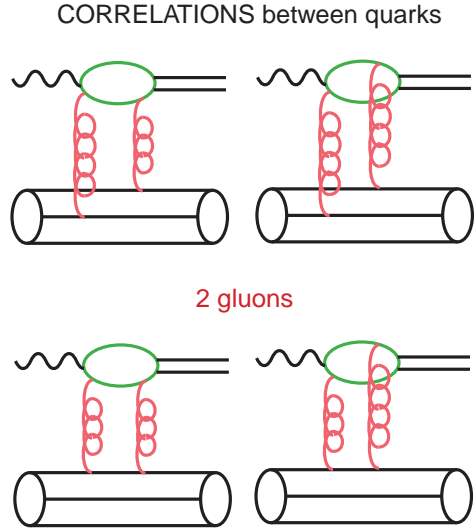,scale=0.4}
\end{minipage}
\begin{minipage}[t]{16.5 cm}
\caption{(Color on line) The two gluon exchange graphs. Top: with quark correlations in the nucleon. Bottom: without these correlations in the nucleon.
\label{2gluons_graph}}
\end{minipage}
\end{center}
\end{figure}

This picture works best when the $\phi$ meson is emitted at the most forward angles (that drive the total cross section). At higher $-t>1$ GeV$^2$, the Pomeron exchange contribution falls down faster than the experimental data as can be seen in Figure~\ref{phi_zeus}. The data are better reproduced by  modeling ~\cite{do89,me95,jml} the Pomeron in terms of the exchange of two gluons as depicted in Figure~\ref{2gluons_graph}. When the two gluons couple to the same constituent quark in the nucleon (bottom part), the interference between their coupling to the same constituent quark or to different quarks in the $\phi$ meson induces a node in the cross section. This node disappears when the coupling to different correlated quarks in the nucleon is allowed (top part). The two gluon contribution matches the Pomeron contribution at the lowest $-t$, but provides more strength at high $-t$. 

It turns out that, in the high energy limit, the Pomeron and the two gluon amplitudes have the same functional form and the same structure at low $-t$. The treatment of the quark loop in the vector meson is the same in the two approaches. The Pomeron Regge pole drives the dynamics of its exchange. Since it is a perturbative-like calculation there is no Regge factor in the two gluon exchange amplitude: the dynamics relies upon the shape of the effective gluon propagator and the detail of the constituent quark wave function of the nucleon. While the two approaches leads to similar results at low $-t$, the two gluon exchange approach is more suited at high $-t$.  A detailed summary of these two approaches is given in section\ref{sec:pomeron-2g}.

The curves in Figure~\ref{phi_zeus} correspond to the latest version~\cite{cano} of the model: it uses the parameterization~\cite{lei99} of the gluon propagator evaluated on lattice, and the correlated constituent quarks nucleon wave function~\cite{bo96} which reproduces the nucleon form factor. The gluon loop two fold integral is performed numerically. Earlier calculations~\cite{do89,me95,jml} used Gaussian forms, both for the gluon propagator and the nucleon wave function, which resulted in  a one fold integral. In the vector meson quark loop, the quarks are frozen with a constituent mass equal to the half of the vector meson mass. Their coupling to the vector meson is determined from the relevant radiative decay constant.

\begin{figure}[tbhp]
\begin{center}
\begin{minipage}[t]{8 cm}
\epsfysize=8.0cm
\epsfig{file=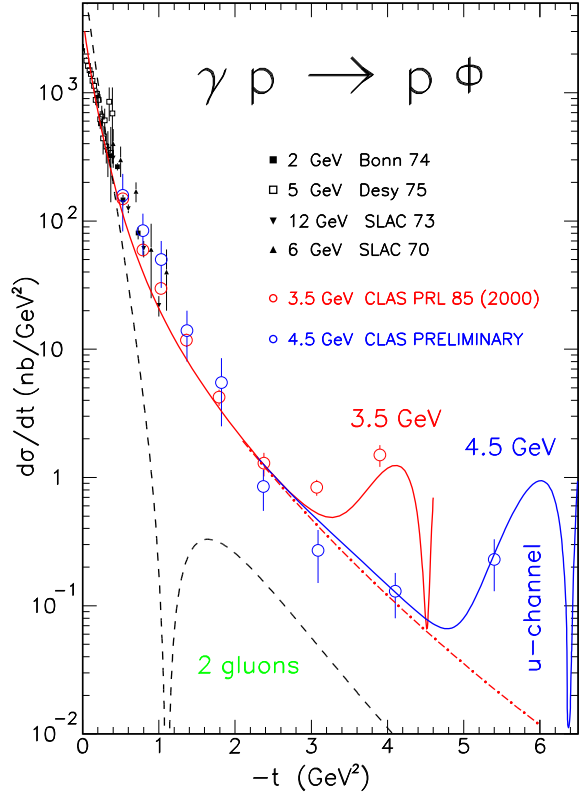,scale=0.4}
\end{minipage}
\begin{minipage}[t]{16.5 cm}
\caption{(Color on line) The differential cross section of the $p(\gamma,\phi)p$ reaction measured at JLab. Dashed line: 2 gluons exchange without quark correlations. Dash-dotted line: with quark correlations in the nucleon. Solid lines: u-channel Regge exchanges included.
\label{phi_JLab}}
\end{minipage}
\end{center}
\end{figure}

Thanks to the high intensity of the Continuous Electron Beam Facility (CEBAF), the $-t$ distribution has been extended to the highest values accessible at Jefferson Laboratory (JLab). In  Figure~\ref{phi_JLab} the $\phi$ production cross sections at E$_\gamma =$ 3.5~GeV (W = 2.73~GeV)~\cite{phiprl} and E$_\gamma =$ 4.5~GeV (W = 3.05~GeV) follow the prediction of the two gluons exchange model supplemented, at the highest $-t$, by the reggeized nucleon exchange in the u-channel. Consistently with the shape  of the angular distribution of the photo-production of $\omega$ meson  (Figure~\ref{omega_JLab}), whose the Regge structure is the same at backward angles,  the use of a non-degenerated nucleon Regge trajectory produces the node outside the range of experimental data. The analysis~\cite{mcc} of the tensor polarization of the emitted $\phi$ meson opened a window on the strange content of the nucleon, by constraining  the $\phi NN$ coupling constant and determining its sign. The experimental data at E$_\gamma =$~4.5~GeV have never been published, and are shown to illustrate the evolution with energy of the u-channel contribution. Clearly, the extension of this experiment at higher energy (E$_\gamma =$ 12~GeV, for instance) will enlarge the domain where two gluons exchange, explicitly taking into account the quark correlations in the nucleon, prevails.

\subsection{\it The $p(\gamma,\omega)p$ reaction \label{sec:omega}}

\begin{figure}[tbhp]
\begin{center}
\begin{minipage}[t]{8 cm}
\epsfysize=8.0cm
\epsfig{file=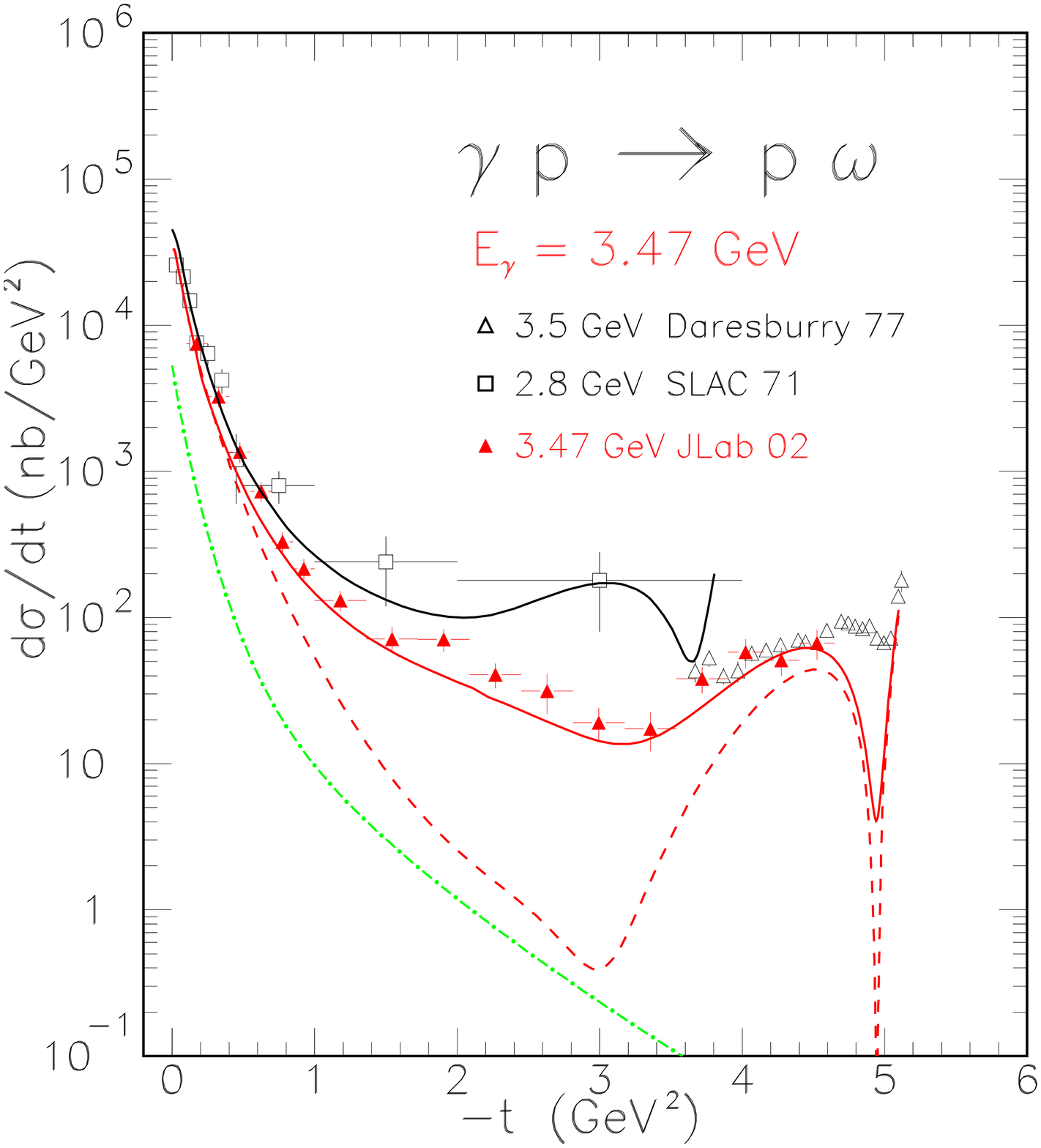,scale=0.4}
\end{minipage}
\begin{minipage}[t]{16.5 cm}
\caption{(Color on line) The differential cross section of the $p(\gamma,\omega)p$ reaction. Dot-dashed line: 2 gluons exchange, including quark correlations. Dashed line: $\pi$, $f_2$ and nucleon exchanges with linear Regge trajectories. Solid lines: with saturating trajectory of the $\pi$. 
\label{omega_JLab}}
\end{minipage}
\end{center}
\end{figure}

\begin{figure}[tbhp]
\begin{center}
\begin{minipage}[t]{8 cm}
\hspace{-1.5 cm}
\epsfysize=8.0cm
\epsfig{file=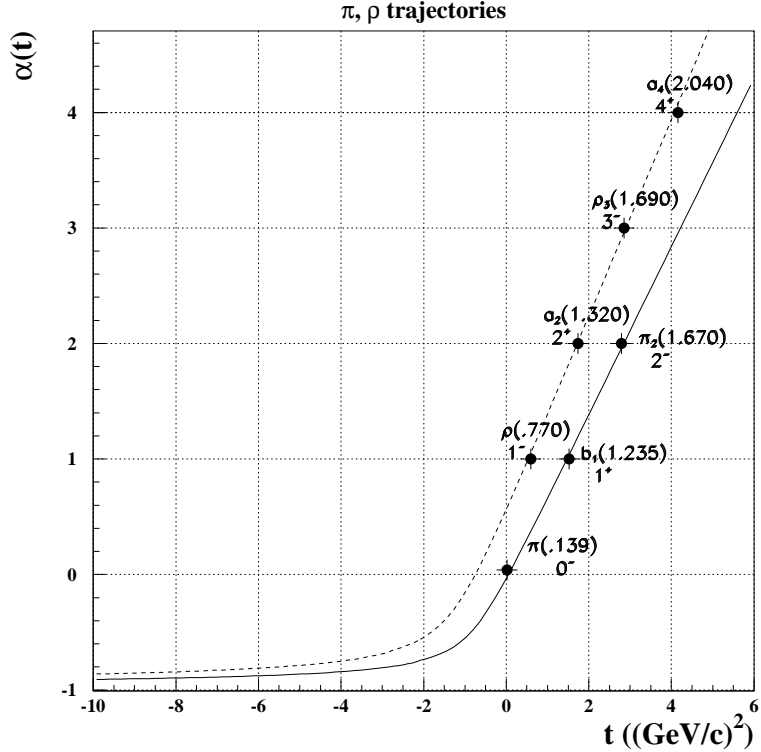,scale=0.7}
\end{minipage}
\begin{minipage}[t]{16.5 cm}
\caption{Saturating Regge trajectories~\cite{gui97,ser94}.
\label{sat_traj}}
\end{minipage}
\end{center}
\end{figure}

In the JLab energy range, t-channel pion Regge exchange dominates the cross section of the $p(\gamma,\omega)p$ reaction~\cite{bat03} at forward angles, while u-channel nucleon Regge exchange dominates at backward angles (Figure~\ref{omega_JLab}). In between, the use of a linear Regge trajectory for the pion misses badly the data. A decent agreement is recovered when the saturating trajectory of the pion~\cite{gui97}, which is plotted in Figure~\ref{sat_traj}, is used. It has been obtained~\cite{ser94} by determining the pion Regge trajectory from a QCD motivated effective interquark potential: its confining part leads to the linear part of the trajectory in the time like region, while its short range one gluon exchange part leads to the saturation of the trajectory $\alpha_{\pi}(t) \approx$~-1 in the space like region. Such a saturation has been also noted~\cite{col84} in nucleon-nucleon and pion-nucleon scattering. It provides a way to reconcile the Regge exchange model with counting rules at high $-t$.

\subsection{\it The $p(\gamma,\rho)p$ reaction \label{sec:rho}}

\begin{figure}[tbhp]
\begin{center}
\begin{minipage}[t]{8 cm}
\epsfysize=8.0cm
\epsfig{file=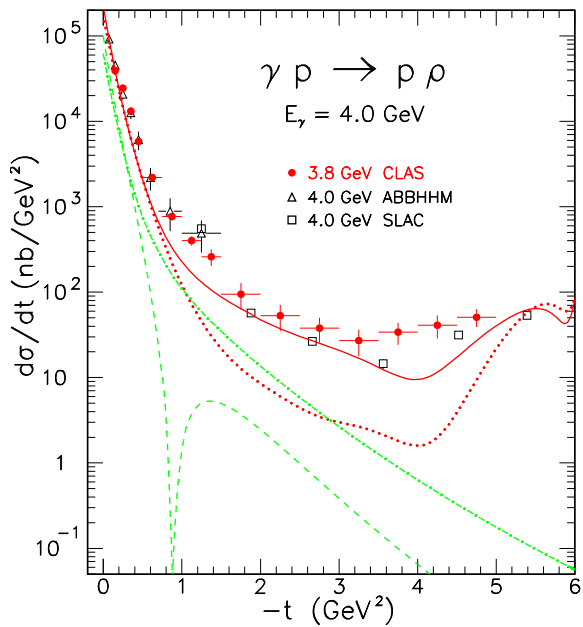,scale=0.4}
\end{minipage}
\begin{minipage}[t]{16.5 cm}
\caption{(Color on line) The differential cross section of the  $p(\gamma,\rho)p$ reaction. Dot-dashed line: 2 gluons exchange with correlations. Dashed line: 2 gluons exchange without correlations.  Dotted line: $\sigma$ and $f_2$ exchange with linear Regge trajectories. Solid line: with saturating trajectory of the $\sigma$. Nucleon and $\Delta$ Regge exchanges in the u-channel are also included in the dotted and full lines. 
\label{rho_JLab}}
\end{minipage}
\end{center}
\end{figure}

Figure~\ref{rho_JLab} shows the differential cross section of the  $p(\gamma,\rho)p$ reaction in the JLab energy range. Besides two gluons exchange, $f_2$ and $\sigma$ Regge exchange contribute significantly at forward angles, while nucleon and $\Delta$ Regge exchanges dominate at backward angles.  The use of a saturating Regge trajectory of the $\sigma$ restores the agreement with the experimental data~\cite{bat01} at high $-t$ and $-u$, in the intermediate angular range ($1\le -t \le 4$ GeV$^2$).

\subsection{\it Real Compton scattering \label{sec:compton}}

\begin{figure}[tbhp]
\begin{center}

\begin{minipage}[t]{8 cm}
\epsfysize=8.0cm
\epsfig{file=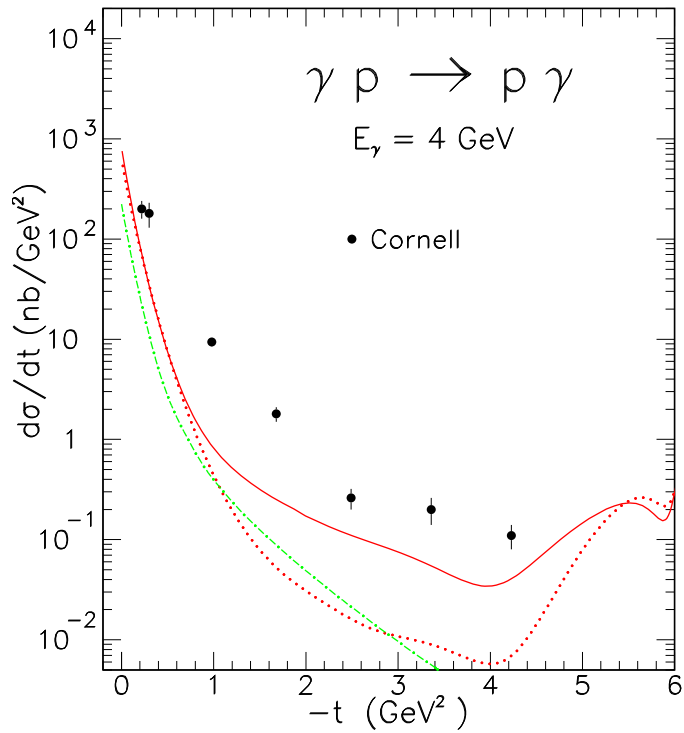,scale=0.4}

\end{minipage}
\begin{minipage}[t]{16.5 cm}
\caption{(Color on line) The differential cross section of the $p(\gamma,\gamma)p$ reaction at $E_\gamma=4$~GeV. The curves are the same as in Figure~\ref{rho_JLab}, but scaled by $4\pi\alpha_{em}/f^2_V$.
\label{compton_JLab}}
\end{minipage}
\end{center}
\end{figure}

As can be seen in Figure~\ref{compton_JLab}, the differential cross section of the $p(\gamma,\gamma)p$ reaction~\cite{sch79} has the same shape as that of the $p(\gamma,\rho)p$ reaction depicted in Figure~\ref{rho_JLab}.  The reason is that the lifetime of the  fluctuation of  a $\nu\approx 4$~GeV outgoing photon into a $\rho$ meson (see left graph in Figure~\ref{dvcs_graphs}) is about $\Delta \tau=2\nu/m_V^2=2.6$~fm, comparable or larger that the size of the nucleon. Therefore the first contribution to the cross section comes from the photo-production of a $\rho$ meson which converts into a real gamma outside the nucleon (left graph in Figure~\ref{dvcs_graphs}): the cross section is simply the $\rho$ meson photo-production cross section scaled by $4\pi\alpha_{em}/f^2_V=0.00361$  where $f_V$ is the radiative decay constant of the $\rho$. The numerical mistake in the evaluation of this scaling factor in~\cite{cano} and~\cite{can03} has been corrected in~\cite{can03E} and subsequent publications. This correction does not affect the spin transfer coefficients, but leads to a theoretical Compton cross section which is lower than the data.
 
\begin{figure}[tbhp]
\begin{center}

\begin{minipage}[t]{8 cm}
\epsfysize=8.0cm
\epsfig{file=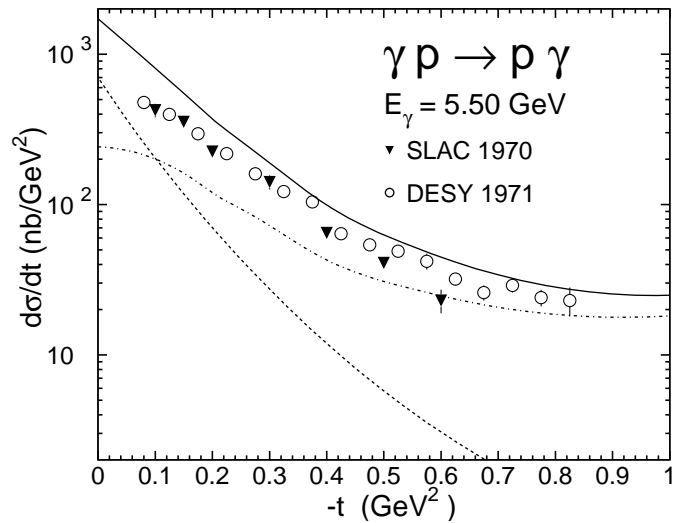,scale=0.4}

\end{minipage}
\begin{minipage}[t]{16.5 cm}
\caption{The real Compton scattering cross section at $\sqrt s =$ 3.35~GeV. Dashed curve: Pole terms. Dot-dashed curve: $\rho$ inelastic cut only. Full curve: both contributions.
\label{real_compton_cut}}
\end{minipage}
\end{center}
\end{figure}

As can be seen in Figure~\ref{real_compton_cut}, the agreement with the data~\cite{and70} and~\cite{bus70} is restored~\cite{lag07} when the conversion of the $\rho$ meson into the final photon occurs in the target nucleon (unitarity inelastic cut in the right part of Figure~\ref{dvcs_graphs}). Unfortunately, the effects of this cut have been quantified only in the range of low $-t$ which has been covered in the existing Virtual Compton Scattering experiments (section~\ref{sec:virt_compton}): The extension to higher $-t$ remains to be done. 
 
The pole model~\cite{can03} leads also to a good description of the experimental spin transfer coefficient~\cite{fan15}, opposite to the prediction of pertubative QCD. However the influence of the rescattering cuts remains to be evaluated here too.

\subsection{\it Charged pion photo-production  \label{sec:pseudo}}

\begin{figure}[tbhp]
\begin{center}

\begin{minipage}[t]{8 cm}
\hspace{-1 cm}
\epsfysize=8.0cm
\epsfig{file=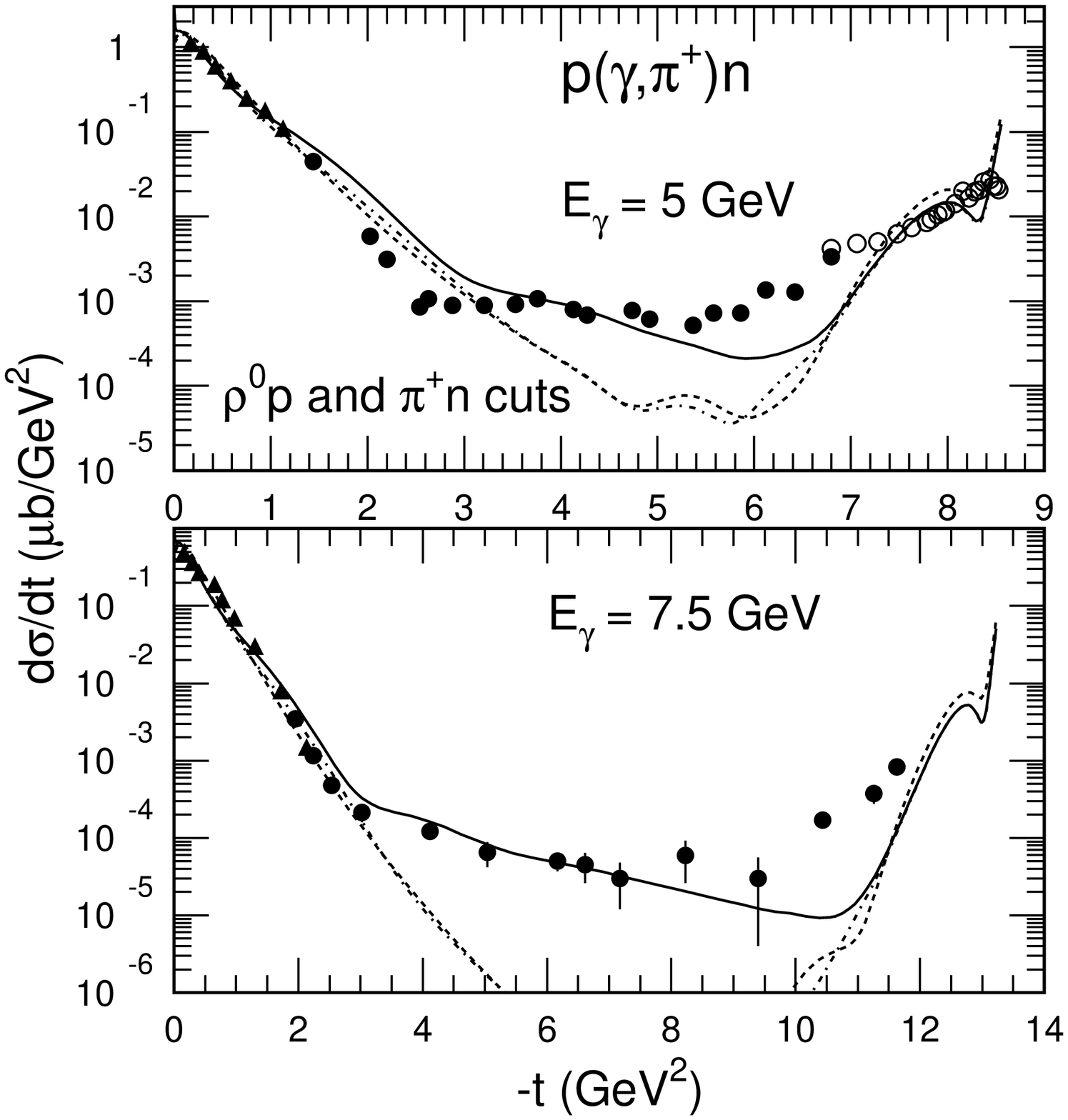,scale=0.5}

\end{minipage}
\begin{minipage}[t]{16.5 cm}
\caption{The differential cross section of the $p(\gamma,\pi^+)n$ reaction at $E_\gamma=$ 5~GeV and 7.5 GeV. Dashed lines: Regge poles. Dash-dotted lines: $\pi^+n$ elastic cut included. Solid lines: $\rho^0 p$ inelastic cut included. The references to the experimental data can be found in~\cite{gui97}.
\label{charged_pion_photo}}
\end{minipage}
\end{center}
\end{figure}

Figure~\ref{charged_pion_photo} shows the differential cross section of $\pi^+$ photo-production at $E_\gamma=$ 5~GeV and 7.5 GeV, in the entire accessible $-t$ range. At low $-t$, the data are well reproduced by the gauge invariant t-channel exchange of the $\pi$ and  $\rho$ Regge poles~\cite{gui97}, with linear trajectories. At the largest $-t$ (lowest $-u$), the exchange of the nucleon and the $\Delta$ Regge poles~\cite{lag10}, in the u-channel, dominates the cross section.
 
\begin{figure}[tbhp]
\begin{center}

\begin{minipage}[t]{8 cm}
\hspace{-4.5 cm}
\epsfysize=8.0cm
\epsfig{file=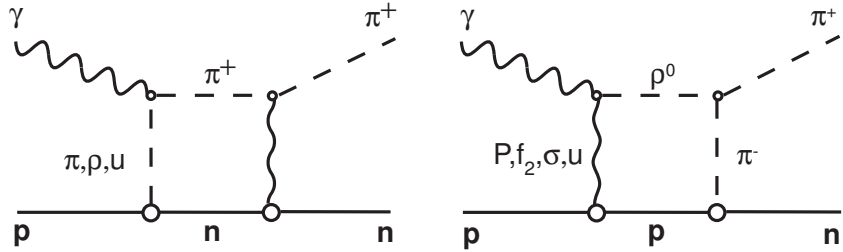,scale=0.6}

\end{minipage}
\begin{minipage}[t]{16.5 cm}
\caption{The unitarity cut contributions in the $p(\gamma,\pi^+)n$ reaction. Left:  $\pi^+n$ elastic cut. Right: $\rho^0 p$ inelastic cut. The label "u" stands for u-channel Regge exchanges.
\label{charged_pion_graphs}}
\end{minipage}
\end{center}
\end{figure}

At large $-t$ and $-u$, the $\rho^0 p$ inelastic cut  (Figure~\ref{charged_pion_graphs}) fairly accounts for the plateau in the differential cross section. The singular part of the rescattering integral is computed numerically and depends upon the on mass shell  elementary amplitudes which are strongly constrained by experiments: see \textit{e.g.} Figure~\ref{rho_JLab} for $\rho^0$ meson photo-production, Figure~3 in~\cite{lag10} for the $\rho^0$ to $\pi^+$ transition. Therefore, the agreement with the data is remarkable, since there are no free parameters.
 
The cut associated with the $\rho^0$ meson propagation survives at the highest photon energies, since the $\rho^0$ production cross section stays almost constant with increasing energy (see Figure~\ref{vector_cross}). It also dominates over the others since the cross section of the  $p(\gamma,\rho^0)p$ reaction is the largest of the various mesons photo-production reaction cross sections, as it is evident from Figure~\ref{vector_cross}.  For instance, the  $\pi^+n$ elastic unitarity cut plays a little role at large $-t$ and $-u$. However, it interferes destructively with the Regge pole amplitudes, due to the dominant absorptive nature of the $\pi^+n$ elastic amplitude at high energy. Such a subtle effect is quantified in Figure~4 of~\cite{lag10}: it helps to reproduce the rising differential cross section a the very forward angles.

Alternately, the plateau at intermediate momentum transfers has also been reproduced~\cite{gui97} by the use of pion and $\rho$ Regge saturating trajectories. Although there is no doubt that Regge trajectories must saturate at large $-t$, the way they have been implemented in the reaction amplitude must be revisited in view of the dominance of the well constrained unitarity rescattering cut contribution (see section~\ref{sec:sat-vs-cuts}).

\subsection{\it Neutral pion photo-production  \label{sec:pseudo_neutral}}

\begin{figure}[tbhp]
\begin{center}

\begin{minipage}[t]{8 cm}
\epsfysize=8.0cm
\epsfig{file=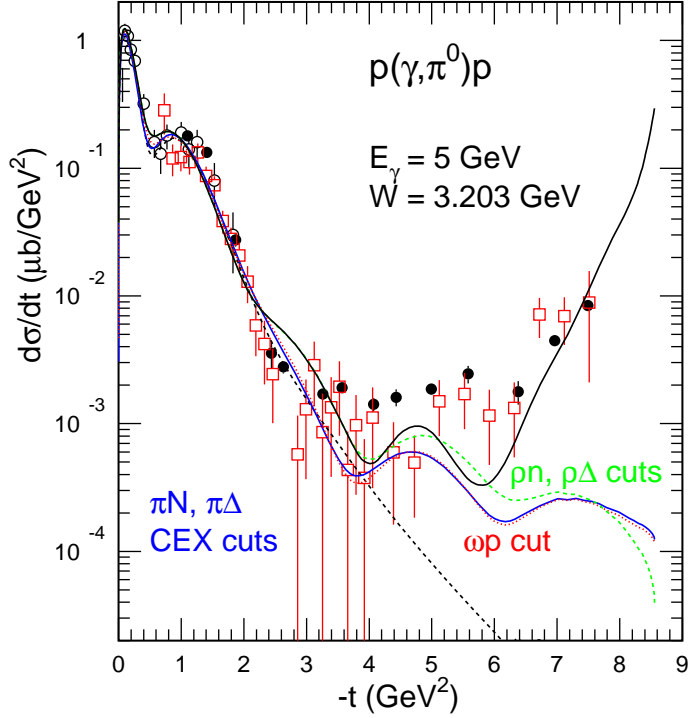,scale=0.5}

\end{minipage}
\begin{minipage}[t]{16.5 cm}
\caption{(Color on line) The differential cross section of the  $p(\gamma,\pi^0)p$ reaction at $E_\gamma=$ 5~GeV. Dashed line: t-channel Regge Poles. Solid black line: including u-channel Regge Poles and unitarity rescattering cuts. Their contributions are split in the labeled curves. Solid blue line: pion Charge Exchange cuts. Dotted red line: $\omega$ cut. Dashed green line: charged $\rho$ cuts. Open squares: JLab data~\cite{kun18}. Open circles: DESY data~\cite{Des68}. Full circles: SLAC data~\cite{Sla76}.
\label{neutral_pion_photo}}
\end{minipage}
\end{center}
\end{figure}

\begin{figure}[tbhp]
\begin{center}

\begin{minipage}[t]{8 cm}
\hspace{-2 cm}
\epsfysize=8.0cm
\epsfig{file=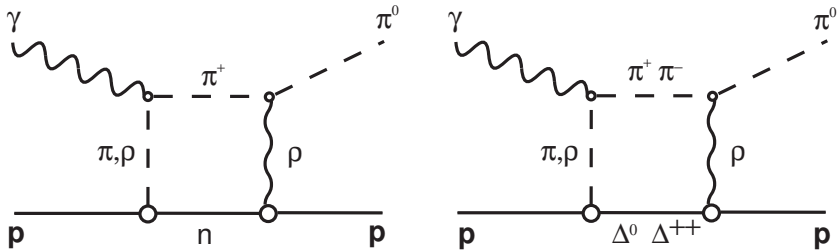,scale=0.6}

\end{minipage}
\begin{minipage}[t]{16.5 cm}
\caption{The Charge Exchange unitarity cuts in the $p(\gamma,\pi^0)p$ reaction.
\label{neutral_pion_CEX_cuts}}
\end{minipage}
\end{center}
\end{figure}

\begin{figure}[tbhp]
\begin{center}

\begin{minipage}[t]{8 cm}
\epsfysize=8.0cm
\hspace{-1.6 cm}
\epsfig{file=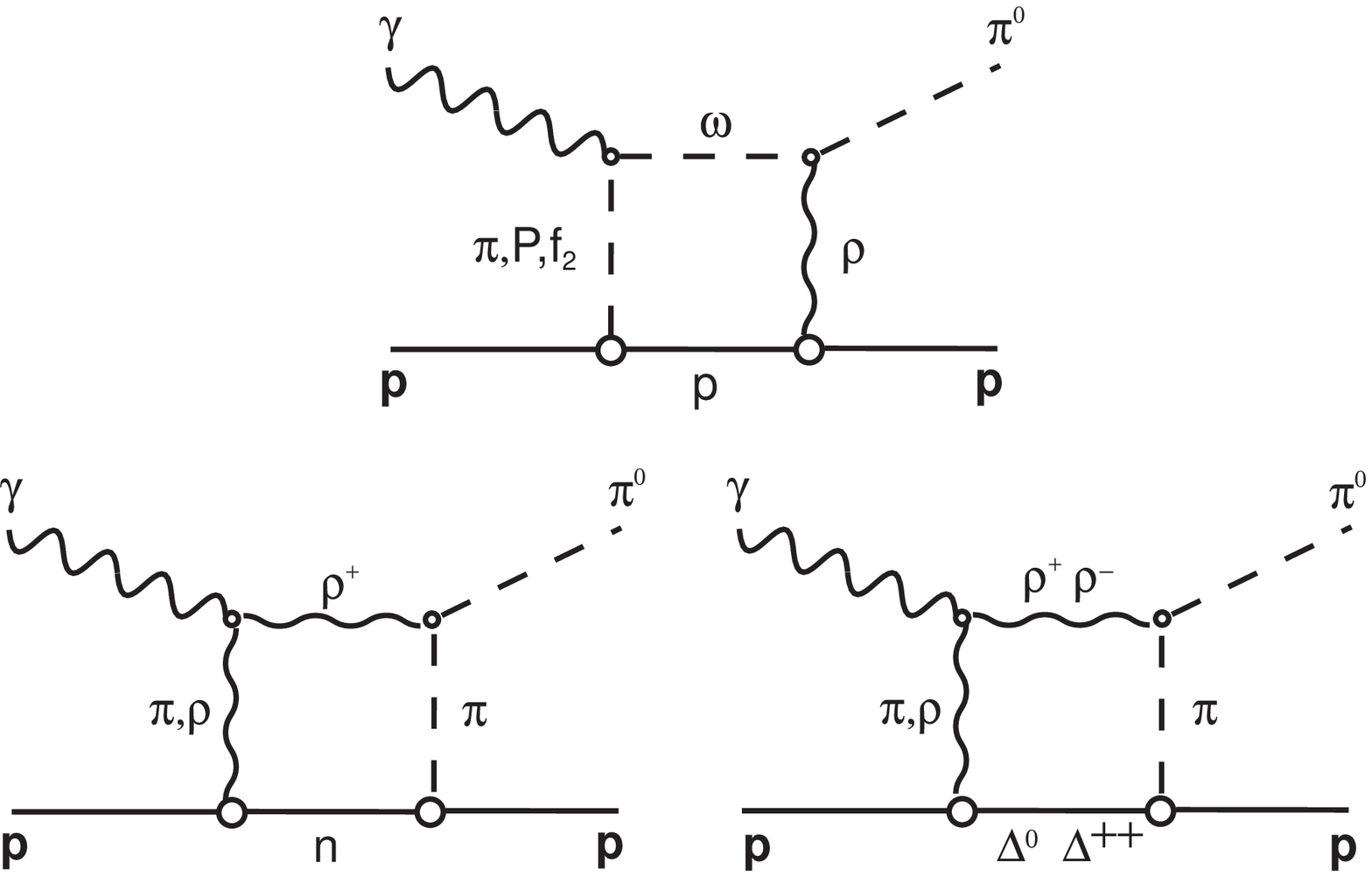,scale=0.6}

\end{minipage}
\begin{minipage}[t]{16.5 cm}
\caption{The vector meson unitarity cuts in the $p(\gamma,\pi^0)p$ reaction.
\label{neutral_pion_vector_cuts}}
\end{minipage}
\end{center}
\end{figure}

None of the inelastic unitarity cuts dominates the neutral pion photo-production cross section. Since the $\rho^0$ meson cannot decays into two $\pi^0$'s, the corresponding cut (which dominates charged pion photo-production) is strongly suppressed. In Figure~\ref{neutral_pion_photo}, Charge Exchange (Figure~\ref{neutral_pion_CEX_cuts}) and vector mesons (Figure~\ref{neutral_pion_vector_cuts}) unitarity cuts contribute equally~\cite{lag11} at intermediate $-t$ and $-u$. At backward angles, u-channel nucleon and $\Delta$ Regge Pole exchange amplitudes are the same as in the charged pion production channel, with trivial changes of the charge and coupling constants. At forward angles the model~\cite{lag11} relies on the exchange in the t-channel of the $\omega$, the $\rho$ and $b_1$ Regge trajectories. Contrary to ref.~\cite{gui97}, where the use of a non degenerated Regge trajectory for the $\omega$ leads naturally to the node at $-t \sim$ 0.5~GeV$^2$, the model~\cite{lag11} uses a degenerated trajectory. The interference with the elastic pion rescattering cut restores the node.  Each description leads to a very similar result at the real photon point, but the model with the interference between the pole and the cut is better suited to the analysis of the virtual photon sector (see section~\ref{sec:virt_neutral_pion}).

\begin{figure}[tbhp]
\begin{center}

\begin{minipage}[t]{8 cm}
\hspace{-2 cm}
\epsfysize=8.0cm
\epsfig{file=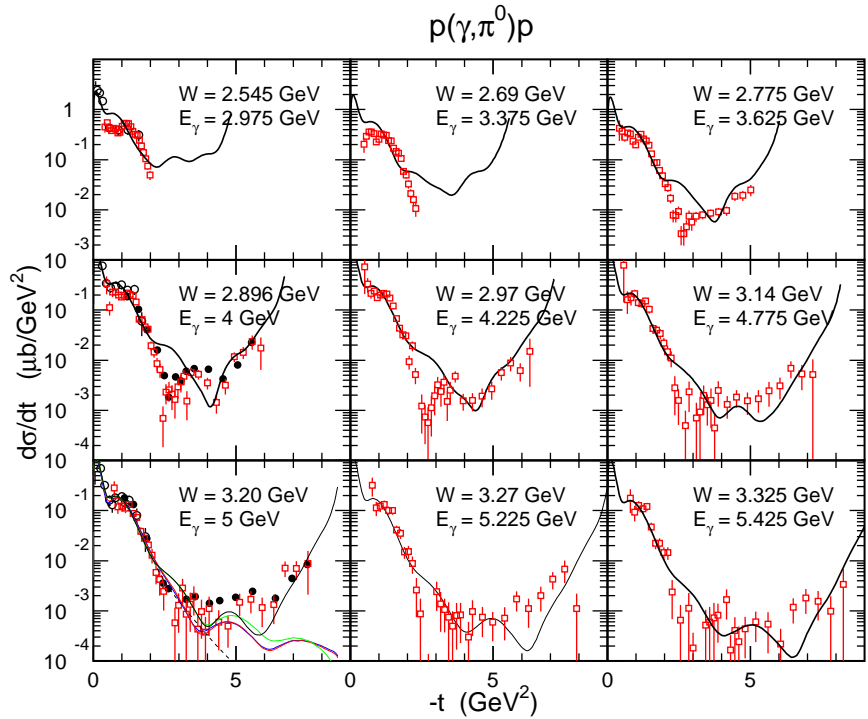,scale=0.5}

\end{minipage}
\begin{minipage}[t]{16.5 cm}
\caption{(Color on line) The differential cross section of the  $p(\gamma,\pi^0)p$ reaction at selected energies. The meaning of the curves and the references to the data are the same as in Figure~\ref{neutral_pion_photo}.
\label{neutral_pion_photo_panel}}
\end{minipage}
\end{center}
\end{figure}

The highest energy ($E_{\gamma}$ above 3~GeV) sample of the latest JLab data~\cite{kun18} are plotted together with older data in Figure~\ref{neutral_pion_photo_panel}. The model describes reasonably well the data, specially at the highest energies accessible at JLab. Clearly the extension to higher energies would allow to enlarge the intermediate $-t$ and $-u$ range and to carry out a more comprehensive study of the various cut contributions.
 
\subsection{\it Kaon photo-production \label{sec:strange}}

\begin{figure}[tbhp]
\begin{center}

\begin{minipage}[t]{8 cm}
\hspace{-1 cm}
\epsfysize=8.0cm
\epsfig{file=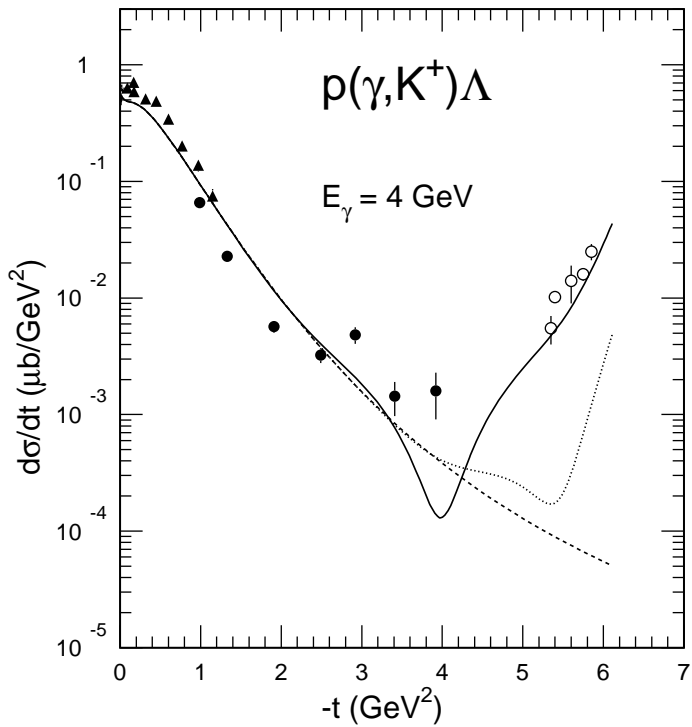,scale=0.5}

\end{minipage}
\begin{minipage}[t]{16.5 cm}
\caption{The differential cross section of the $p(\gamma,K^+)\Lambda$ reaction at $E_\gamma=$ 4~GeV. Dashed line: $K$ and $K^*$ exchanges in the t-channel. Dotted line: with $\Lambda$ and $\Sigma$ exchange in the u-channel. Solid line: with $\Sigma^*$ exchange in the u-channel. References to the data can be found in~\cite{gui97}.
\label{kaon_photo}}
\end{minipage}
\end{center}
\end{figure}

Figure~\ref{kaon_photo} shows the differential cross section of the $p(\gamma,K^+)\Lambda$ reaction. The exchange of the $K^+$ and the $K^{*+}$ meson non saturating Regge trajectories~\cite{gui97} dominates at forward angles, while the exchange of the $\Lambda$, $\Sigma$  and $\Sigma^*(1385)$ baryon Regge trajectories in the u-channel dominates at backward angles. Those u-channel exchange amplitudes have the same form as the nucleon and $\Delta$ amplitudes~\cite{lag10} with trivial adjustments of the coupling constants and propagators.

It seems that there is little room for unitarity cuts. However, the energy may be too low. Also, contrary to the pion channels, there is no obvious candidate for an intermediate state with a sizable production cross section. This is an open issue. An extension of the experiment at higher energies is clearly needed.

At lower energies (below $E_{\gamma}=$ 3 GeV), the model reproduces reasonably well the JLab data~\cite{schu} over the entire angular range. However, those data lie in the resonance region, where a Regge approach is not supposed to work well. In the $\Lambda$ production sector many resonances contribute and the Regge amplitude averages their amplitudes. This is not the case in the $\Sigma$ production sector where the contribution of a few dominant resonances must be taken into account explicitly.
 
\subsection{\it $J/\Psi$ meson photo-production \label{sec:charm}}

Because of the high value of the mass of the charm quark (about half of the mass of the $J/\Psi$ vector meson) the approximate treatment of the quark propagator in the $c\overline{c}$ loop in the two gluon exchange amplitude is better justified here (see the discussion of Figure~\ref{2gluons_graph} above and section~\ref{sec:pomeron-2g}). It also provides us with a new mass scale.

\begin{figure}[tbhp]
\begin{center}
\begin{minipage}[t]{8 cm}
\hspace{-2 cm}
\epsfysize=8.0cm
\epsfig{file=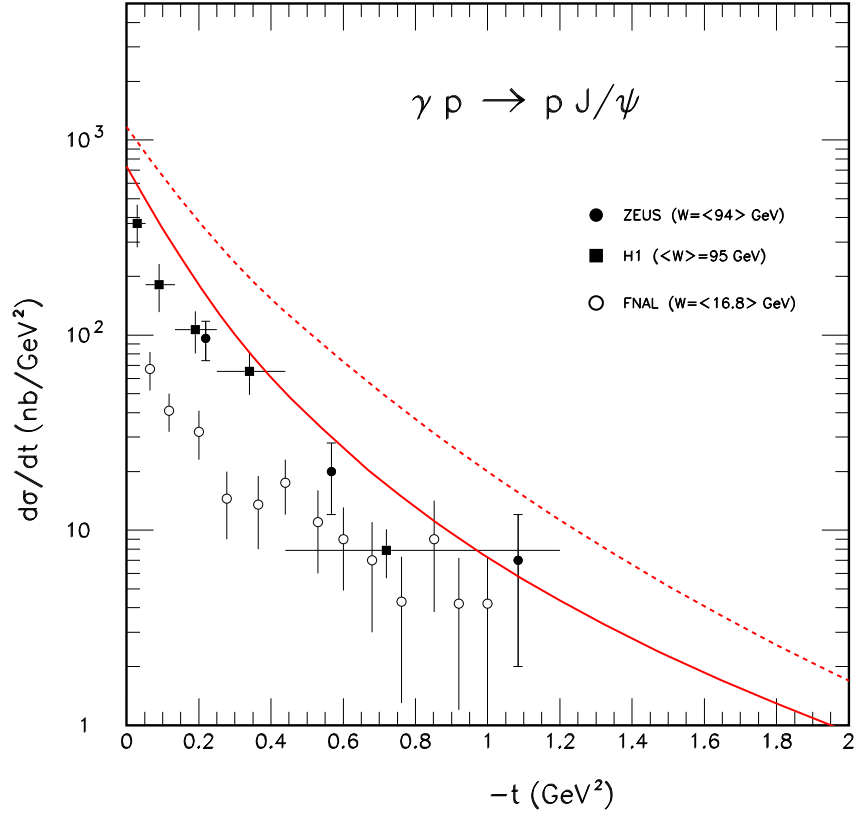,scale=0.6}
\end{minipage}
\begin{minipage}[t]{16.5 cm}
\caption{(Color on line) The differential cross section of the $p(\gamma,J/\Psi)p$ reaction measured at HERA~\cite{Bre00,Adl00} and FNAL~\cite{Bli82}. Dotted line: with the approximate nucleon wave function and gluon propagator used in~\cite{jml}. Solid line: with the realistic wave function of the nucleon and the gluon propagator from lattice used in~\cite{cano}.
\label{jpsi_dsdt_cano}}
\end{minipage}
\end{center}
\end{figure}

\begin{figure}[tbhp]
\begin{center}
\begin{minipage}[t]{8 cm}
\hspace{-2 cm}
\epsfysize=8.0cm
\epsfig{file=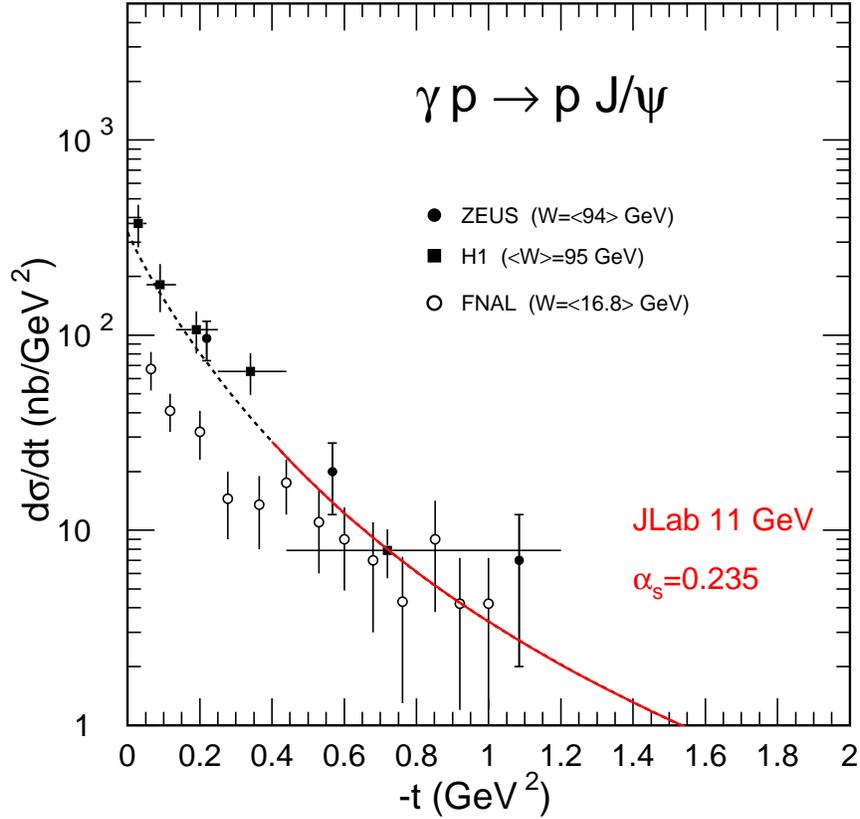,scale=0.6}
\end{minipage}
\begin{minipage}[t]{16.5 cm}
\caption{(Color on line) The differential cross section of the $p(\gamma,J/\Psi)p$ reaction at JLab. The kinematics domain accessible at $E_{\gamma}=$ 11~GeV is shown by the red solid line. The  dashed line shows the range accessible at HERA~\cite{Bre00,Adl00} and FNAL~\cite{Bli82}. Compared to the curve in Figure~\ref{jpsi_dsdt_cano}, the strong coupling constant $\alpha_s$ has been slightly adjusted to best fit the HERA data.
\label{jpsi_dsdt_an0235}}
\end{minipage}
\end{center}
\end{figure}

Figure~\ref{jpsi_dsdt_cano} compares the latest version of the two gluon exchange model~\cite{cano} to the  differential cross section of the $p(\gamma,J/\Psi)p$ reaction investigated at HERA~\cite{Bre00,Adl00} and FNAL~\cite{Bli82}. On the one hand, it shows the sensitivity of the model to the choice of the nucleon wave function and the gluon propagator: the more realistic correlated quark nucleon wave function~\cite{bo96} together with the lattice gluon propagator~\cite{lei99}, that have been used in~\cite{cano}, leads to a very good agreement with HERA data ($W=$ 94~GeV), contrary to the approximate Gaussian form used in~\cite{jml}. On the other hand, the FNAL data ($W=$ 16.8~GeV) exhibit a slower decrease but match the HERA data above $-t=$ 0.4~GeV$^2$ (within the error bars) where the model is best suited. As can be seen in Figure~\ref{jpsi_dsdt_an0235}, this high $-t$ domain is selected at JLab12 between the threshold and $W=$ 4.64~GeV ($E_{\gamma}=$ 11~GeV), due to the large value of $|t|_{min}\geq$~0.4~GeV$^2$.

It has also been proposed~\cite{bro01} that three gluon exchange may dominate the cross section near threshold. Such a prediction have been confirmed by the data recently released by the Hall D collaboration~\cite{Ali19} at JLab.

All these issues will be addressed in the next few years by ongoing experiments at JLab.

\section{The virtual photon sector}

\subsection{\it Vector mesons electro-production \label{sec:virt_vec}}

\begin{figure}[tbhp]
\begin{center}
\begin{minipage}[t]{8 cm}
\epsfysize=8.0cm
\epsfig{file=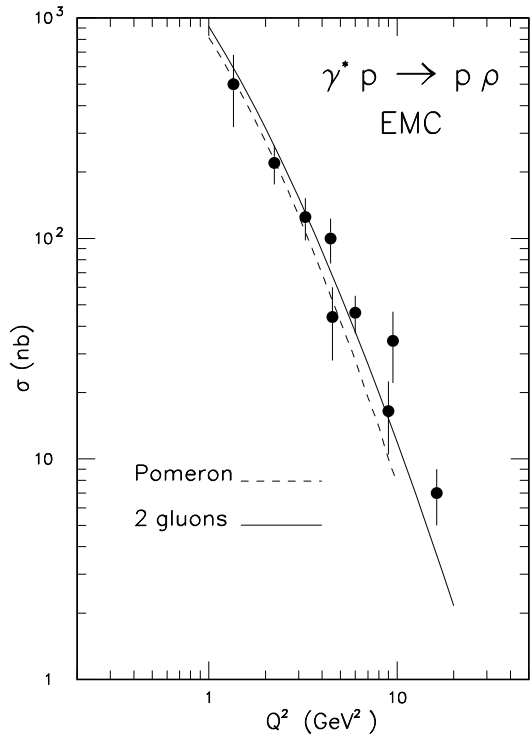,scale=0.4}
\end{minipage}
\begin{minipage}[t]{16.5 cm}
\caption{The variation with $Q^2$ of the total cross section of the  $p(\gamma^*,\rho^0)p$ reaction measured at CERN. The  dashed line shows the Pomeron exchange  prediction. The solid line shows the two gluon exchange prediction.
\label{rho_EMC}}
\end{minipage}
\end{center}
\end{figure}

The $Q^2$ dependency is already built in the two gluons exchange amplitude. Without any extra parameters, it fairly accounts for the  $\rho^0$ meson electro-production cross section determined at CERN energies by EMC collaboration~\cite{Aub85}, as can be seen in Figure~\ref{rho_EMC}. I refer the reader to~\cite{me95} for more details and references to experimental data.  

\begin{figure}[tbhp]
\begin{center}
\begin{minipage}[t]{8 cm}
\epsfysize=8.0cm
\epsfig{file=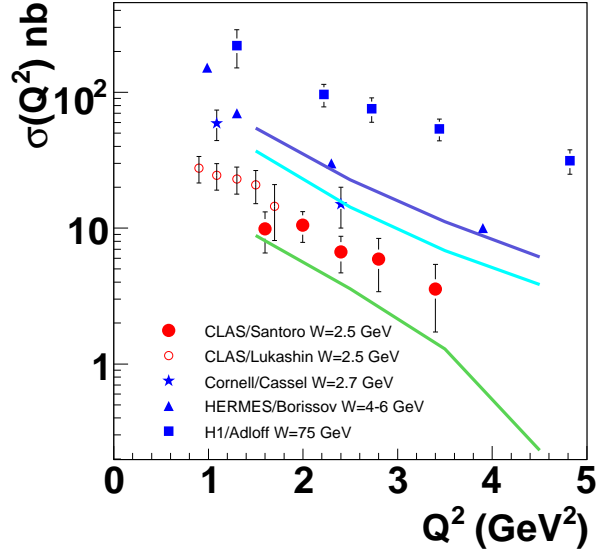,scale=0.4}
\end{minipage}
\begin{minipage}[t]{16.5 cm}
\caption{(Color on line) The variation with $Q^2$ of the total cross section of the  $p(\gamma^*,\phi)p$ reaction measured at JLab, CORNELL, HERMES and HERA. The curves are the prediction of the 2-gluon exchange model at $W=$  2.9, 2.45 and 2.1 GeV (top to bottom). See~\cite{san08} for references to the data.
\label{tot_santoro}}
\end{minipage}
\end{center}
\end{figure}

\begin{figure}[tbhp]
\begin{center}
\begin{minipage}[t]{8 cm}
\epsfysize=8.0cm
\epsfig{file=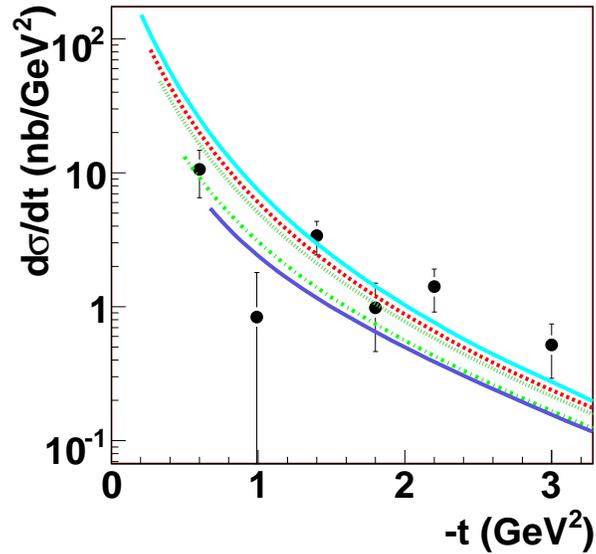,scale=0.4}
\end{minipage}
\begin{minipage}[t]{16.5 cm}
\caption{(Color on line) The variation with $-t$ of the differential cross section of the  $p(\gamma^*,\phi)p$ reaction measured at JLab. The curves are computed at $W=$ 2.5~GeV and $Q^2=$ 1.6, 2.1, 2.6, 3.8 and 5 GeV$^2$ (top to bottom).
\label{dsdt_santoro}}
\end{minipage}
\end{center}
\end{figure}

The latest version~\cite{cano} of the two gluon exchange model also leads to a fair description of the sparse available data in the $\phi$ meson electro-production. Figure~\ref{tot_santoro} shows the evolution of the total cross section with $Q^2$ in the HERA high energy range ($<W>=$ 90~GeV), the HERMES intermediate energy range ($<W>=$ 5 GeV ) and the JLab energy range~\cite{san08} close to threshold ($<W>=$ 2.4~GeV). Since the data correspond to a large bin in $W$, the curves correspond to the mean value and the two limits of the bin. The model reproduces also fairly well the differential cross section which has been measured at JLab (Figure~\ref{dsdt_santoro}). Again the curves correspond to the range of virtuality in the bin.  

Such an agreement is highly non trivial, since no extra parameters have been adjusted when entering the virtual photon sector. It thus confirms both the $-t$ and the $Q^2$ dynamics that are built into the $q\overline q$ quarks loop and the two gluons loop.

In contrast to the description of the $\phi$ production data, electromagnetic form factors have to be added to the Regge pole amplitudes that contribute significantly to the production of $\omega$ and $\rho$ at low energies.

\begin{figure}[tbhp]
\begin{center}
\begin{minipage}[t]{8 cm}
\hspace{-1 cm}
\epsfysize=8.0cm
\epsfig{file=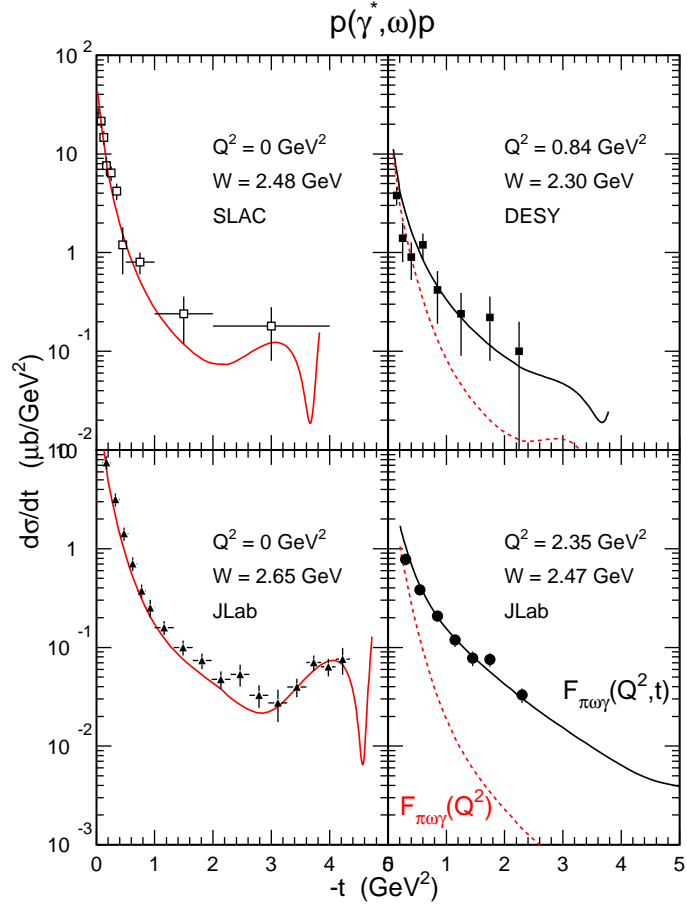,scale=0.5}
\end{minipage}
\begin{minipage}[t]{16.5 cm}
\caption{(Color on line) The differential cross section of the  $p(\gamma^*,\omega)p$ reaction at $W\sim$ 2.5 GeV. Left panels: real photons; the curves and the data are the same as in Figure~\ref{omega_JLab}. Right panels: virtual photons.  Dashed line: canonical pion electromagnetic form factor. Solid lines: $t$-dependent form factor. See~\cite{lag04} for references to the data.
\label{dsdt_ome_virt}}
\end{minipage}
\end{center}
\end{figure}

\begin{figure}[tbhp]
\begin{center}
\begin{minipage}[t]{8 cm}
\epsfysize=8.0cm
\epsfig{file=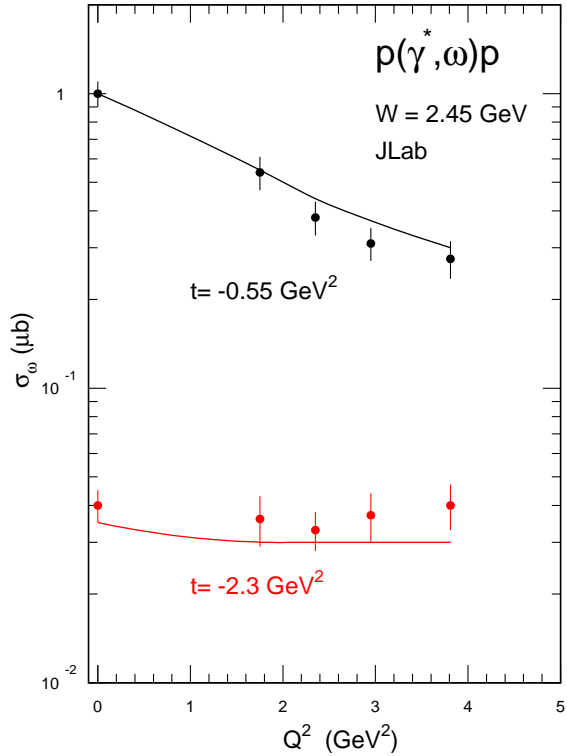,scale=0.4}
\end{minipage}
\begin{minipage}[t]{16.5 cm}
\caption{(Color on line) The $Q^2$ dependency of the differential cross section of the  $p(\gamma^*,\omega)p$ reaction at constant $-t$.
\label{q_dep_t_ome}}
\end{minipage}
\end{center}
\end{figure}

Figure~\ref{dsdt_ome_virt} summarizes the evolution of the differential cross section of the $p(\gamma^*,\omega)p$ reaction with the photon virtuality $Q^2$. When the canonical pion electromagnetic form factor
\begin{equation}
F_{\pi \omega \gamma}(Q^2)= \frac{1}{1+\frac{Q^2}{\Lambda_{0}^2}}
\end{equation}
with $\Lambda_{0}^2= 0.462$~GeV$^2$, is introduced at the $\pi \omega \gamma$ vertex of the $\pi$-exchange amplitude which reproduces real photon data (section~\ref{sec:vector}) in the left panels, one obtains the dashed curves in the right panels. They reproduce the evolution of the virtual photon cross section at low momentum transfer, but fall short by more than an order of magnitude at large momentum transfer. Here, the  experimental cross sections~\cite{mor05} are nearly independent of  the virtuality $Q^2$ of the photon, as demonstrated in Figure~\ref{q_dep_t_ome}. This points towards a coupling to a more compact object: The agreement is restored (solid curves) when a t-dependence on is given to the pion form factor~\cite{lag04}. It is natural to relate it to the way the pion saturating Regge trajectory, $\alpha_{\pi}(t)$~\cite{gui97}, approaches its asymptote $-1$:
\begin{equation}
F_{\pi \omega \gamma}(Q^2,t)= \frac{1}{1+\frac{Q^2}{\Lambda_{\pi}^2(t)}}
\label{ff1}
\end{equation}
with
\begin{equation}
\Lambda_{\pi}^2(t)= \Lambda_{0}^2 \times \left(\frac{1+\alpha_{\pi}(0)}{1+\alpha_{\pi}(t)}\right)^2
\label{ff2}
\end{equation}

When $t\rightarrow -\infty$, $\alpha_{\pi}(t)\rightarrow -1$, and $F_{\pi \omega \gamma}(Q^2,t)$ becomes independent of $Q^2$ at large $-t$.

Such an ansatz links the evolution of $F_{\pi \omega \gamma}$, from the coupling to a nearly on-shell pion toward the coupling to a point-like far off-shell constituent, with the saturation of the pion Regge trajectory which is already at work in the amplitude at large $-t$ (section~\ref{sec:vector}). While such a t-dependent form factor is given to us by the experiment, it provides us with a quantity which can be used in related channels; however it remains  to be explained in a more fundamental theory. 

The use of the same $-t$ dependency in the $\gamma \sigma \rho$ electromagnetic form factor leads to a fair accounting of the  $p(\gamma^*,\rho)p$ reaction cross section that have been determined in a large kinematics range at JLab. I refer to the corresponding experimental paper~\cite{mor09} for a detailed comparison between the data and the model.

\subsection{\it Charged pion electro-production \label{sec:virt_charged_pion}}

\begin{figure}[tbhp]
\begin{center}
\begin{minipage}[t]{8 cm}
\hspace{-0.9 cm}
\epsfysize=8.0cm
\epsfig{file=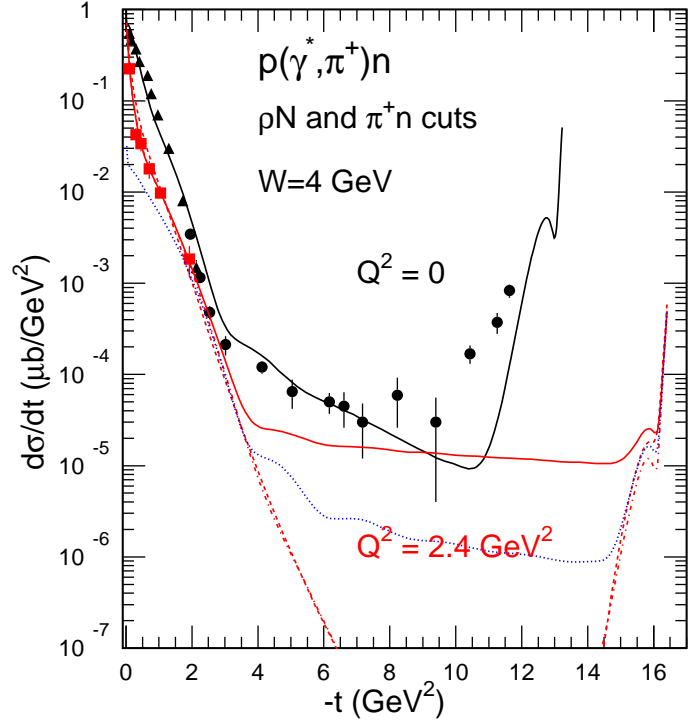,scale=0.5}
\end{minipage}
\begin{minipage}[t]{16.5 cm}
\caption{(Color online) The differential cross section of the  $p(\gamma^*,\pi^+)n$  reaction when  $W=$ 4 GeV. Real photon point: black data and curve as in Figure~\ref{charged_pion_photo}. Virtual photon  in the HERMES~\cite{air08} kinematics: $Q^2=$ 2.4 GeV$^2$. Experiment: red squares. Dashed line: t- and u-channel Regge poles. Dot-dashed line: $\pi^+$n elastic cut included. Solid line: $\rho^0$p inelastic cut included. Dotted line: Transverse cross section only.
\label{piplus_hermes}}
\end{minipage}
\end{center}
\end{figure}

The analysis of charged pion electro-production needs also a t-dependent form factor (equations~\ref{ff1} and~\ref{ff2}) in the Regge pole amplitudes as well as in the $\rho^0$p rescattering amplitude. Figure~\ref{piplus_hermes} compares the Hermes data~\cite{air08}, at $W=$ 4 GeV and $Q^2=$ 2.4~GeV$^2$, to such a prediction as well as to the real photon data (from figure~\ref{charged_pion_photo}). While the Transverse part (dotted line) is strongly suppressed at the lowest $-t$, as expected from the behavior of the canonical electromagnetic form factor of the pion, it becomes comparable to the real photon cross section at $-t\sim$ 2~GeV$^2$, as predicted by the use of the t-dependent form factor. In addition, the pion pole dominates the Longitudinal cross section at the lowest $-t$. The extension of such a measurement in the entire $-t$ range would be worthwhile at JLab12 in order to check the contribution of the unitarity cuts in the virtual photon sector. (See Appendix~B for the definition of the Transverse and the Longitudinal cross sections).

Such a systematic study has been carried out at lower energies (JLab6, $W\sim$ 2.3 GeV). The data~\cite{par13} confirm the need of t-dependent form factors. I refer to this experimental paper for a comprehensive comparison between the model and the data. 

%The predictions~\cite{van98} (known as VGL) of the  model~\cite{gui97}, without cuts and without $-t$ dependency of the electromagnetic form factors, satisfactorily agree with the data recorded before the  JLab-HERMES era at low $-t$ and low $Q^2$.  

An interesting alternative approach~\cite{kas11} leads also to a fair agreement with the world experimental data at low $-t$. It takes into account baryonic resonances in the s-channel, or u-channel, amplitudes that are   necessary to restore the gauge invariance of the pion pole. It uses a dual connection between the exclusive hadronic form factors and inclusive deep inelastic structure functions. To what extent this approach is suited to larger $-t$ is an open problem.

\subsection{\it Neutral pion electro-production \label{sec:virt_neutral_pion}}

\begin{figure}[tbhp]
\begin{center}
\begin{minipage}[t]{8 cm}
\hspace{-1.5 cm}
\epsfysize=8.0cm
\epsfig{file=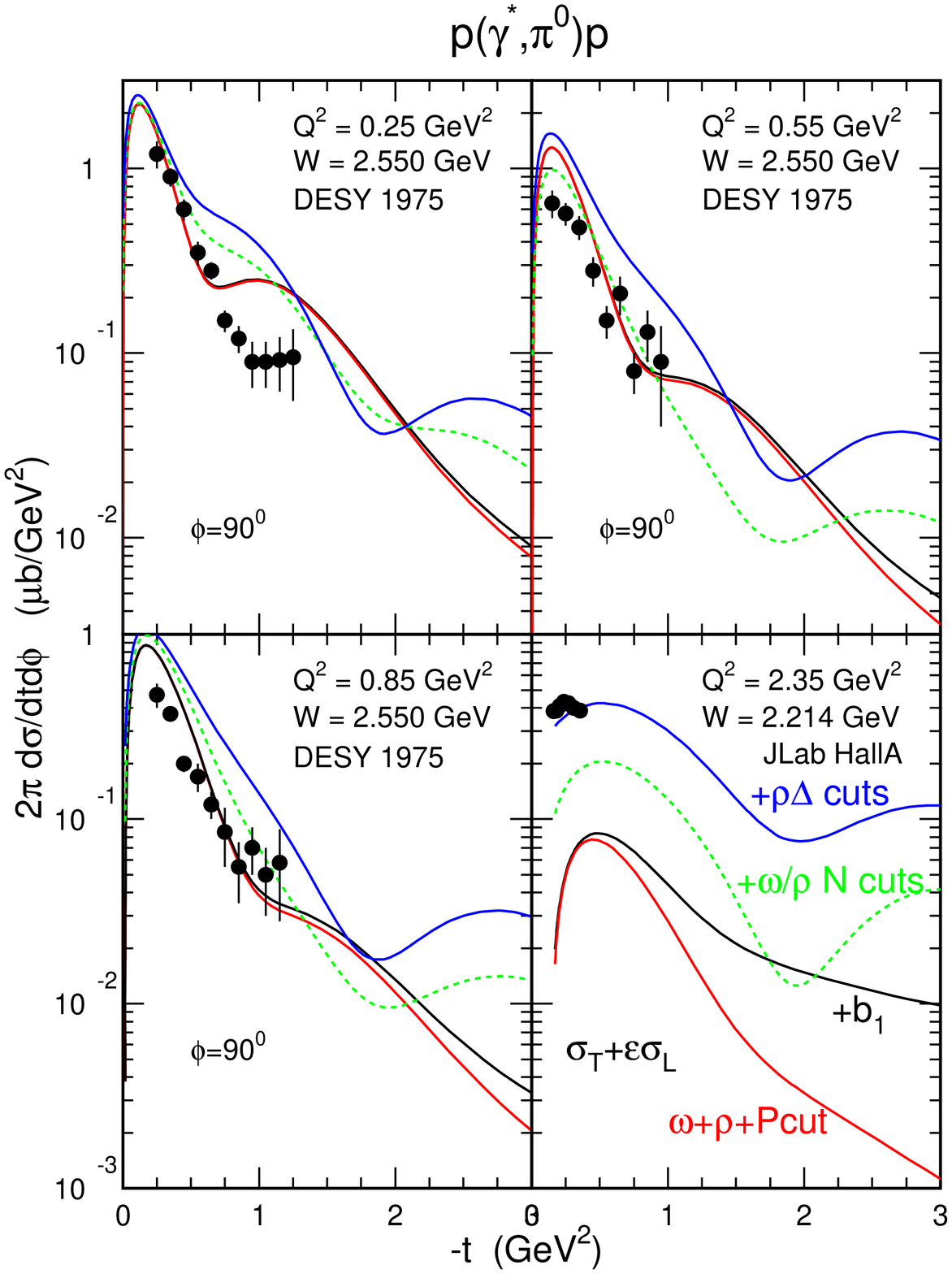,scale=0.6}
\end{minipage}
\begin{minipage}[t]{16.5 cm}
\caption{(Color online) The differential cross section of the $p(\gamma^*,\pi^0)p$ reaction measured at DESY~\cite{Bra78}, top row and bottom left, and JLab~\cite{Cam10}, bottom right. The basic Regge Pole model with the Pomeron cut corresponds to the red (without $b_1$ pole) and the black (with $b_1$ pole)  solid curves. The dashed curves (green) in addition take into account the contribution of the $\pi$  charge exchange scattering cuts and the inelastic $\omega p$ and $\rho^+ n$ cuts. The (blue) full line curves take also into account the contribution of the inelastic $\rho^{\pm} \Delta$ cuts. 
\label{pizero_hallA}}
\end{minipage}
\end{center}
\end{figure}

When the virtuality $Q^2$ of the photon increases, two things happen: Firstly, the dip, around $-t=$ 0.5~GeV$^2$ in the real photon differential cross section (see \textit{e.g.} Figure~\ref{neutral_pion_photo}), disappears already at low virtuality: this cannot be explained by simply supplementing the real photon amplitudes with a single form factor. The problem has been solved~\cite{lag11} by giving a different $Q^2$ dependency to the form factor attached to the $\omega$ Regge pole amplitude and the form factor attached to the elastic unitarity cut. As can be seen in Figure~\ref{pizero_hallA}, such a procedure allows to move the node to higher values of $-t$ and to achieve a good understanding of the DESY data~\cite{Bra78} at low virtuality of the photon (red and black solid lines).

\begin{figure}[tbhp]
\begin{center}
\begin{minipage}[t]{8 cm}
\hspace{-2.5 cm}
\epsfysize=8.0cm
\epsfig{file=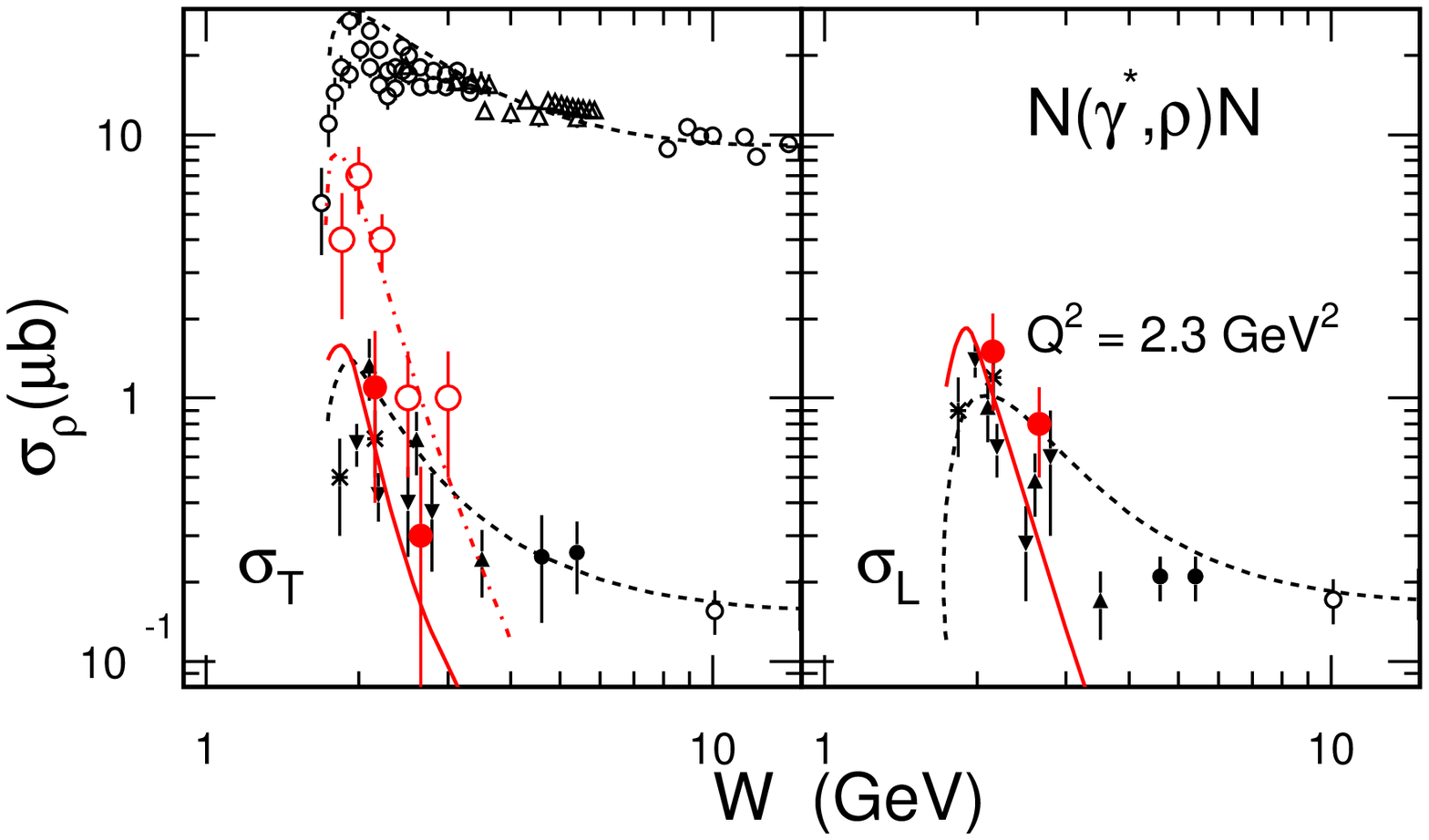,scale=0.7}
\end{minipage}
\begin{minipage}[t]{16.5 cm}
\caption{(Color on line) The comparison between the total cross sections of the $p(\gamma,\rho^0)p$ (black) and $p(\gamma,\rho^+)n$ (red) reactions at $Q^2=$ 0 (top) and $Q^2=$ 2.3~GeV$^2$ (bottom). The open red circles and the dash-dotted red curve correspond to the photo production of $\rho^-$, while the filled red circles and the solid red curves correspond to the electro-production of $\rho^+$. The left panel shows the Transverse cross section, while the right panel shows the Longitudinal one. The references to experiments are given in~\cite{lag11}. 
\label{charged_rho_xsection}}
\end{minipage}
\end{center}
\end{figure}

 Secondly, this pole model misses, by about an order of magnitude, the data~\cite{Cam10} measured at JLab (Hall A) around $Q^2=$ 2.3~GeV$^2$ (bottom right of Figure~\ref{pizero_hallA}). It turns out that the inelastic unitarity cuts associated with charged $\rho$ meson propagation dominate the cross section here and lead to a good understanding of the data. The reason is that the cross section of the electro-production of charged rho mesons, which corresponds to only 10\% of the cross section of electro-production of neutral rho mesons at the real photon point, becomes comparable at $Q^2\sim$ 2.3~GeV$^2$, as can be seen in Figure~\ref {charged_rho_xsection}.

\begin{figure}[tbhp]
\begin{center}
\begin{minipage}[t]{8 cm}
\hspace{-3.2 cm}
\epsfysize=8.0cm
\epsfig{file=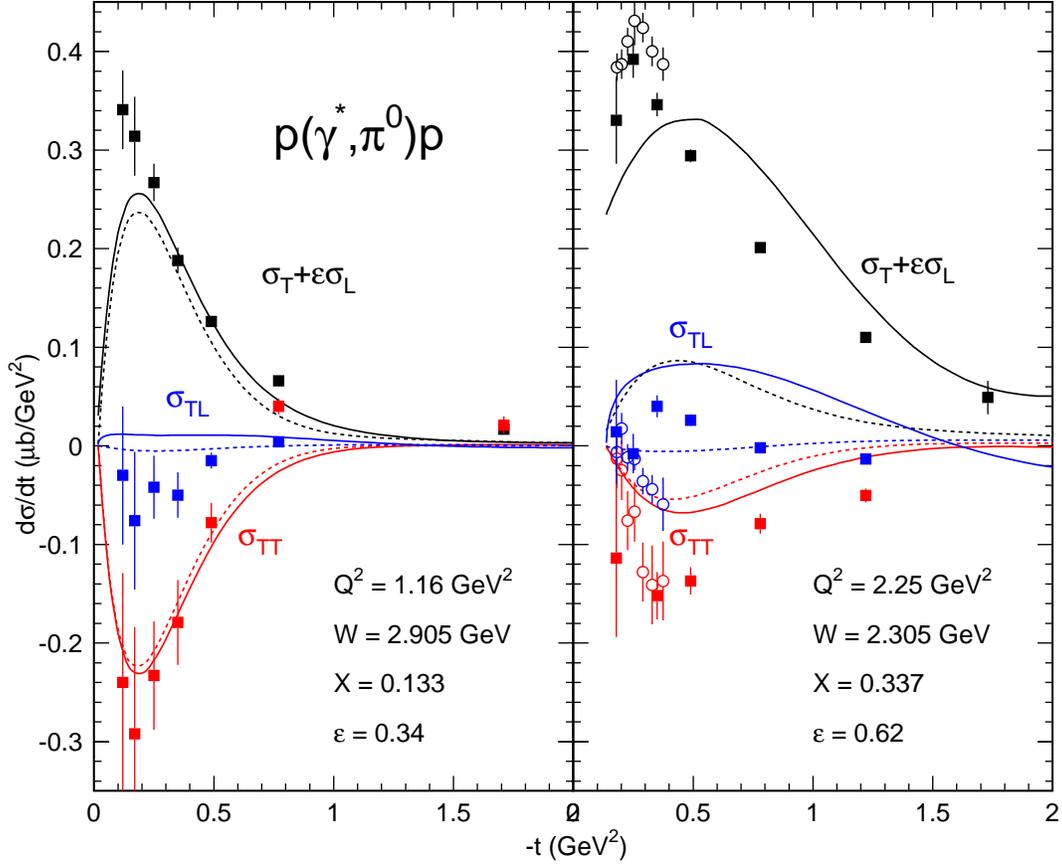,scale=0.5}
\end{minipage}
\begin{minipage}[t]{16.5 cm}
\caption{(Color on line) The differential cross section of the $p(\gamma^*,\pi^0)p$ reaction measured at JLab by the CLAS collaboration~\cite{bed14} (full squares) and Hall A collaboration~\cite{Cam10} (open circles). Black curves and points: unpolarized cross sections. Red curves and points: Transverse-Tranverse cross sections. Blue curves and points: Transverse-Longitudinal Cross sections. Solid lines: full model with cuts. Dashed line: without inelastic cuts.
\label{CLAS_pizero_elec}}
\end{minipage}
\end{center}
\end{figure}

The model reproduces also fairly well the data recorded at JLab (Hall B), over an extended range  in $Q^2$, $W$ and $-t$. I refer the reader to Figure~20 of the experimental paper~\cite{bed14}, for a comprehensive comparison of the model with the data. Figure~\ref{CLAS_pizero_elec} emphasizes two kinematics: lowest virtuality $Q^2= 1.16$~GeV$^2$, but highest energy $W=2.9$~GeV; central virtuality $Q^2= 2.25$~GeV$^2$, and energy $W=2.3$~GeV. This latter kinematics is very close to the JLab (Hall A) kinematics~\cite{Cam10} in Figure~\ref{pizero_hallA}. The definition of the unpolarized ($\sigma_L +\epsilon \sigma_T$) and interference cross sections ($\sigma_{TT}$ and $\sigma_{TL}$) are given in Appendix B. As far as the unpolarized and Transverse-Tranverse cross sections are concerned the two experiments got similar results. The slight difference (about 20\%) between the unpolarized cross sections follows the expected energy variation of the theoretical cross section: compare the curves in Figure~\ref{pizero_hallA} (bottom right) and Figure~\ref{CLAS_pizero_elec} (right) which have been computed for the corresponding kinematics (as summarized in the body of the Figures). It turns out that the Transverse-Longitudinal experimental cross sections are comparable in magnitude but opposite in sign. From the experimental papers it is not possible to understand the origin of this inconsistency.  

Since the modeling of the elementary amplitudes of the charged vector meson production and the $\rho$ to $\pi$ Charge Exchange reaction are constrained by independent experiments, the contribution of the $\rho^+ n$  inelastic cut is on solid ground. Due to the lack of dedicated experiments, the contributions of the $\rho^+ \Delta^0$ and $\rho^- \Delta^{++}$ cuts are less constrained, but are based on reasonable assumptions.

Since the charged vector meson production amplitude is driven by the exchange of Reggeons, the contribution of the corresponding cuts is expected to decrease when the total energy $W$ increases, in contrast with the cuts associated with the propagation of the neutral vector meson  which survive at high energies, since the production amplitude is driven by the exchange of the Pomeron (see \textit{e.g.} Figure~\ref{vector_cross}).

The extension of the measurement to higher energies (at JLab12, for instance) would allow to check these conjectures. 

\subsection{\it Virtual Compton scattering \label{sec:virt_compton}}

\begin{figure}[tbhp]
\begin{center}
\begin{minipage}[t]{8 cm}
\hspace{-4.25 cm}
\epsfysize=8.0cm
\epsfig{file=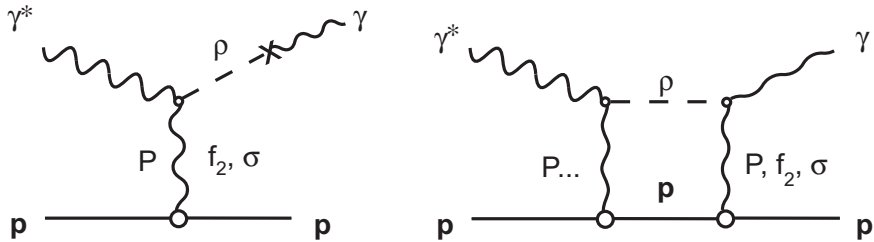,scale=0.6}
\end{minipage}

\begin{minipage}[t]{16.5 cm}
\caption{The relevant graphs in the $\gamma{^\star} p \rightarrow p \gamma$ reaction. Left: Poles + direct conversion. Right: $\rho$-nucleon unitarity cut. 
\label{dvcs_graphs}}
\end{minipage}
\end{center}
\end{figure}

Deeply Virtual Compton Scattering (DVCS) is advocated as the flagship experiment to determine the Generalized Parton Distributions (GPD) of the nucleon. At sufficiently high virtuality $Q^2$, there is no doubt~\cite{Co97,Co99} that the DVCS amplitude factorizes into the perturbative QCD coupling of the photon to a current quark and the GPD. However, unitarity rescattering cuts (Figure~\ref{dvcs_graphs}) contribute significantly~\cite{lag07} in the range of energy and momentum transfer that has been accessible in current experiments.

\begin{figure}[tbhp]
\begin{center}
\begin{minipage}[t]{8 cm}
\hspace{-2 cm}
\epsfysize=8.0cm
\epsfig{file=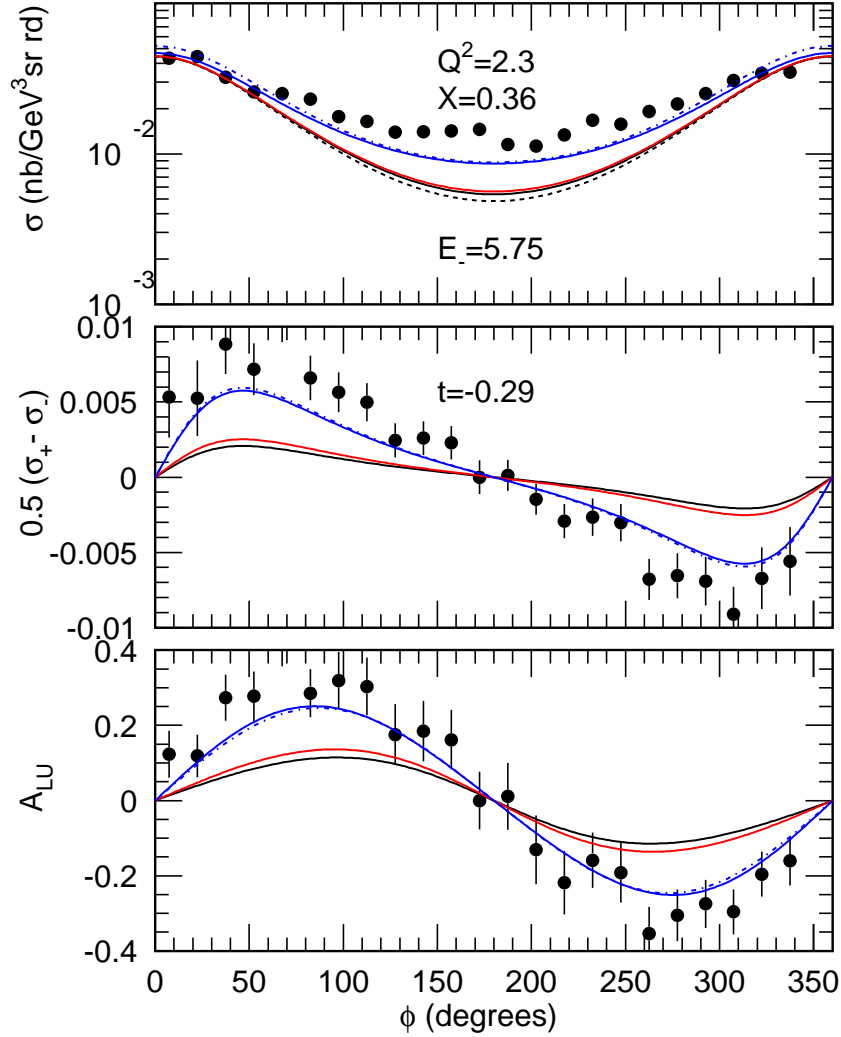,scale=0.6}
\end{minipage}
\begin{minipage}[t]{16.5 cm}
\caption{ (Color on line) The JLab Hall A DVCS cross sections and beam asymmetry are plotted against the azimuthal angle of the outgoing real photon. Top panel: unpolarized cross section, $d\sigma/dE_ed\Omega_edtd\phi$. Middle panel: Difference between the electron helicity dependent cross sections. Bottom panel: beam spin asymmetry. Black dashed curve: Bethe-Heitler contribution. Black solid curves: Regge pole contributions included. Red solid curves: coupling to $\rho$-nucleon channel included. Blue solid curves: coupling to diffractively produced intermediate states included. Blue dash-dotted: principal part of the rescattering integral included.
\label{dvcs_phidep_hallA}}
\end{minipage}
\end{center}
\end{figure}

When added to the pure QED Bethe-Heitler amplitude they reproduce the pioneering experimental results of JLab~\cite{def14}, as can be seen in Figure~\ref{dvcs_phidep_hallA}, and Hermes\cite{Ai01,Ai07} (see Figure~7 in~\cite{lag07}).

The coupling with vector meson production channels represents the major part of the DVCS amplitude at reasonably low $Q^2$ (up to about $3$~GeV$^2$). The coupling to the elastic $\rho p$ channel is on solid ground, since it relies on on-shell matrix elements that are driven by  measured $\rho$ photo- and electro-production cross sections. It provides us with a reliable lower limit of the DVCS cross sections and observables. The coupling with the inelastic channels is constrained by unitarity and leads to a fair account of all the DVCS observables. This conjecture has to be kept in mind in any attempt to access and determine the GPD at the low virtuality $Q^2$ currently available at JLab or Hermes. The available energy is too low ($\sqrt{s}<$ 2.5~GeV at JLab) to sum over the complete basis of hadronic intermediate states, and safely rely on a dual description at the quark level. 

At higher $Q^2$, the available energy increases (at fixed Bj\"orken-$x$) and a partonic description may become more justified: Above which $Q^2$ is an open issue. However, the contribution of the unitarity cut associated with the propagation of $\rho^0$ meson will survive even at the highest energies, since its production cross section not only dominates over the other channels, but also stays constant (see Figure~\ref {vector_cross}).

Any sensible program, that is aimed at determining the GPD with the DVCS reaction, cannot ignore the very relevant contribution of the various hadronic unitarity cuts and should include a concurrent determination of the transverse and longitudinal cross sections of the elastic and inelastic production of vector mesons.

\section{Players in the game \label{sec:players}}

Let us now go back to each of the players in the game that we have met in the preceding sections.

\subsection{\it Regge Poles in the $t$-channel \label{sec:t-channel}}

At forward angles the reaction amplitude is driven by Feynman graphs in which the lightest mesons are exchanged according to the spin-isospin structure of the vertices. While this description has proven to be efficient at low energies (see {\it e.g.}~\cite{physrep}), it fails at high energies since the Feynman amplitude increases as $s^{J}$, being $J$ the spin of the exchanged particle, due to the derivative nature of the vertices. Not only this is not supported by experiments, but also this does not satisfy unitarity.

\begin{figure}[tbhp]
\begin{center}
\begin{minipage}[t]{8 cm}
\hspace{-1.5 cm}
\epsfysize=8.0cm
\epsfig{file=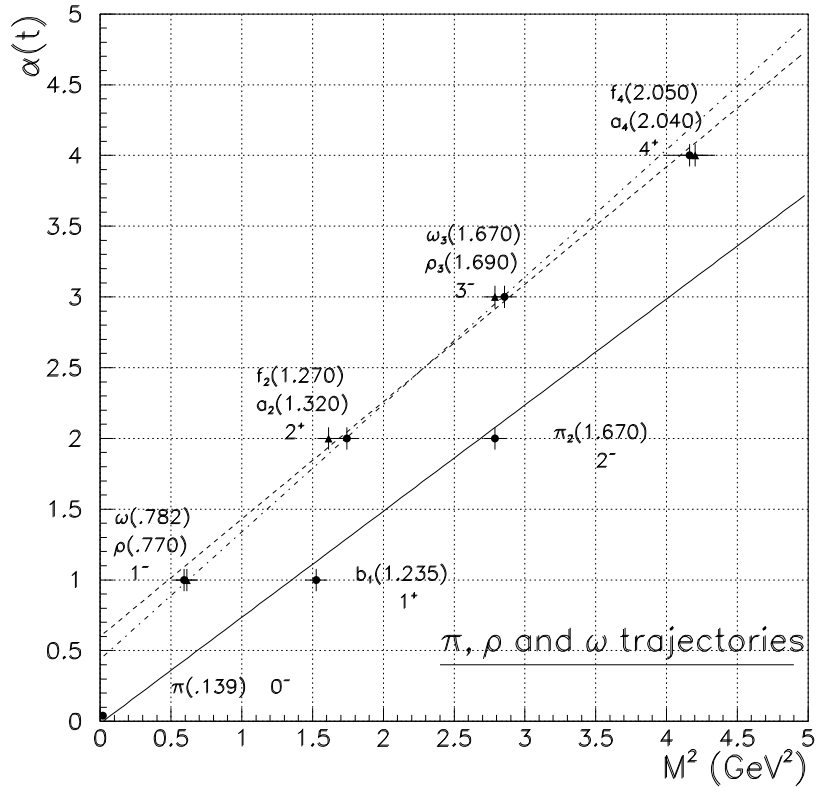,scale=0.6}
\end{minipage}

\begin{minipage}[t]{16.5 cm}
\caption{The Regge trajectories associated with the $\pi$ (solid line), the $\rho$ (dashed line) and the $\omega$ (dot-dashed line) families. 
\label{lin_traj}}
\end{minipage}
\end{center}
\end{figure}

Regge poles allow to reconcile particle exchanges and the correct energy dependency of the corresponding amplitudes which respects unitarity. In the late fifties early sixties, on the basis of the analytic and unitarity properties of the S-matix and in the limit of high energies, T.~Regge~\cite{reg59, reg60}  showed that the reaction amplitude is driven by a few poles, and a few cuts, in the complex angular momentum plane. Each of these Regge poles is related to a family of hadrons which have the same quantum numbers but differ by their spin. In the time-like region ($t >$~0) each family lies on a linear Regge trajectory $\alpha(t)$ which relates their spin $J$ to the square of their mass (which coincides with $t$):
\begin{equation} 
\alpha(t)=J= \alpha_0 +\alpha' t
\label{lin_traj_eq}
\end{equation}
where the slope $\alpha'$ is close to $0.9$~GeV$^2$ for all trajectories and the intercept $\alpha_0$ is specific to each trajectory, as can be seen in Figure~\ref{lin_traj}.

The Regge amplitude takes the generic form:
\begin{equation}
{\cal A}_{Regge}=\beta(t,J_0) {\cal P}_{Regge}
\label{Regge_generic_amp}
\end{equation}
where $\beta(t,J_0)$ stands for the vertices spin-momentum dependent coupling with the exchanged particle with the lightest mass in the trajectory, and spin $J_0$, while ${\cal P}_{Regge}$ is the Regge pole:
\begin{equation}
{\cal P}_{Regge}= \left(\frac{s}{s_0}\right)^{\alpha(t)-J_0}
\frac{\pi \alpha'}{\sin(\pi \alpha(t))}
\frac{{\cal S}+e^{-i\pi\alpha(t)}}{2}
\frac{1}{\Gamma(1-J_0+\alpha(t))}
\label{Regge_pole}
\end{equation}

In this expression, $s_0$ is a mass scale which is conventionally taken  as $s_0=1$~GeV$^2$ in all channels, except in the Pomeron exchange channel (see section~\ref{sec:pomeron-2g}), and the gamma function $\Gamma$ suppresses the poles in the space-like scattering region  ($t <0$). The Regge propagator reduces to the Feynman propagator ($1/(t-m^2)$) when one approaches the first pole on the trajectoriy  (i.e. $t\rightarrow m^2$). This means that farther one goes from the pole, the more the result of the Regge model will differ from conventional Feynman diagram based models. On the one hand, the term $s^{-J_0}$ in the propagator compensates the term $s^{J_0}$ which drives the vertex amplitude at high energy, and the reaction amplitude~(\ref{Regge_generic_amp}) behaves as $s^{\alpha(t)}$. Since the intercept $\alpha_0$ of all hadron trajectories is lower than unity, the Regge cross sections decrease with energy and do not violate the unitarity bound. On the other hand, the extrapolation of the linear trajectories in the space-like region allows to reproduce the characteristic exponential decrease of the experimental cross sections up to $-t\sim 2$~Gev$^2$ (above the trajectories may become non-linear as we shall see in section~\ref{sec:sat-vs-cuts}).

In the late sixties early seventies, this scheme was used to fit the helicity amplitudes of the reactions available in the few GeV energy range. On the contrary, in~\cite{gui97} (and subsequent works) we started from the Feynman amplitudes (with the full spin and momentum dependencies) where we replaced the Feynman propagator by the corresponding Regge propagator. Therefore the predictions of the Regge model are basically parameter free, since the structure of the matrix elements relies on the the well known symmetries and conservation laws, and since the various coupling constants have been independently determined in the numerous and precise studies and analyses, in the resonance region for instance. 

In addition two subtle improvements have been documented in~\cite{gui97}: i) the use of degenerated trajectories with either constant phase or rotating phase; ii) the implementation of gauge invariance in $\pi^{\pm}$ production channels.

The signature ${\cal S}= \pm 1$ insures that each trajectory connects particles with even or odd spin. In the space-like region, it induces zeros in the amplitude which coincide with the experimental nodes in certain channels (for instance, in the $\pi^0$ photo-production channel), but which are absent in other channels (for instance the $\pi^{\pm}$ photo-production channels). It turns out that the trajectories with opposite signature go by pair. In Figure~\ref{lin_traj} the $\pi$ (even spins) and the $b_1$ (odd spins) trajectories are almost the same. This is also the case for the $\rho$ (odd spins) and the $a_2$ (even spins) trajectories. Adding or subtracting the corresponding Regge amplitudes gets rid of the zeros in the space-like region, and leads to degenerate trajectories with either a rotating ($e^{-i\pi\alpha(t)}$) or a constant ($1$) phase. The actual degeneracy scheme is driven~\cite{gui97} by  the $G$~parity of the photon-meson coupling. For instance, the $\pi^{\pm}$ production cross sections read, in a schematic notation:
\begin{eqnarray}
\label{eq:pip}
&&\frac{d\sigma}{dt}(\gamma p\to\pi^+ n)\propto
\mid(\pi+{b_1})+(\rho+a_2)\mid^2 \;,\\
&&\frac{d\sigma}{dt}(\gamma n\to\pi^- p)\propto
\mid-(\pi-{b_1})+(\rho-a_2)\mid^2 \;,
\label{eq:pim}
\end{eqnarray}
More explicitly, Eqs.(\ref{eq:pip}) and (\ref{eq:pim}) are written as~:
\begin{eqnarray}
\label{eq:pip-phase}
&&\frac{d\sigma}{dt}(\gamma p\to\pi^+ n)\propto
\mid \mathcal{M}_{\pi}\,e^{-i\pi\alpha_\pi(t)}+
\mathcal{M}_{\rho}\,e^{-i\pi\alpha_\rho(t)}\mid^2 \;,\\
&&\frac{d\sigma}{dt}(\gamma n\to\pi^- p)\propto
\mid -\mathcal{M}_{\pi}\cdot 1+\mathcal{M}_{\rho}\cdot 1\mid^2 \;,
\label{eq:pin-phase}
\end{eqnarray}
where $\mathcal{M}_{\pi}$ and $\mathcal{M}_{\rho}$ are the 
$\pi^+$ production amplitudes (without the Regge phase) 
corresponding with the exchange of 
a $\pi$ or $\rho$ trajectory respectively. 
$G$-parity considerations thus lead to a rotating phase for the 
$\pi$ and $\rho$ trajectories in the \rpip process and to 
a constant phase in the \rpim process. 

It turns out that the $t$-channel $\pi$ exchange diagram alone does not satisfy electromagnetic gauge invariance by itself. In~\cite{gui97}, the current conservation was restored in a minimal way by adding the electric part of the $s$-channel proton exchange ($\pi^+$ production) or $u$-channel proton exchange ($\pi^-$ production). This leads to the following gauge invariant reggeized $\pi$-exchange  current operators :
\beqn
J_{\pi}^{\mu}(\gamma p \to n\pi^{+}) \to && 
-i\{\sqrt{2}\}e\frac{f_{\pi NN}}{m_{\pi}} \;
\nonumber\\
&&\cdot\bar{N}_{f}[(q-p_{\pi})^{\mu}
\not{q}\gamma^{5}-2m_N (t-m^2_\pi)\gamma^5
\frac{(\not p_s+m_N)}{s-m^2_N}\gamma^\mu]N_{i}
{\mathcal{P}^\pi_{Regge}}\;, \nonumber\\
\label{mivt2}\\
J_{\pi}^{\mu}(\gamma n \to p\pi^{-}) \to && 
i\{\sqrt{2}\}e\frac{f_{\pi NN}}{m_{\pi}}
\nonumber\\
&&\cdot\bar{N}_{f}[(q-p_{\pi})^{\mu}
\not{q}\gamma^{5}+2m_N (t-m^2_\pi) \gamma^\mu
\frac{(\not p_u+m_N)}{u-m^2_N}\gamma^5]N_{i}
{\mathcal{P}^\pi_{Regge}}\;. \nonumber\\
\label{mivt3}
\eeqn
The second term in Eq.(\ref{mivt2}) (resp. Eq.({\ref{mivt3}})) can be interpreted as the electric part of the $s$-channel (resp. $u$-channel) nucleon exchange contribution when using Pseudo-Scalar $\pi NN$ coupling. We have kept for the $s$-channel diagram the minimum necessary term to restore gauge invariant, i.e. the electric part coupling. This interpretation does of course not imply that one reggeizes all $s$-channel (resp. $u$-channel) exchanges in addition to $t$-channel exchanges which would violate duality. Being a Born term, this particular $s$-channel ($u$-channel) ``diagram'' is an indissociable part of the $t$-channel $\pi$-exchange and is imposed by charge conservation. Indeed a gauge can be found \cite{jones80} where the contribution from the $\pi$-exchange $t$-channel diagram vanishes and where consequently the nucleon pole graph produces a $\pi$-pole in the cross section! It is therefore a natural choice to reggeize it by multiplying with ${\mathcal{P}^{\pi}_{Regge}}\,.\,{(t-m_\pi^2)}$ as this is exactly the same factor which enters when reggeizing the $t$-channel pion exchange diagram. 

This implementation of gauge invariance is one of the two new features of ref.~\cite{gui97} compared to various previous works on pion photo-production. The introduction of the $s$ ($u$) -channel diagram has an immediate effect on the forward differential cross-section of the reaction \rpip (\rpim)~: it creates a sharp rise at very forward angles, in lieu of the node at $t=0$ of the $\pi$-exchange diagram alone. Together with the use of degenerate trajectories whose the phases are determined by Eqs.~\ref{eq:pip-phase} and~\ref{eq:pin-phase}, it is the key to the success of the Regge model to  account for the ratio of the $\pi^+$ and $\pi^-$ cross sections (Figure~10 of~\cite{gui97}), as well as the various spin observables.

\begin{figure}[tbhp]
\begin{center}
\begin{minipage}[t]{8 cm}
\hspace{-0.75 cm}
\epsfysize=8.0cm
\epsfig{file=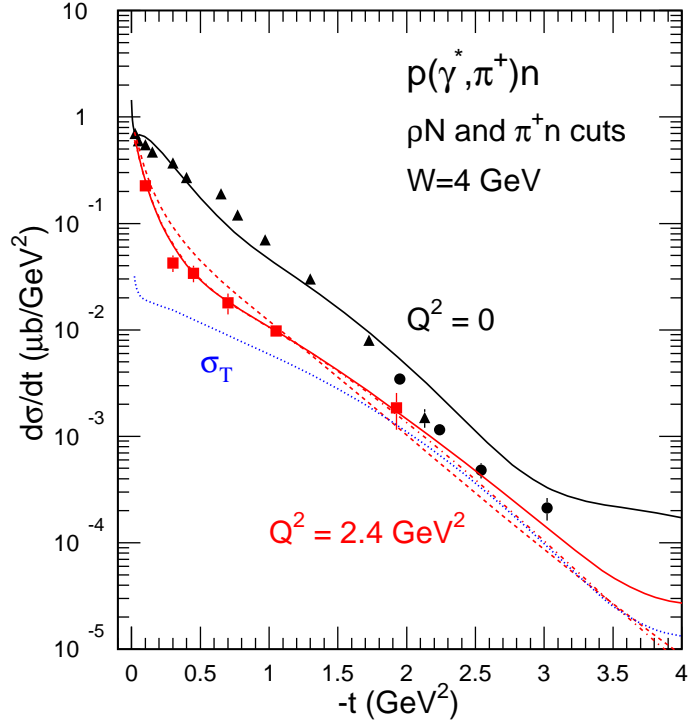,scale=0.5}
\end{minipage}

\begin{minipage}[t]{16.5 cm}
\caption{ (Color on line) The differential cross section of the $\pi^+$ electro-production at forward angle. Real photon point: black data and curve as in Figure~\ref{charged_pion_photo}. Virtual photon  in the HERMES~\cite{air08} kinematics: $Q^2=$ 2.4 GeV$^2$. Experiment: red squares. Red dashed line: t- and u-channel Regge poles. Red dot-dashed line: $\pi^+$n elastic cut included. Red solid line: $\rho^0$p inelastic cut included. Blue dotted line (labeled $\sigma_T$) : Transverse cross section only. 
\label{zoom_hermes}}
\end{minipage}
\end{center}
\end{figure}

In the virtual photon sector, the Longitudinal pion pole amplitude alone does not vanish at the most forward angles and dominates over the Transverse one. As an example, Figure~\ref{zoom_hermes}  is a zoom of Figure~\ref{piplus_hermes}. The transverse cross section (dotted line) exhibits the same behavior as the real photon cross section; it is simply smaller consistently with the $Q^2$ behavior of the electromagnetic form factor. The pion pole drives the shape of the Longitudinal cross section (dashed line). The cuts (see next sections) bring the model in almost perfect agreement (full line) with the Hermes data. As reminded in Appendix B,  gauge invariance relates the Coulomb and the third component of the spatial part of the current. The Longitudinal cross section is proportional to the square of the third component of the current, while the Transverse cross section reduces to the real photon cross section at the real photon point. 

The vector meson exchange amplitudes are gauge invariant by themselves. Their expressions are given in~\cite{gui97}. In the $\pi^0$ production channel the choice of a non degenerate trajectory for the the $\omega$ exchange naturally reproduces the node in the cross section around $-t=0.5$~GeV$^2$. However, as we shall see in section~\ref{sec:el-cuts}, this node can also be obtained from the interference between a degenerate trajectory and the $\pi^0$ elastic rescattering cut. 

I refer the reader to~\cite{gui97} for a more complete description of the Regge amplitudes and the actual values of the coupling constants, as well as a comprehensive comparison of the model with available experiments. The expressions and the coupling constants of $\pi$, $f_2$ and $\sigma$ meson exchange amplitudes in the vector meson production channels are given in~\cite{jml} and~\cite{cano}.

\subsection{\it Regge Poles in the $u$-channel \label{sec:u-channel}}

\begin{figure}[tbhp]
\begin{center}
\begin{minipage}[t]{8 cm}
\hspace{-1.75 cm}
\epsfysize=8.0cm
\epsfig{file=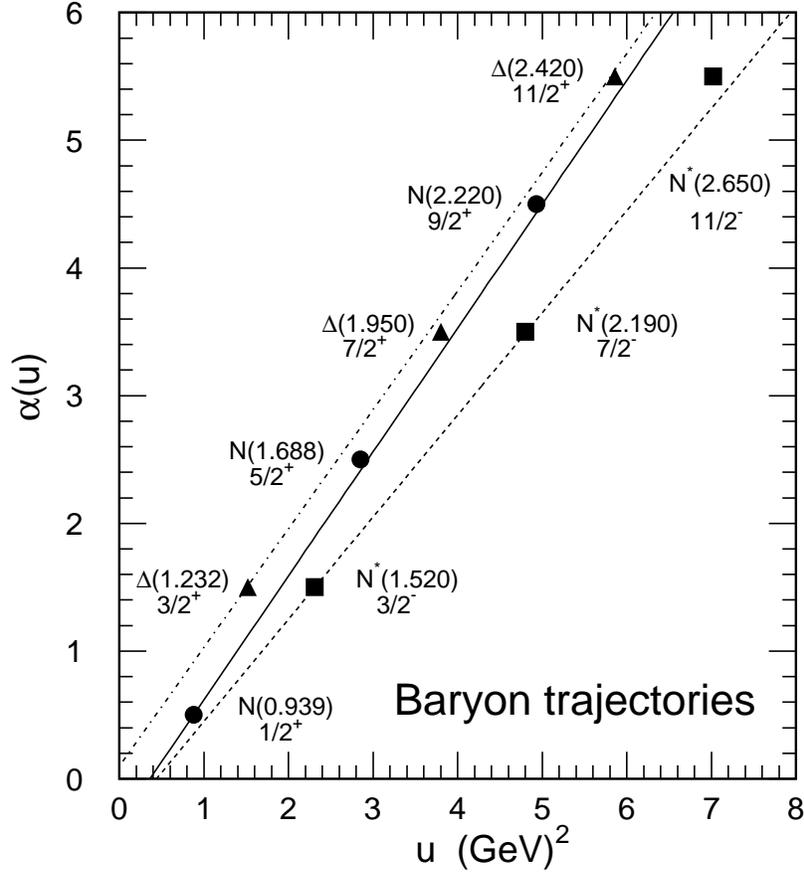,scale=0.6}
\end{minipage}

\begin{minipage}[t]{16.5 cm}
\caption{The Regge trajectories associated with the $N$ (solid line and circles), the $D_{13}$ (dashed line and squares) and the $\Delta$ (dot-dashed line and triangles) families. 
\label{lin_traj_u}}
\end{minipage}
\end{center}
\end{figure}

At backward angles the reaction amplitude is driven by Feynman graphs in which baryons are exchanged, in the $u$-channel, according to the spin-isospin structure of the vertices. As can be seen in Figure~\ref{lin_traj_u}, each baryon family (same quantum numbers, but different spin and mass) lies on an almost linear Regge trajectory:
\begin{equation} 
\alpha(u)=J= \alpha_0 +\alpha' u
\label{lin_traj_eq_u}
\end{equation}

As in the $t$-channel we start from the Feynman amplitude, which incorporates the full spin-momentum dependent vertex function $\beta(u,J_0)$, but where the Regge propagator ${\cal P}_{Regge}$ is used instead of the Feynman propagator:
\begin{equation}
{\cal A}_{Regge}=\beta(u,J_0) {\cal P}_{Regge}
\label{Regge_generic_amp_u}
\end{equation}
\begin{equation}
{\cal P}_{Regge}= \left(\frac{s}{s_0}\right)^{\alpha(u)-J_0}
\frac{\pi \alpha'}{\sin(\pi \alpha(u))}
\frac{{\cal S}+e^{-i\pi\alpha(u)}}{2}
\frac{1}{\Gamma(1-J_0+\alpha(u))}
\label{Regge_pole_u}
\end{equation}

Only the exchange of the nucleon is allowed in the $\gamma p\rightarrow p\omega$ and $\gamma p\rightarrow p\phi$ reactions. The spatial part of the amplitude is as  follows~\cite{jml}:
\begin{eqnarray}
{\cal T}_N = i\frac{e\;\mu_pg_V(1+\kappa_V)}{2m}
                  \left( \chi_f\left|\vec{\sigma}\cdot \vec{k}\times \vec{\epsilon} \;
                   \vec{\sigma}\cdot \vec{P_V}\times \vec{\epsilon_V} \right| \chi_i \right)  						\nonumber \\
    \left(\frac{s}{s_1}\right)^{\alpha_N(u)-\frac{1}{2}}                                                                                                              
    \frac{\pi \alpha'_N(1- e^{-i\pi (\alpha_N(u)+\frac{1}{2})})}
           {2\sin(\pi(\alpha_N(u)+\frac{1}{2}))\Gamma (\alpha_N(u)+\frac{1}{2})}    	       
\end{eqnarray} 
where the non degenerate nucleon trajectory (with $\alpha_N(u) = -0.37 + \alpha'_N u$ and $\alpha'_N = 0.98$ GeV$^{-2}$) has been chosen in order to produce a node in the cross section at the place it has been observed at backward angles~\cite{cli77}. In this expression, $e$ is the electric charge and $\mu_p$ is the magnetic moment of the proton, $\vec{k}$ is the momentum and $\vec{\epsilon}$ is the polarization of the incoming photon, while $\vec{P_V}$ is the momentum and $\vec{\epsilon_V}$ is the polarization of the emitted vector meson. The coupling constants of the vector meson with nucleon are $g_V$ and $\kappa_V$. 

\begin{figure}[tbhp]
\begin{center}
\begin{minipage}[t]{8 cm}
\hspace{-1.5 cm}
\epsfysize=8.0cm
\epsfig{file=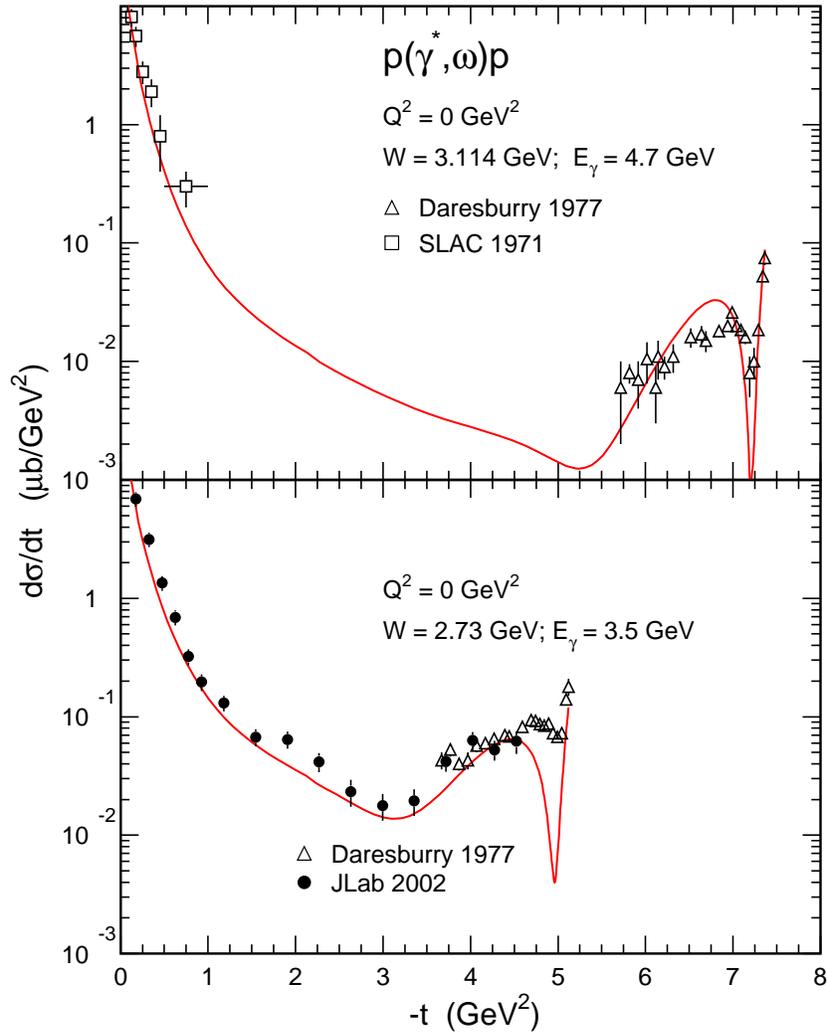,scale=0.6}
\end{minipage}

\begin{minipage}[t]{16.5 cm}
\caption{(Color on line) The differential cross section of the $\gamma p\rightarrow p\omega$ reaction at $E_{\gamma}=$~3.5 and 4.7~GeV. The curve includes also the contributions of meson Regge pole exchanges, as well as the Pomeron exchange, which dominate at low and intermediate $-t$. The full dots come from~\cite{bat03}. the empty triangles come from ~\cite{cli77}. The empty boxes come from~\cite{bal73}. 
\label{omega_back}}
\end{minipage}
\end{center}
\end{figure}

As shown in Figure~\ref{omega_back}, this amplitude leads to a good accounting of the backward angle cross-section in the $\omega$ channel where all the coupling constants are fixed. Namely, we use $g_{\omega}=15$ and $\kappa_{\omega}=0$ consistently with the analysis of nucleon-nucleon scattering. The agreement is particularly good at the highest energy, where Regge ideas are best suited. The softening of the node at the lowest energy may indicate a possible contribution of cuts (see next section). 

In the $\phi$ channel a good accounting of the rise of the cross section at the most backward angles is achieved (Figure~\ref{phi_JLab}) with the choice $g_{\phi} (1+\kappa_{\phi})=3$ consistently with the analysis~\cite{jaf89}  of the nucleon form factors.

In the $\gamma p\rightarrow \pi^+ n$ reaction the nucleon exchange amplitude takes the form~\cite{lag10}:
\begin{eqnarray}  
\cal{T}_N&=& -i\; e \mu_n g_{\pi} \sqrt{2}  \frac{\sqrt{(E_i+m)(E_f+m)}}{2m} {\cal P}_N^R(u)
\nonumber \\
&& \left( \lambda_f \left |
\vec{\sigma} \cdot \vec{k_{\gamma}} \times \vec{\epsilon}
 \;\;\vec{\sigma} \cdot \left[ 
 \frac{\vec{k_{\pi}}-\vec{p_i}}{\sqrt{m^2+ (\vec{k_{\pi}}-\vec{p_i})^2}+m} 
\right .\right .\right .    \nonumber \\ &&
 \left . \left . \left . + \frac{\vec{p_i}}{E_i+m} \right]
\right | \lambda_i \right )
\label{uN}
\end{eqnarray}
Where $(E_i, \vec{p_i})$ and $(E_f, \vec{p_f})$ are the four momenta of the target proton and the final neutron, respectively. Where $\vec{k_{\gamma}}$ is the momentum of the ingoing photon and $\vec{\epsilon}$ is its polarization. Where $(E_{\pi}, \vec{k_{\pi}})$ is the four momentum of the outgoing pion. The four momentum transfer in the $u$-channel is $u= (k_{\pi}-p_i)^2$. The magnetic moment of the neutron is $\mu_n=$ -1.91, and the pion nucleon coupling constant is $g^2_{\pi}/4\pi=$~14.5.

To be consistent with the node in the backward angle cross section of the $p(\gamma,\omega)p$ channel, where only the proton can be exchanged in the $u$-channel, we use the same non degenerated Regge propagator
\begin{eqnarray}  
{\cal P}_N^R&=& \left (\frac{s}{s_{\circ}} \right)^{\alpha_N -0.5}
 \alpha'_N \Gamma (0.5-\alpha_N)
\frac{1- e^{-i\pi(\alpha_N+0.5)}}{2}
 \label{prop_N}
\end{eqnarray}
where $s_{\circ}=$ 1~GeV$^2$ and where the nucleon trajectory is
$  \alpha_N= -0.37 +\alpha'_N \; u$,
with $\alpha'_N=$~0.98.

The $u$-channel amplitudes of the $\gamma n\rightarrow \pi^- p$ and $\gamma p\rightarrow \pi^0 p$ reactions have the same expression with  trivial changes of the charge and well known coupling constants. 

In the pion production channels the $\Delta$ Regge trajectory can be exchanged as well, but contributes to the cross section only at the level of $~10\%$. I refer the reader to~\cite{lag10} for a detailed expression of the corresponding amplitude.

\subsection{\it Unitarity cuts \label{sec:cuts}}

The unitarity of the S-matrix relates the imaginary part of the reaction amplitude of a given channel to the sum of the on the mass shell production and rescattering amplitudes of every channels to which it can couple.  The large cross section of rho production channels compensates the suppression of the loop integral and boosts the contribution of the corresponding inelastic cuts. The contribution of the elastic cut is less important but reveals itself through the purely destructive interference with the dominant pole amplitudes.

\subsubsection{\it Elastic unitarity cuts \label{sec:el-cuts}}

The first graph in Fig.~\ref{charged_pion_graphs} depicts the elastic rescattering of the pion. The singular part of the corresponding rescattering amplitude takes the form:
\begin{eqnarray}
 T_{\pi N}&=& -i\frac{p'_{c.m.}}{16\pi^2}\frac{m}{\sqrt{s}}
 \int  d{\Omega} \left[T_{\gamma p\rightarrow \pi^+ n}(t_{\gamma})
 T_{\pi^{+} n \rightarrow \pi^{+} n}(t_{\pi})
 \right]
 \nonumber \\ && 
 \label{pi_el}
\end{eqnarray}
where $p'_{c.m.}=\sqrt{(s-(m_{\pi}-m)^2)(s-(m_{\pi}+m)^2)/4s}$  is the on-shell momentum  of the intermediate nucleon, for  the c.m. energy $\sqrt s$. The two fold integral runs over the solid angle $\Omega$ of the intermediate nucleon, and is done numerically. The four momentum transfer between the incoming photon and the intermediate $\pi$ is $t_{\gamma}=(k_{\gamma}-P_{\pi})^2$, while the four momentum transfer between the intermediate $\pi$ and the outgoing pion is $t_{\pi}=(k_{\pi}-P_{\pi})^2$. The summation over all the spin indices of the intermediate particles is meant.

The pion photoproduction amplitude is the same as in the previous sections~\ref{sec:t-channel} and~\ref{sec:u-channel}: it incorporates the pole terms in the $t$-channel as well as in the $u$-channel. The pion nucleon elastic scattering amplitude is purely absorptive:
 \begin{eqnarray}
T_{\pi^{+} n \rightarrow \pi^{+} n}=-\frac{\sqrt{s}\; p'_{c.m.}}{m} 
(\epsilon_{\pi} + i)\sigma_{\pi^-p} \exp[\frac{\beta_{\pi}}{2}t_{\pi}]
\label{scat_pin}
\end{eqnarray}
Above $\sqrt{s}\sim 2$~GeV, the total cross section stays constant at the value $\sigma_{\pi^-p}= 30$~mb~\cite{PDG}, and the fit of the differential cross section at forward angles leads to a slope parameter $\beta_{\pi}=6$~GeV$^{-2}$~\cite{La72}. At high energy the ratio between the real and imaginary part of the amplitude is small~\cite{PDG} and is set to $\epsilon_{\pi}=0$.

Under those assumptions, the sum of the Regge pole amplitude and the elastic $\pi N$ cut takes the form:
 \begin{eqnarray}
 T&=& T_{\gamma p\rightarrow \pi^+ n}(t)
 \nonumber \\ 
 && -\frac{p'\,^2_{c.m.}}{16\pi^2}\sigma_{\pi^-p}
 \int  d{\Omega} T_{\gamma p\rightarrow \pi^+ n}(t_{\gamma})
 \exp[\frac{\beta_{\pi}}{2}t_{\pi}]
\label{scat_sum}
\end{eqnarray}
where $t=(k_{\pi}-k_{\gamma})^2$ is the overall four momentum transfer. It clearly shows the purely destructive interference, that is expected for an absorptive rescattering. It reduces (by about a factor two) the contribution of the $u$-channel Regge poles at the very backward angles (Figure~\ref{charged_pion_photo}). The effect is less important at the very forward angles, simply because the $t$-channel Regge contribution is more than an order of magnitude larger than the $u$-channel one.

\begin{figure}[hbt]
\begin{center}
\begin{minipage}[t]{8 cm}
\hspace{-1.5 cm}
\epsfysize=8.0cm
\epsfig{file=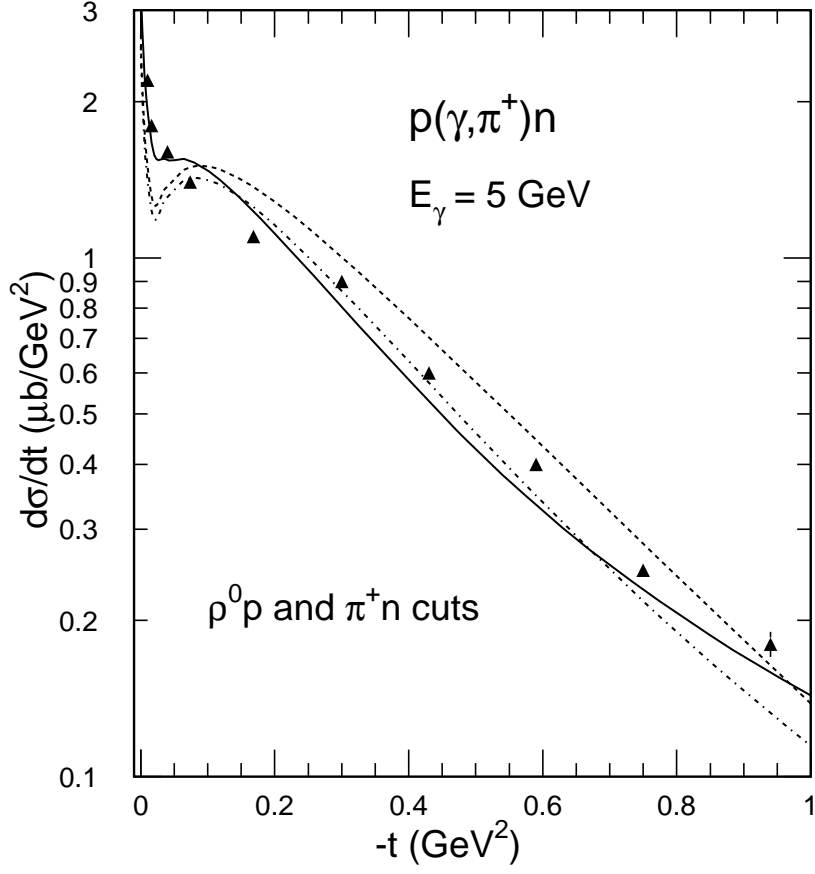,scale=0.6}
\end{minipage}

\begin{minipage}[t]{16.5 cm}
\caption[]{The cross section of the $p(\gamma,\pi^+)n$ reaction at forward angles. Basic Regge pole model: dotted lines. Elastic pion rescattering cut included: dash-dotted lines. Inelastic $\rho^{\circ}N$ also included: full lines.  Filled triangle: ref~\cite{AnXY}}.
\label{zoom}
\end{minipage}
\end{center}
\end{figure}

Figure~\ref{zoom} is a zoom of Figure~\ref{charged_pion_photo} that shows how the delicate interference between the pole terms and the cuts reproduces the $\pi^+$ cross section at forward angles. The rise of the cross section at the very forward angles comes from the interference between the $t$-channel $\pi$ and $\rho$ exchange and the $s$-channel nucleon exchange that is necessary to restore gauge invariance in the basic Regge model (section~\ref{sec:t-channel}). The $\pi N$ elastic cut brings the model very close to the data at moderate $-t$, while the $\rho N$ inelastic cut (next section) fills in the minimum around $-t=$ 0.02~GeV. In the virtual photon sector too the elastic cut brings the model in perfect agreement with the Hermes $\pi^+$ electro-production data (see Figure~\ref{zoom_hermes}). 

It offers also an elegant way to understand the disappearance of the node of the $\pi^0$ photo-production  cross section (see Figure~\ref{neutral_pion_photo}) around $-t=0.5$ Gev$^2$ when one enters the virtual photon sector (see Figure~\ref{pizero_hallA}). To a good approximation the pion elastic scattering amplitude in Eg.~\ref{pi_el} is driven by the exchange of the Pomeron (as we will see in section~\ref{sec:pomeron}). Therefore it is possible to express the cut amplitude as an effective Regge pole~\cite{Do02}. In this scheme, the $\omega$ Regge exchange amplitude takes the form:
\begin{eqnarray}
{\cal T}_{\omega} &=& {\cal T}_F \times (t-m_{\omega}^2)\times
\frac{g_{\omega NN}}{g_{\omega NN}^{GLV}}
\nonumber \\
&&\left( e^{-i\pi \alpha_{\omega}(t)}
\left(  \frac{s}{s_0} \right)^{\alpha_{\omega}(t)-1} 
\alpha'_{\omega} \Gamma(1-\alpha_{\omega})F_{\omega}(Q^2)
\right. \nonumber \\ && \left. 
- e^{-i\pi \delta_{c}(Q^2)}
\left(  \frac{s}{s_0} \right)^{\alpha_{c}(t)-1} 
\alpha'_{c} \Gamma(1-\alpha_{c})F_{c}(Q^2) G_c(t)
\right)  
\nonumber \\
\label{T_Regge}
\end{eqnarray}
where $t=(k_{\omega}-k_{\gamma})^2$ is the four momentum transfer and $m_{\omega}$ is the mass of the $\omega$ meson. The Feynman amplitude ${\cal T}_F$ has the same spin-momentum structure as in the GLV~\cite{gui97} scheme; simply the $g_{\omega NN}$ coupling constant will be re-fitted to experiment. The $\omega$ Regge trajectory is the same as in the GLV scheme:
\begin{eqnarray}
\alpha_{\omega}(t)&=& \alpha_{\omega}(0)+ \alpha'_{\omega} t
= 0.44 +0.9t
\label{omega_traj}
\end{eqnarray}
while the intercept and the slope of the effective trajectory of the cut take the form:
\begin{eqnarray}
\alpha_{c}(0)&=& \alpha_{\omega}(0)+ \alpha_{P}(0)-1\;\;\;\;\;\, = 0.44
\nonumber \\
\alpha'_c
&=& (\alpha'_{\omega}\times\alpha'_{P})/(\alpha'_{\omega}+\alpha'_{P})
= 0.2
\label{cut_traj}
\end{eqnarray}
where the intercept and the slope of the Pomeron Regge trajectory are respectively $\alpha_P(0)=1$ and $\alpha'_P= 0.25$ (see section~\ref{sec:pomeron})

\begin{figure}[hbtp]
\begin{center}
\begin{minipage}[t]{8 cm}
\hspace{-1. cm}
\epsfysize=8.0cm
\epsfig{file=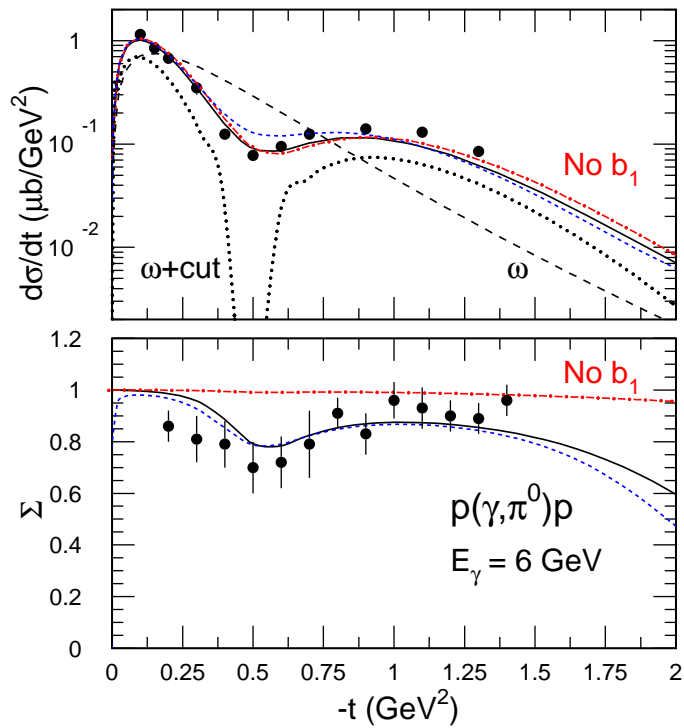,scale=0.5}
\end{minipage}

\begin{minipage}[t]{16.5 cm}
\caption[]{(Color on line) The cross section of the $p(\gamma,\pi^0)p$ reaction~\cite{An70} (top) and the photon asymmetry~\cite{An71} (bottom). The dashed line curve corresponds to the  $\omega$ Regge Pole amplitude with a degenerated trajectory. The dotted line curve includes the Pomeron cut. The red dot-dashed line curves include the $\rho$ exchange, while the black solid lines curves include also the $b_1$ exchange.  The blue short-dashed line curves include all the inelastic cuts.}
\label{P_cut_6GeV}
\end{minipage}
\end{center}
\end{figure}

It is possible to reproduce the GLV non-degenerated $\omega$ amplitude (Figure~12 in~\cite{gui97}) with the following choice (dotted line curve in Fig.~\ref{P_cut_6GeV}):
\begin{eqnarray}
 g_{\omega NN}^2/4\pi&=& 4.9
\nonumber \\
G_c (t)&=& 3.7 e^{2t}
\label{P_cut}
\end{eqnarray}
It is worth pointing out that the $\omega NN$ coupling constant is about half of the GLV one ($g_{\omega NN}^2/4\pi=$ 17.9), in better agreement with the range of values that are determined in the analysis of low energy $NN$ scattering data. It is also almost the same as the value ($g_{\omega NN}^2/4\pi=$ 6.44)  needed in the analysis of the $p(\gamma,\eta)p$ reaction~\cite{lag05}. In this channel, there is no node and the $\eta N$ scattering cross section is much lower than the $\pi N$ one: The use of a degenerated $\omega$ Regge amplitude alone, with no elastic cut, is more justified. The exchange of the $\rho$ meson degenerated trajectory, with the same coupling constants as in GLV, brings the model close to the unpolarized cross section. The exchange of the $b_1$ meson contributes little to the cross section but is essential to reproduce the photon asymmetry. Note that an axial-tensor coupling is used at the $b_1NN$ vertex, contrary to the axial-vector coupling which was wrongly used in~\cite{gui97} (I refer the reader to~\cite{lag11} for details).

Such a treatment offers us with the most obvious way of shifting the minimum of the cross section when the virtuality $Q^2$ of the photon increases (Figure~\ref{pizero_hallA}). Using slightly different cut off masses in the electromagnetic form factors of the poles and the Pomeron cut, $F(Q^2)=1/(1+Q^2/{\Lambda ^2})$, modifies their relative importance when the photon becomes virtual. The following choice leads to a good accounting of the DESY data~\cite{Bra78} at $Q^2=$ 0.85~GeV$^2$: $\Lambda_{\omega}^2=$ 0.325~GeV$^2$, $\Lambda_{\rho}^2=$ 0.400~GeV$^2$, $\Lambda_{b_1}^2=$ 1.~GeV$^2$,  $\Lambda_{c}^2=$ 0.300~GeV$^2$ and $\delta_c(Q^2)= -0.46 Q^2/0.85$. The agreement is good too at $Q^2=$ 0.55~GeV$^2$, but it is not possible to get rid of the second maximum in $t$ when $Q^2=$ 0.22~GeV$^2$. 

Note that this the only place in this review where the couplings of a cut have been adjusted to reproduce the experiments. But this is justified by the extreme sensitivity of the interference between the pole and the elastic cut amplitudes, whose the functional expressions are nevertheless well defined. 

The extrapolation at $Q^2=$ 2.3~GeV$^2$ misses the recent JLab data~\cite{Cam10} and~\cite{bed14}, but the agreement is restored when inelastic rescattering cuts are taken into account (next section). The agreement is less good at moderate virtuality, but no attempt has been made yet to adjust the electromagnetic form factors of the pole and elastic cut, when the contribution of inelastic cut is included in the amplitude.

\subsubsection{\it Inelastic unitarity cuts \label{sec:inel-cuts}}

The second graph in Fig.~\ref{charged_pion_graphs} depicts the production of the $\rho$ meson followed by the reabsorption of one of its decay pions by the nucleon. 
In charged pion production reactions, the corresponding rescattering amplitude takes the form:
\begin{eqnarray}  
{\cal T}_{\rho N}&=& \int \frac{d^3\vec{p}}{(2\pi)^3}
\frac{m}{E_p}
 \frac{1}{P^2_{\rho}-m^2_{\rho}+i\epsilon}
%\nonumber\\ &&  \times 
T_{\gamma p\rightarrow \rho^{0}p}T_{\rho^{0} p \rightarrow \pi^{+} n}
\end{eqnarray}
where the integral runs over the three momentum $\vec{p}$ of the intermediate nucleon, of which the mass is $m$ and the energy is $E_p=\sqrt{p^2+m^2}$. The four momentum  and the mass of the intermediate $\rho$ are respectively $P_{\rho}$ and $m_{\rho}$. The integral can be split into a singular part, that involves on-shell matrix elements, and a principal part~$\cal{P}$:
\begin{eqnarray}
 T_{\rho N}&=& -i\frac{p_{c.m.}}{16\pi^2}\frac{m}{\sqrt{s}}
 \int  d{\Omega} \left[T_{\gamma p\rightarrow \rho^{0} p}(t_{\gamma})
 T_{\rho^{0} p \rightarrow \pi^{+} n}(t_{\pi})
 \right]
 \nonumber \\ && 
 +\cal{P}
 \label{sing}
\end{eqnarray}
where $p_{c.m.}=\sqrt{(s-(m_{\rho}-m)^2)(s-(m_{\rho}+m)^2)/4s}$  is the on-shell momentum  of the intermediate proton, for  the c.m. energy $\sqrt s$. The two fold integral runs over the solid angle $\Omega$ of the intermediate proton. The four momentum transfer between the incoming photon and the $\rho$ is $t_{\gamma}=(k_{\gamma}-P_{\rho})^2$, while the four momentum transfer between the $\rho$ and the outgoing pion is $t_{\pi}=(k_{\pi}-P_{\rho})^2$. The summation over all the spin indices of the intermediate particles is meant. 

I neglect the principal part. The singular part of the integral relies entirely on on-shell matrix elements and is parameter free as long as one has a good description of the production and the absorption processes. Again the integral is done numerically.

For photoproduction of vector mesons, $T_{\gamma p\rightarrow \rho^{0} p}$, the Regge model~\cite{jml,cano} which reproduces the world set of data in the entire angular range (see for instance section~\ref{sec:rho}, and reference~\cite{bat01}) is used. The model takes into account the exchange of the Pomeron, the $f_2$ and $\sigma$ mesons in the $t$-channel, as well as the exchange of the nucleon and the Delta in the $u$-channel.

\begin{figure}[hbtp]
\begin{center}
\begin{minipage}[t]{8 cm}
\hspace{-1.5 cm}
\epsfig{file=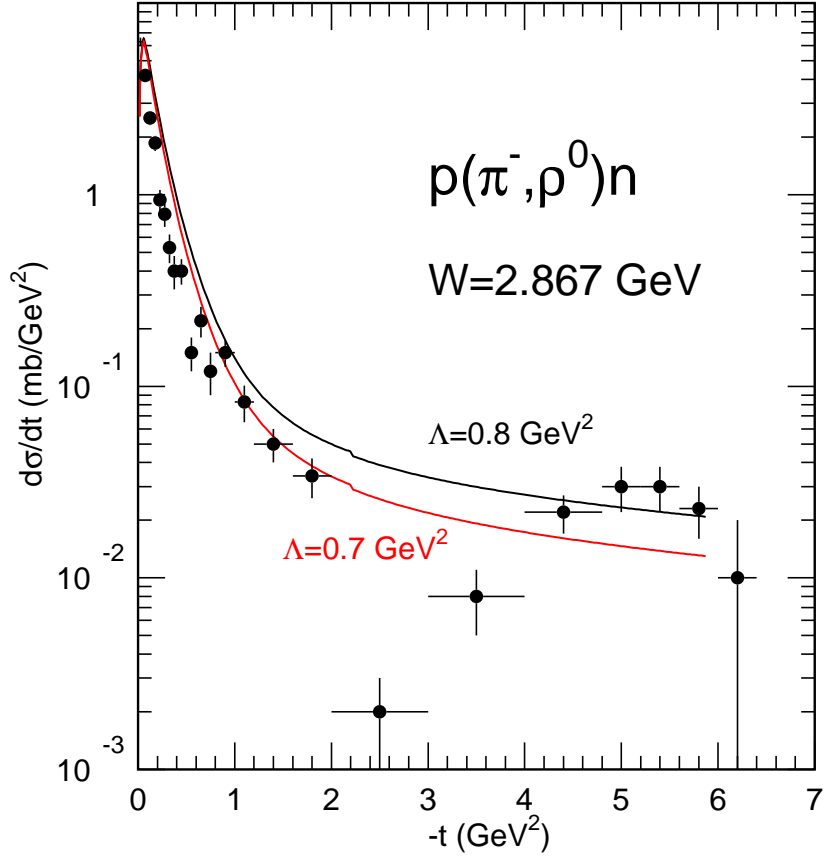,scale=0.6}
\end{minipage}

\begin{minipage}[t]{16.5 cm}
\caption[]{(Color on line) The cross section of the reaction $p(\pi^-,\rho^{\circ})n$. The experimental data are from ref.~\cite{PiroXX}. }
\label{rho_pi}
\end{minipage}
\end{center}
\end{figure}

For the reabsorption of the $\rho$ meson, $T_{\rho^{0} p \rightarrow \pi^{+} n}$, a pion exchange Regge description nicely reproduces the data in the energy range that is considered in this study (Figure~\ref{rho_pi}). The corresponding matrix element takes the form:
\begin{eqnarray}
T_{\rho\pi}&=& \sqrt{2} g_{\pi} g_{\rho} \frac{\sqrt{(E_p+m)(E_f+m)}}{2m}
{\cal P}_{\pi}^{R}(t_{\pi}) F_1(t_{\pi})
 \nonumber \\ && 
\left[ \vec{\epsilon_{\rho}} \cdot \left( \vec{k_{\pi^+}} -\vec{p}+\vec{p_f}\right)
-\epsilon^{\circ}_{\rho} \left( E_{\pi^+}-E_p+E_f \right)  \right]
 \nonumber \\ && 
\left( \lambda_f \left|\vec{\sigma} \cdot 
\left[  \frac{\vec{p}}{E_p+m} -\frac{\vec{p_f}}{E_f+m}  \right]
 \right| \lambda \right)
 \nonumber \\ && 
 \label{rho_to_pi}
\end{eqnarray}
where $(\epsilon_{\rho}^{\circ},\vec{\epsilon_{\rho}})$ is the intermediate $\rho$ polarization, and where $(E_f,\vec{p}_f)$ and $(E_{\pi^+},\vec{k}_{\pi^+})$ are the four momenta of  the  outgoing neutron and pion respectively. The pion nucleon coupling constant is $g^2_{\pi}/4\pi=$~14.5, and the rho decay constant is $g^2_{\rho}/4\pi=$~5.71 as deduced from the $\rho$ width.  ${\cal P}_{\pi}^{R}$ is the pion Regge propagator, with a degenerated non rotating
saturating trajectory, and $ F_1$ is the nucleon Dirac form factor (with $\Lambda=0.7$~GeV$^2$):
\begin{eqnarray}
F_1(t)= \frac{1-\frac{2.79t}{4m^2}}
{(1-\frac{t}{4m^2})(1-\frac{t}{\Lambda})^2}
 \label{F1}
\end{eqnarray}

Up to $-t=$ 2~GeV$^2$, the model reproduces fairly well the dominant peak but it misses the dip at $-t=$ 2.5~GeV$^2$. Above, the cross section of the $\rho^{0} p \rightarrow \pi^{+} n$ reaction is two orders of magnitude smaller than at forward angles. It contributes less to the rescattering integral and an average description of the elementary channel cross section is sufficient here.

\begin{figure}[hbt]
\begin{center}
\begin{minipage}[t]{8 cm}
\hspace{-0.6 cm}
\epsfig{file=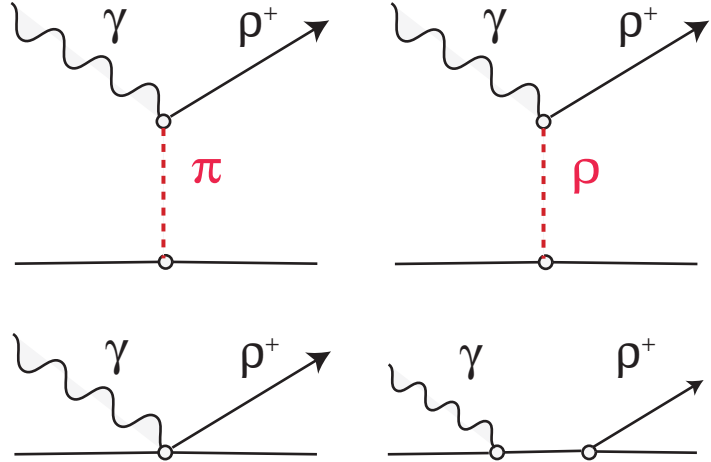,scale=0.5}
\end{minipage}

\begin{minipage}[t]{16.5 cm}
\caption[]{(Color on line) The graphs of the reaction $p(\gamma,\rho^+n$). Top: meson exchange graphs. Bottom: contact and nucleon exchange gauge graphs.   }
\label{charged_rho_graph}
\end{minipage}
\end{center}
\end{figure}

In neutral pion production reactions, the coupling to the $\rho^0 N$ intermediate state is forbidden (isospin does not allow the decay of the $\rho^0$ into two $\pi^0$). At the real photon point, the contributions of the other possible inelastic cuts ($\pi N$ charge exchange, $\omega N$, $\rho^+p$, $\rho^{\pm}\Delta$) have been found to be small~\cite{lag11}, consistently with the experimental data (see Figure~\ref{neutral_pion_photo}). On the contrary, the contributions of the $\rho^+n$, $\rho^{\pm}\Delta$ cuts become dominant in the virtual photon sector. As we have already seen in section~\ref{sec:virt_neutral_pion}, the cross section of the electro-production of charged rho mesons becomes comparable to the cross section of the electro-production of neutral rho mesons at $Q^2\sim$ 2.3~GeV$^2$,  (Figure~\ref {charged_rho_xsection}). The amplitude of the $\rho^+n$ cut takes the same form as Equation~\ref{sing}. The amplitude of the elementary reaction $\rho^+ n \rightarrow \pi^0 p$ is trivialy related to the amplitude of the reaction $\rho^0 p \rightarrow \pi^+n$ by changing the charges and the isospin couplings. The amplitude of the elementary reaction $\gamma p \rightarrow \rho^+ n$ relies on a Regge model which incorporates the pion exchange pole as well as the $\rho$ exchange pole supplemented by the contact and the s-channel nucleon exchange necessary to restore gauge invariance (Figure~\ref{charged_rho_graph}). I refer the reader to~\cite{lag11} for a more detailed account.

\subsection{\it Saturating trajectories vs cuts \label{sec:sat-vs-cuts}}

In the highest transverse momentum $p_T$ (large $-t$ and $-u$) range, which has been reached to day in exclusive reactions, inelastic cuts provide a robust interpretation  of the plateau which has been observed in the angular distributions around $90^{\circ}$. Their contributions are almost parameter free, as they are based on on-shell reaction amplitudes that are fixed by independent experiments. They leave little room for other possible more exotic processes. 

For instance, we have proposed~\cite{gui97} an alternative explanation of the flattening of the angular distribution of the $\gamma N \rightarrow \pi^{\pm} N$ reactions based on the use of the saturating Regge trajectories of Figure~\ref{sat_traj}. This is a way  to reconcile, at higher momentum transfer, 
the Regge exchange model with experiment and the counting rules~\cite{bro}.

The analysis of nucleon-nucleon and pion-nucleon scattering, led to infer~\cite{col84} that the Regge trajectories saturate at -1 for $-t\to \infty$, consistently with the expectation of the behavior of the QCD interaction of the exchanged two quarks at asymptotic momentum transfer. A prescription inspired from~\cite{col84} has been applied  to the pion  photo-production reactions~\cite{gui97}. At high momentum transfers, the hard scattering amplitude $\mathcal{M}_{hard}$ factorizes into  
\begin{equation}
\mathcal{M}_{hard}=\mathcal{M}_{soft}\cdot F_{\pi}(t)\cdot F_N(t)
\label{moddur}
\end{equation}
where $\mathcal{M}_{soft}$ is the amplitude of the Regge exchange model 
with saturating Regge trajectories   and $F_{\pi}(t)$ and $F_N(t)$ are the transition form factors of the two outgoing particles.
 
For meson trajectories, this saturation follows~\cite{ser94}  a QCD motivated effective interquark potential of the form:
\begin{equation}
\mathcal{V}(r)=-\frac{4}{3}\frac{\alpha_s}{r}+\kappa r+\mathcal{V}_0\;,
\label{iqpot}
\end{equation}
where the first term represents the short distance part
of the quark-antiquark interaction and corresponds to one gluon exchange, 
whereas the term which is linear in $r$, leads 
to confinement ($\kappa$ is known as the ``string tension") 
and $\mathcal{V}_0$ is a constant term in the potential.
\newline
\indent
The eigenvalue equation ($E^2 = t$ as function of the angular momentum $\alpha$) 
of the potential of Eq.~\ref{iqpot} was inverted in Ref.~\cite{ser94} and 
the resulting equation for $\alpha$ was analytically continued into 
the scattering region ($t < 0$). This leads to the $\pi$ and $\rho$ Regge trajectories shown in Figure~\ref{sat_traj}. They become linear as $t\to+\infty$ which reflects the bound  state spectrum but  saturate at -1 for $t\to -\infty$ which reflects the hard scattering (pQCD) region. This treatment quantifies the evolution of the interaction between the two exchanged quarks which have been schematically depicted in Figure~\ref{space_time}: at low $-t$, on the left of the figure, they exchange many gluons, while at asymptotic high $-t$, on the right,  they exchange one gluon only.
 
The resulting hard scattering amplitude of Eq.~\ref{moddur} with 
these saturating trajectories 
has a $t$-dependence which is much flatter than 
the exponential $t$-dependence of the ``soft'' model. 
This leads to a plateau for the differential cross section at large transverse momentum transfers $p_T$. 
Moreover the differential cross section satisfies the counting rules 
as shown in~\cite{col84} 
because of the saturation of the trajectories and because of the scaling behavior of the form factors of the outgoing particles. Let us also remark that the high momentum transfer amplitude of Eq.~\ref{moddur} is normalized  by the low momentum transfer amplitude. The resulting cross sections are shown in  Figure~1 of~\cite{gui97}. They are very similar to the cross sections, which include the contribution of the cuts, presented in Figure~\ref{charged_pion_photo} in this review.

The concept of saturating trajectories, as well as their determination in~\cite{ser94}, are on solid ground and can be safely used as guide to implement the dependency against $-t$ of the electromagnetic form factors that has been used in sections~\ref{sec:virt_vec}  and~\ref{sec:virt_charged_pion}, for instance. However, we may question the way we have implemented them in the practical treatment~\cite{gui97} of the amplitude. On the one hand, the particular choice of the hadronic form factors in Eq.~\ref{moddur} should be revisited. For instance, we have used the Dirac form factor Eq.~\ref{F1} of the nucleon, but we have set $F_{\pi}=1$. Using a more realistic (but still badly known at the highest momentum) pion transition form factor will certainly decrease the effect of the saturation. On the other hand, we may have not yet reached high enough $p_T$ for the factorization in Eq.~\ref{moddur} to be justified. 

So, in the range of energies $W$ and transverse momentum $p_T$ which have been covered so far in exclusive reactions, the physics resides in the contribution of the cuts. In the light quark sector, the cut associated with the propagation of the $\rho^0$ meson dominates since the large production cross section of the $\rho^0$ compensates the cost to pay in the rescattering integral. Its contribution is expected to survive at higher energies, in the charged pion production as well as in Virtual Compton Scattering, since the $\rho^0$ production cross section stays almost constant with energy (see Figure~\ref{vector_cross}). The cuts associated with the propagation of the charged vector mesons $\rho^{\pm}$ dominate neutral pion electro-production, due to the peculiar dependency against the virtuality $Q^2$ (see Figure~\ref{charged_rho_xsection}). But their importance is expected to decrease with the energy $W$, since the production cross section is driven by the exchange of Reggeons. 

In the heavy (strange and charm) quark sectors, cuts are not expected to play a significant role. On the one hand, there is no channel with large enough cross section to couple with; on the other hand, quark exchange mechanisms are suppressed, due to the different nature of the quark components of the nucleon and the meson. As we shall see in the next section, gluon exchange mechanisms dominate and are more amenable to a partonic description.

In between, the production cross sections of vector mesons made of light quarks are dominated by gluon exchanges at high energies, but receive also contributions from quark exchange at low energies $W< 10$~GeV (see $\it e.g.$ sections~\ref{sec:omega} and~\ref{sec:rho}). However no obvious cut dominates, and the way saturating trajectories have been taken into account in~\cite{gui97}  provides an elegant ``QCD inspired" parameterization of the quark exchange  part of the amplitudes (see figures~\ref{omega_JLab} and~\ref{rho_JLab}). Here the vector meson transition form factor does not need to be the same as the pion transition form factor in Eq.~\ref{moddur}, while the nucleon form factor is the same.

\subsection{\it The Pomeron and its two gluon realization \label{sec:pomeron-2g}}

\subsubsection{\it Pomeron exchange \label{sec:pomeron}}

The Pomeron occupies a singular place in the family of Regge poles. Its trajectory has been deduced from the analysis of forward cross section of proton-proton as well as meson-nucleon scattering up to the highest energies (see~\cite{Do02} for a review):

\begin{equation}
\alpha_P = 1+ \epsilon' + \alpha'_P t
\label{P_traj_equ}
\end{equation} 

Its slope $\alpha'_P = 0.25$~GeV$^{-2}$ is about four times smaller than the slope of the other Reggeon trajectories, and was deduced from the angular distributions at forward angles. Its intercept $1+\epsilon'$ is slightly greater than 1, and the choice of $\epsilon'= 0.08$ leads to a good accounting of the slight rise of the proton-proton total cross section, which behaves as $s^{\alpha_P(0)-1}= s^{\epsilon'}$, up to energies which has been reached before~\cite{Do02}  and even at~\cite{ant18} the Large Hadron Collider (LHC at CERN). Note that at these high energies (about $\sqrt s=13$~TeV) the cross section is still lower, by about an order of magnitude, than the Froissart's unitary bound. While we do not know yet which mechanisms will restore unitarity at asymptotic energies, we can safely rely on Pomeron exchange in the energy range covered in this review.

\begin{figure}[tbhp]
\begin{center}
\begin{minipage}[t]{8 cm}
\hspace{-1.65 cm}
\epsfysize=8.0cm
\epsfig{file=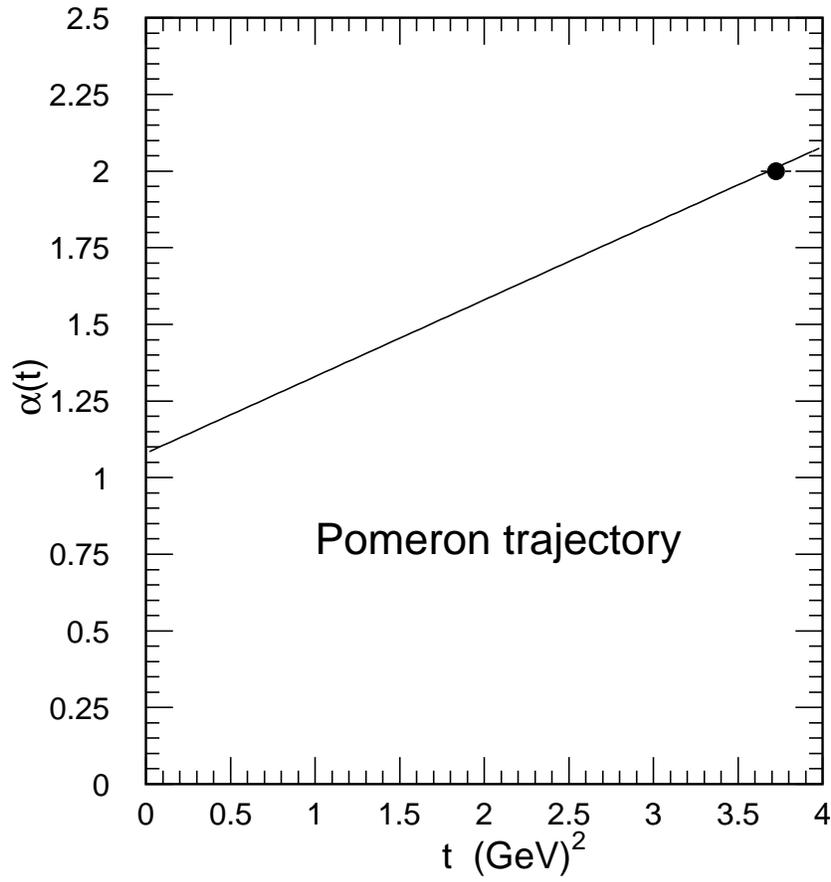,scale=0.6}
\end{minipage}

\begin{minipage}[t]{16.5 cm}
\caption{The Regge trajectories associated with the Pomeron. The $2^{++}$ glueball candidate~\cite{aba94} lies on the trajectory $\alpha(t)= 1.08 + 0.25t$. 
\label{Pom_traj}}
\end{minipage}
\end{center}
\end{figure}

Contrary to the trajectories of other Reggeons, the Pomeron trajectory has not been determined from a family of particles, but directly determined from scattering experiments. It turns out however that a $2^{++}$ glueball candidate~\cite{aba94}, with a mass around 1930~MeV, lies on the Pomeron trajectory in Figure~\ref{Pom_traj}. This is remarkable and suggests~\cite{Do02} that the Pomeron is related to a  family of glueballs, of which the other members remain to be found.

This analysis  of nucleon-nucleon  scattering  at  high  energy  leads~\cite{La90}  to  infer that the Pomeron couples to the valence quark as an isoscalar photon and that the corresponding quark-quark interaction is~\cite{Do87}~:
\begin{eqnarray}
\beta_0^2\overline{u}\gamma_{\mu}u\;\overline{u}\gamma_{\mu}u
\left(\frac{s}{s_0}\right)^{\alpha(t)-1}
\exp \left[-i\frac{\pi}{2}\alpha(t) \right]
\label{pom1}
\end{eqnarray}
where $\beta_0^2=4\; GeV^{-2}$ and
where   the   parameters   of   the   Pomeron   Regge   trajectory
$\alpha(t)=1+\epsilon'+\alpha't$ are
$\alpha'=0.25\;GeV^{-2}$, $\epsilon'=0.08$ and $s_0=1/\alpha'=4\;GeV^2$.
Within this model, the differential cross-section of elastic
nucleon-nucleon scattering reduces to~:
\begin{equation}
\frac{d\sigma}{dt}=\frac{1}{4\pi}\left[3\beta_0F_1(t)\right]^4
\left(\frac{s}{s_0}\right)^{2\alpha(t)-2}
\label{crossNN}
\end{equation}
where the factor $3$ takes into account that three valence quarks must
recombine into a nucleon, whose  structure results in the isoscalar
form factor $F_1(t)$, Eq.~\ref{F1}.

\begin{figure}[tbhp]
\begin{center}
\begin{minipage}[t]{8 cm}
\hspace{-3.5 cm}
\epsfig{file=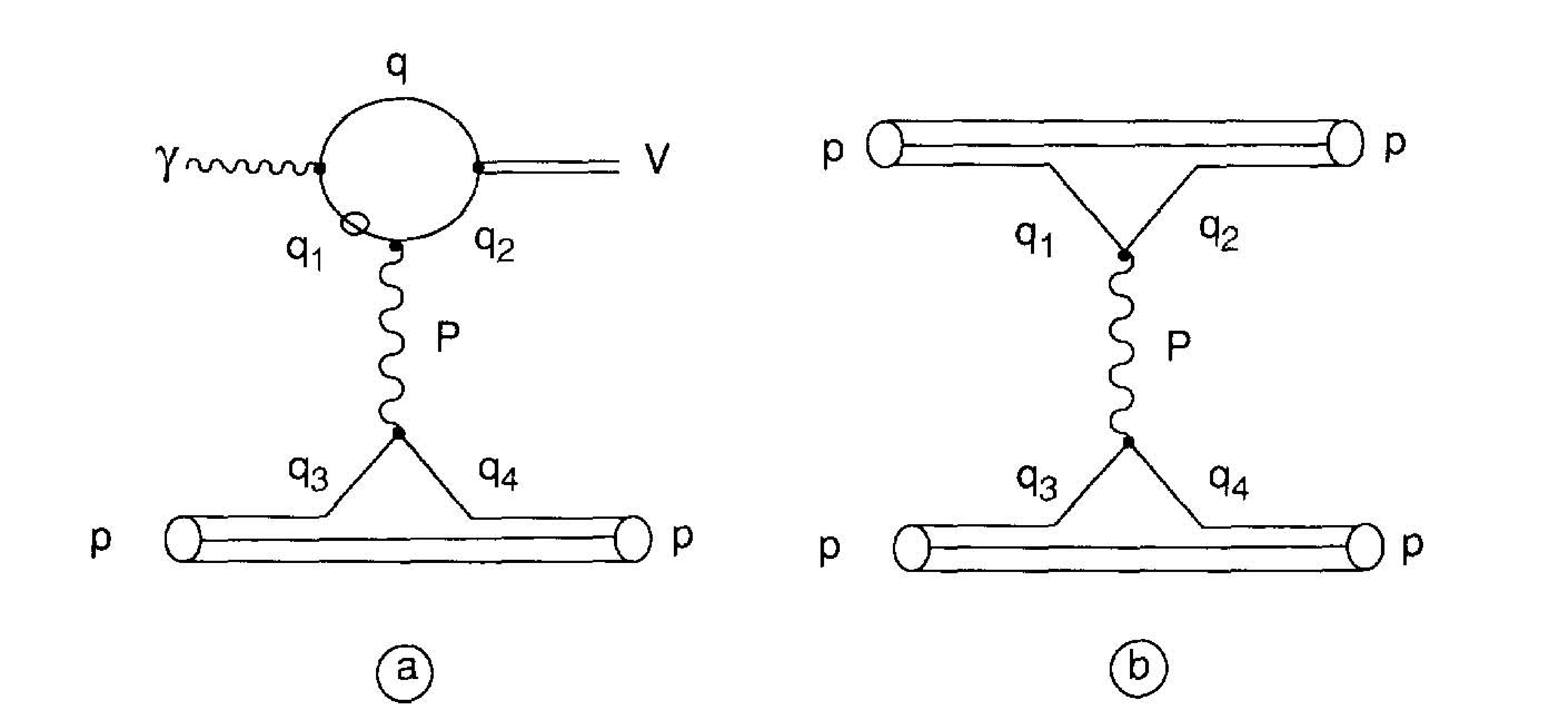,scale=0.3}
\end{minipage}

\begin{minipage}[t]{16.5 cm}
\caption{ Pomeron   exchange   graphs   in  exclusive  vector  meson
photoproduction~(a) and elastic nucleon-nucleon scattering (b). The quark marked with a circle is far off-shell. 
\label{grahs_Pom}}
\end{minipage}
\end{center}
\end{figure}

In  the  graph  which  describes  the electro-production of vector mesons
(Figure~\ref{grahs_Pom}), the  loop integral  in the  upper part,  is computed assuming that the two quark legs, $q$ and $q_2$, which recombine into the vector
meson, are almost on-shell and  share equally  the four-momentum  of the  vector meson ($q\simeq q_2  \simeq P_V/2$).   The  third quark  leg $q_1$, which  is
exchanged between the photon and the Pomeron, is far off-shell and its  propagator is factored out of the loop integral.  The denominator of its propagator is~:
\begin{equation}
q_1^2-m_q^2=-\Sigma^2=-\frac{1}{2}[Q^2+m_V^2-t]
\label{off}
\end{equation}
assuming  that  the  quark  mass  is  $m_q\simeq  m_V/2$.    After   the
integration in the quark loop, the vertex function of the  vector meson,
including the attached quark propagator, reduces to:
\begin{equation}
\frac{1}{2\sqrt{6}}\gamma\cdot \epsilon_V  f_V (\gamma\cdot  P_V +  m_V)
\end{equation}
where  $\epsilon_V$  and  $P_V$  are  respectively   the
polarization vector and the four-momentum of the vector meson, and where
its  decay  constant  $f_V$  is  related  to  its  radiative decay width
$\Gamma_{e^+e^-}$~\cite{PP92}       as       follows:
\begin{equation}
f_V^2e_q^2=\frac{3m_V}{8\pi\alpha_{em}^2}\Gamma_{e^+e^-}
\end{equation}

The lower part of the graph  is nothing but the exchange of  the Pomeron
between a  quark and  a nucleon. The
amplitude  can  be   cast  in  the   simple  form:     
\begin{equation}
\left[3F_1(t)\;\beta_0\right]
\;s\;\left(\frac{s}{s_0}\right)^{\alpha(t)-1}\;
\left[\frac{\mu_0^2}{-q_1^2+m_q^2+\mu_0^2}\;\beta^0\right]
\end{equation} 
It is made of three parts.  The first is the effective
coupling of the Pomeron to an  isoscalar nucleon.  The second is  the
Pomeron  Regge  trajectory.    And  the  last factor is the effective
coupling of the Pomeron to the off-shell quark $q_1$.

Gathering all these pieces together and performing the necessary trace
evaluation lead to the amplitude~\cite{Do88}:
\begin{equation}
M=3F_1(t)\frac{4\sqrt{6}m_Ve_qf_V \beta_0^2\mu_0^2}
{(Q^2+m_V^2-t)(2\mu_0^2+Q^2+m_V^2-t)}\;\epsilon\cdot\epsilon_V
\;s\;\left(\frac{s}{s_0}\right)^{\alpha(t)-1}
\exp \left[-i\frac{\pi}{2}\alpha(t) \right]
\label{MPomeron}
\end{equation}
where $\epsilon$ is the photon polarization factor. When squared this
amplitude leads to the cross-section:
\begin{equation}
\frac{d\sigma}{dt}=\frac{81m^3_V \beta^4_0\mu^4_0 \Gamma_{e^+e^-}}
{\pi\alpha_{em}}\left(\frac{s}{s_0}\right)^{2\alpha (t)-2}
\left[\frac{F_1(t)}{(Q^2+m_V^2-t)(2\mu_0^2+Q^2+m_V^2-t)}\right]^2
\label{DL}
\end{equation}

\subsubsection{\it Two gluon exchange \label{sec:2gluons}}

\begin{figure}[tbhp]
\begin{center}
\begin{minipage}[t]{8 cm}
\hspace{-4 cm}
\epsfig{file=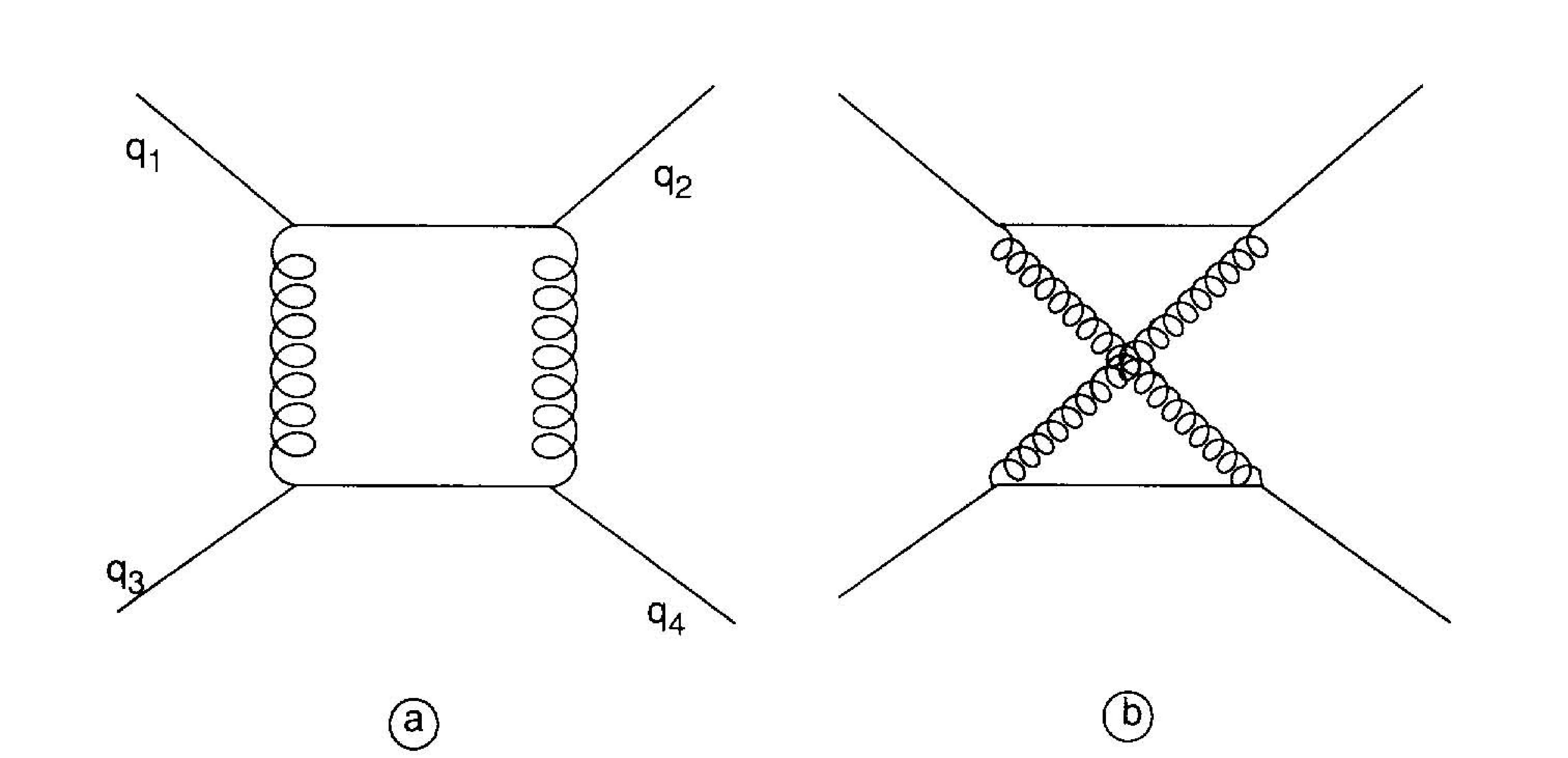,scale=0.3}
\end{minipage}

\begin{minipage}[t]{16.5 cm}
\caption{Direct (a) and crossed (b) two gluon exchange graphs in
quark-quark scattering.
\label{grahs_2gluon_crossed}}
\end{minipage}
\end{center}
\end{figure}
In the  large energy  limit ($\nu\rightarrow  \infty$), the  crossed two
gluon exchange graph  contribution cancels the  real part of  the direct
two  gluon  exchange  in  quark-quark  scattering (Figure~\ref{grahs_2gluon_crossed}).  Only the
imaginary part of the direct exchange graph, where the two  intermediate
quarks  are  on-shell,  survives.    Considering  only  abelian  gluons,
Landshoff and Nachtmann~\cite{La87} demonstrated that this  contribution
to the quark-quark scattering amplitude  reduces to a form very  similar
to the Pomeron exchange amplitude (Eq.~\ref{pom1})~:
\begin{equation}
\overline{u}\gamma_{\mu}u\;\overline{u}\gamma_{\mu}u
\frac{1}{8\pi^2}\int d^2\vec{l}
\left[4\pi\alpha_n D\left([\vec{l}+\frac{1}{2}\vec{q_T}]^2\right)\right]
\left[4\pi\alpha_n D\left([\vec{l}-\frac{1}{2}\vec{q_T}]^2\right)\right]
\end{equation}
where $l^2=-\vec{l}^2$ is the square of the  transverse momentum of the
exchanged gluon and $\vec{q_T}^2=-t$. When an exponential form 
\begin{eqnarray}
\alpha_n\;D(l^2)=\frac{\beta_0}{\sqrt{\pi}\lambda_0}
\exp\left(\frac{l^2}{\lambda^2_0}\right)  \hskip 0.3 cm, \hskip
1 cm
l^2\leq 0
\label{prop}
\end{eqnarray}
is used for the gluon propagator this amplitude reduces to:
\begin{equation}
\overline{u}\gamma_{\mu}u\;\overline{u}\gamma_{\mu}u
\frac{1}{8\pi}\int dl^2
\left[4\pi\alpha_n D(l^2+\frac{1}{4}t)\right]^2
\end{equation}
The value of the range parameter $\lambda^2_0=2.7$~GeV$^2$, which is used in this review as well as in~\cite{jml} and~\cite{me95}, corresponds to a correlation length $a\approx 0.19$~fm (see appendix C of~\cite{me95}).

When this two-gluon exchange amplitude is used in 
nucleon-nucleon elastic scattering, the cross-section at $t=0$ fixes the
norm of the gluon propagator:
\begin{equation}
2\pi\int_{-\infty}^0 dl^2\left[\alpha_n D(l^2)\right]^2=\beta_0^2
\label{norm}
\end{equation}

The vector meson production amplitude, corresponding to the coupling
of the two gluons to the same quark of the vector meson (bottom left graph in Figure~\ref{2gluons_graph}), takes the form:
\begin{eqnarray}
M&=&3F_1(t)\frac{2\sqrt{6}m_Ve_qf_V}{(Q^2+m_V^2-t)}\;s\;
\epsilon\cdot\epsilon_V
\sqrt{\frac{\alpha_s(\Sigma^2)}{\alpha_n}} \nonumber \\
&&\frac{1}{8\pi}\int dl^2
\left[4\pi\alpha_n D(l^2+\frac{1}{4}t)\right]^2
\label{2g:exch}
\end{eqnarray}
where $\Sigma^2$ stands for the degree of off-shellness of the quark
which connects the photon and one gluon, as given by Eq.~\ref{off}.

Using the norm  condition (Eq.~\ref{norm}), Eq.~\ref{2g:exch}  reduces
to  the  Pomeron  exchange  amplitude  (Eq.~\ref{MPomeron}), without the
phenomenological  form-factor  $\mu_0^2/(\mu_0^2+m_q^2-q_1^2)$  at   the
quark-Pomeron          vertex,           in          the           limit
$\frac{\alpha_s(\Sigma^2)}{\alpha_n}\rightarrow 1$.

When each gluon couples to a different quark of the vector
meson (bottom right  graph in Figure~\ref{2gluons_graph}), the denominator of the propagator of the far off-shell quark is~:
\begin{equation}
-4l^2+Q^2+m_V^2
\end{equation}
and the corresponding amplitude is~:
\begin{eqnarray}
M&=&-3F_1(t)\;2\sqrt{6}m_Ve_qf_V\;s\;
\epsilon\cdot\epsilon_V
\sqrt{\frac{\alpha_s(\Sigma^2)}{\alpha_n}} \nonumber \\
&&\frac{1}{8\pi}\int dl^2
\left[4\pi\alpha_n D(l^2+\frac{1}{4}t)\right]^2
\frac{1}{(Q^2+m_V^2-4l^2)}
\end{eqnarray}
which combined with Eq.~\ref{2g:exch} leads to~:
\begin{eqnarray}
M&=&3F_1(t)\;2\sqrt{6}m_Ve_qf_V\;s\;
\epsilon\cdot\epsilon_V
\sqrt{\frac{\alpha_s(\Sigma^2)}{\alpha_n}} \nonumber \\
&&\frac{1}{8\pi}\int dl^2
\left[4\pi\alpha_n D(l^2+\frac{1}{4}t)\right]^2
\left[\frac{1}{(Q^2+m_V^2-t)}-\frac{1}{(Q^2+m_V^2-4l^2)}\right]
\nonumber \\
\mbox{ }
\end{eqnarray}
and to  the  cross-section:
\begin{eqnarray}
\frac{d\sigma}{dt}&=&
\frac{81 m_V^3 \Gamma_{e^+e^-}}{256\pi^3\alpha_{em}}
\left[F_1(t)\right]^2 \nonumber \\
&&\left[ \sqrt{\frac{\alpha_s(\Sigma^2)}{\alpha_n}}
\int dl^2\frac{-4l^2+t}{(m_V^2+Q^2-4l^2)(m_V^2+Q^2-t)}
\left[4\pi \alpha_nD(l^2+\frac{1}{4}t) \right]^2\right]^2
\label{amp}
\end{eqnarray}

Donnachie and Landshoff~\cite{Do88}, and then Cudell~\cite{Cu90}, showed
an  analogy  between  the  phenomenological  Pomeron  and  the two gluon
exchange  approaches.    Obviously,  since  it  is  a  perturbative like
calculation, there is  no Regge factor  in the two  gluon result.   When
this Regge factor  is removed from  the Pomeron exchange  amplitude, the
similarity  between  the  two  cross-sections  is striking when the loop
integral   in   Eq.~\ref{amp}    is   performed   in    an   approximate
way~\cite{Cu90}. If the squared  mass $Q^2$ of the virtual  photon,
or the mass $m_V$ of the vector meson is
large  enough,  one  may  neglect  the  dependence  against $l^2$ in the
denominator of  the integrand  of Eq.~\ref{amp},  and the  cross section
takes the simple form~:
\begin{eqnarray}
 \frac{d\sigma}{dt}&=&\frac{81 m_V^3 \beta_0^4\Gamma_{e^+e^-}}
{4\pi\alpha_{em}} \; \left[ F_1(t) \right ]^2 \nonumber \\
&&\left[ \sqrt{\frac{\alpha_s(\Sigma^2)}{\alpha_n}}
\frac{2\lambda_0^2+t}{(m_V^2+Q^2+2\lambda_0^2)
(m_V^2+Q^2-t)}\;\exp\left(\frac{t}{2\lambda_0^2}\right)\;
\right]^2
\label{cross}
\end{eqnarray}
At  $t=0$,  and  in  the  limit  $\alpha_s/\alpha_n \rightarrow 1$, this
cross-section  coincides   with  the   Pomeron  exchange   cross-section
(Eq.~\ref{DL}), provided  that $\lambda_0$  is identified  with $\mu_0$.
Thus, in  this limit,  the two  approaches are  almost identical for low
values of $t$.  The coupling to two different quarks in the vector meson (bottom right  graph in Figure~\ref{2gluons_graph}) provides  us  with  a  more microscopic explanation of the phenomenological form factor which is used in the Pomeron exchange cross-section.

Eq.~\ref{cross} shows explicitely that the cross-section exhibits a  dip at $t=-2 \lambda_0^2$.  This  dip results from  the  cancellation  between  the  direct  two  gluon exchange graph, where the two gluons couple  to the same quark in the  vector meson, and  the crossed  graph where  each gluon  couples to a different quark.   This behavior  also remains  in the  complete expression of the cross-section (Eq.~\ref{amp}). The results shown in~\cite{jml} and~\cite{me95} have been obtained by performing explicitly the integral in Eq.~\ref{amp} and not using Eq.~\ref{cross}. The effective $\alpha_n$ and the strong running $\alpha_s$ coupling constants were set as:
\begin{eqnarray} 
\alpha_n &=& 0.3 \nonumber \\
\alpha_s(\Sigma^2) &=& 1.396/\ln(\frac{\Sigma^2}{0.04})
\end{eqnarray} 

Two improvements have been implemented in~\cite{jml}. On the one hand, the full expression of the coupling between the two gluons and the photon and the vector meson has been taken into account, instead of its leading part at high energy $s\epsilon_V \epsilon$. The amplitude is explicitly gauge invariant. On the other hand, the coupling of the exchanged gluons with different quarks in the nucleon (top graphs in Figure~\ref{2gluons_graph}) has been taken into account. Under the assumption that the gluons couple to valence quarks equally sharing the longitudinal momentum of the nucleon ($x_q=1/3$), the correlation function can be expressed~\cite{Gu77,Cu94,HuXY} as the isosclar form factor at a different argument $F_1(3l^2+\frac{t}{4})$. The resulting amplitude takes the form:
\begin{eqnarray}
{\cal M}_{2g} = \frac{6\sqrt{6}m_V e_q f_V }{8\pi } 
                       [2p\cdot q \epsilon_V \cdot \epsilon +2\epsilon_V\cdot p \epsilon \cdot (q-P_V)
                                                                                                       \nonumber \\
 -2\epsilon_V \cdot q\epsilon \cdot p +\frac{Q^2+m^2_V-t}{p\cdot q} \epsilon_V \cdot p\epsilon \cdot p]
                                                                                                       \nonumber \\
\int {\rm d}l^2 [4\pi \alpha_s D(l^2+\frac{1}{4}t)] ^2    \protect{[F_1(t)- F_1(3l^2+\frac{t}{4})]}
                                                                                                       \nonumber \\
\left( \frac{1}{Q^2+m_V^2-t} - \frac{1}{Q^2+m_V^2-4l^2} \right)
\label{2g}
\end{eqnarray}

Each of the two last terms in eq.~\ref{2g} corresponds to the propagator of the quark which is off-shell in the vector meson loop, when the two gluons couple either to the same quark or to two different quarks. The contribution of the quark correlations in the nucleon does not modify the cross section at low momentum transfer, but get rid of the node in the uncorrelated cross section and bring it close to the experimental cross section at high $-t$. Note that the off-shell ratio $\alpha_s(\Sigma^2)/\alpha_n$ has been set to 1 in Eq.~\ref{2g}. 

Two last improvements have been implemented in~\cite{cano}. Going beyond the simple assumption that the quarks equally share the longitudinal momentum of the nucleon, the correlation function has been evaluated using a more realistic nucleon wave function~\cite{bo96}, which incorporates the dependency on the fraction of the longitudinal momentum of the nucleon $x_q$ carried by each quark. The two fold integral runs now on $x_q$, besides $l^2$, and has been computed numerically. Also the gluon dressed propagator~\cite{lei99} deduced from a lattice calculation has been used (together with $\alpha_n=0.28$) instead of the simpler gaussian parameterization~Eq.~\ref{prop} (which includes $\alpha_n$). 

The node in the uncorrelated part of the cross section in Figure~\ref{phi_JLab} moves from $-t= 2.3$~GeV$^2$~\cite{jml} to $-t= 1.2$~GeV$^2$, but the correlated part restores the agreement with the experiments at high $-t$. I refer the reader to~\cite{cano} for more details and a more complete presentation. It is this last version of the model that has been used in this review.

\subsubsection{\it The Partonic Non-Perturbative Approach  \label{sec:PNP_Approach}}

\begin{figure}[tbhp]
\begin{center}
\begin{minipage}[t]{8 cm}
\epsfysize=8.0cm
\epsfig{file=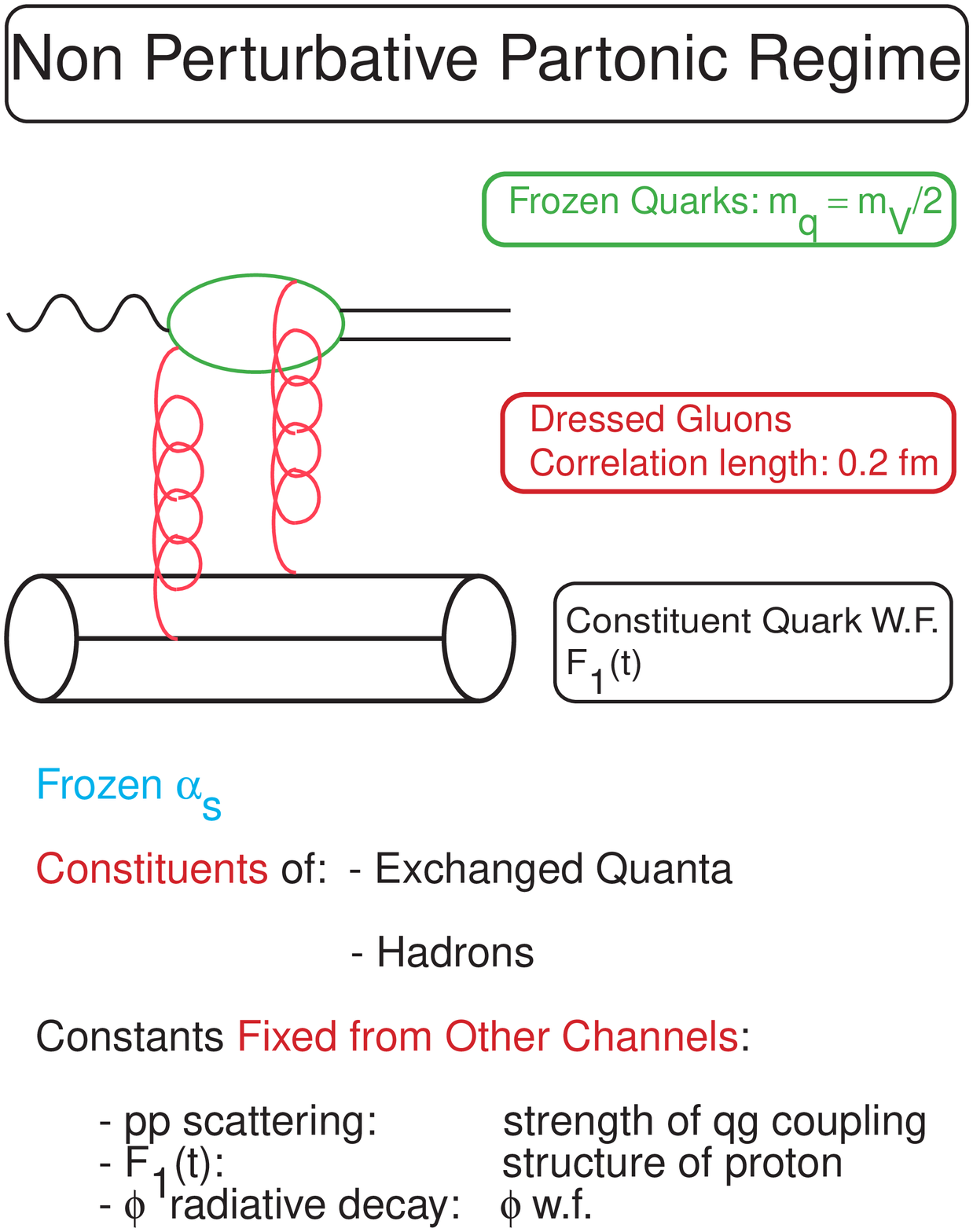,scale=0.45}
\end{minipage}

\begin{minipage}[t]{16.5 cm}
\caption{(Color on line) A sketch of the Partonic Non-Perturbative Approach.
\label{PNP_sketch}}
\end{minipage}
\end{center}
\end{figure}

Such a finding is important as it tells us that, in the intermediate range of momentum  transfer (let's say $1\leq -t\leq 10$ GeV~$^2$), large angle exclusive vector meson production cross sections can be understood  at the level of effective parton degrees of freedom: dressed quark and gluon propagators, constituent quark wave functions of the nucleon and of the meson, etc\ldots. As summarized in Figure~\ref{PNP_sketch}, such a ``Partonic Non-Perturbative Approach" relates different aspects of the structure of hadronic matter which are constrained by specific experiments. 

At low momentum transfer (up to $-t\approx 2$~GeV$^2$), the cross section is driven by their integral properties: any nucleon wave function which reproduces the nucleon form factor leads to similar results; the differences which may be due to gluon propagators can be reabsorbed in a reasonable value of $\alpha_n$ (for the amplitude based on the lattice  propagator we have used $\alpha_n=0.28$). At higher momentum transfer, the cross section becomes more sensitive to the details of the wave function (giving access to quark correlations) and the shape of the gluon propagator.

\section{Conclusion}

After this journey in the field of exclusive photo- and electro-production of mesons, it appears that the current quarks hide themselves in the nucleon and that unitarity constraints prevent the photon to directly see them. On the one hand, unitarity regularizes the Feynman pole amplitudes that drive the peripheral interactions at forward and backward angles: Regge poles prevent them to exceed the unitarity bound at high energies. By treating in a compact form the exchange of families of mesons (or baryons) they provide us with a good effective representation at forward (or backward) angles.

On the other hand, unitarity s-channel rescattering cuts provide the strength at intermediate angles (large $-t$ and $-u$), where head-on collision were expected to reveal the quark degrees of freedom. In the light quark sector, these cuts contribute even at forward angles in the neutral pion electro-production and Deeply Virtual Compton Scattering, where they cannot be neglected. Any sensible analysis of DVCS should take into account their contribution.

In fact, the energy available in the center of mass at JLab or Hermes is too low to sum over the complete basis of all the possible intermediate hadronic states and safely replace them by a dual basis of current quark states. Only a few hadronic cuts are relevant: they deserve a dedicated treatment.

Amplitudes for the production of vector mesons made of light quarks are less sensitive to cuts: there is no channel with higher cross section to couple with. At high energies they are driven by gluon exchange mechanisms, while they receive contributions of saturating Regge trajectories at lower energies (let say $W<10$~GeV). 

Also, in the production of vector mesons involving heavy quarks, the contribution of such unitarity cuts is suppressed: the different flavors of the quarks in the meson and in the nucleon prevent quark exchange mechanisms and favor gluon exchanges. These channels open a window for the "Partonic Non-Perturbative Approach" discussed in the present review: It is  based on constituent quarks and dressed gluon propagators and has been successfully tested at low $-t$ up to the HERA energy range as well at high $-t$ in the JLab6 energy range.  The study of its evolution at higher  momentum transfer may open the way to approach the current quarks domain. In this respect, photo- and electro-production of the $\phi$ meson at large $-t$ should be actively studied at JLab12, and at higher energies.

Charm production near threshold becomes accessible at JLab12. The large value of  $|t|_{min}$, as well as the large mass of the charm quark,  make pertinent the "Partonic Non-Perturbative Approach". It also opens the way to the study of new mechanisms, such as \textit{e.g.} three gluon exchanges.

As a last word, let me emphasize that Nature organizes itself in a complex way from simple structures. We have visited an efficient architecture which tries to span a bridge between  many aspects of Hadronic Matter.  It remains to a new generation of young researchers to polish it.

\section{Acknowledgments}

Let me thank L. Cardman and D. Richards for their warm scientific hospitality at JLab after my retirement from CEA-Saclay. Jefferson Science Associates operates Thomas Jefferson National Facility for the United States Department of Energy under contract DE-AC05-06OR23177. The works reported in this review have greatly benefited from  a fruitful collaboration, over the years, with R. Mendez-Galain, F. Cano, M. Guidal and M. Vanderhaeghen. Continuous discussions with G. Audit and M. Gar\c{c}on have triggered many idea and the way to conduct dedicated experiments.

\section{Appendix A}

More detailed expressions of the amplitudes of the various channels, and further discussions, can be found in the following list the references:

\begin{itemize}
\item Pomeron and 2-gluons exchange amplitudes: \cite{jml}, \cite{me95}, \cite{cano};
\item t-channel Reggeon amplitudes: \cite{gui97}. Their expression in terms of $\sigma$ matrices and Pauli spinors can be found in the Appendix of~\cite{lag06}. The $b_1$ exchange amplitude has been corrected in~\cite{lag11};
\item u-channel Reggeon amplitudes: \cite{jml}, \cite{lag10};
\item Unitarity rescattering cuts: \cite{lag07}, \cite{lag10}, \cite{lag11}.

\end{itemize}
 
For the interested reader, an introduction to Regge or two gluon exchange models, as well as a review of previous works, can be found in~\cite{gui97} and~\cite{me95} respectively. A more general introduction to hard scatterings and Pomeron can be found in~\cite{col84},~\cite{Do02} and~\cite{colmart84}.

\section{Appendix B}

The cross-section of the $p(e,e'V)p$ reaction is related to the cross-section of the virtual photon reaction $p(\gamma_v,V)p$ as follows~:
\begin{equation}
\frac{d\sigma}{dE'd\cos\theta_ed\Phi_edt}=
\Gamma_v\frac{d\sigma}{d\Phi_edt}
\end{equation}
where the flux of the virtual photons is~:
\begin{equation}
\Gamma_v=\frac{\alpha_{em}}{2\pi^2}\frac{E'}{E}
\frac{\mid\vec{k}\mid}{1-\epsilon}\;\frac{1}{-q^2}
\end{equation}
The polarization of the virtual photon is~:
\begin{equation}
\epsilon=\left [1-2\frac{\vec{k}^2}{q^2}\tan^2\frac{\Theta_e}{2}
\right]^{-1}
\end{equation}
The energy of the incident electron is $E$, while the energy and the
angles of the scattered electron are respectively $E'$, $\Theta_e$ and
$\Phi_e$ (as measured with respect to the meson emission plane, with the convention that $\Phi_e=0$ when the outgoing electron and meson lie on the same side with respect to the virtual photon direction). The energy, the momentum and the squared mass of the virtual photon are respectively $\nu$, $\vec{k}$ and $q^2=-Q^2= \nu^2 -\vec{k}^2$

The virtual photon absorption cross section takes the form~\cite{lag11}:
\begin{eqnarray}
\frac {d\sigma} {dt d\Phi_e}=\frac{d}{2\pi dt}\left [\sigma_T
+\epsilon \sigma_L +\epsilon \cos 2\Phi_e \; \sigma_{TT}
+\sqrt{2\epsilon (\epsilon +1)}\cos \Phi_e \;
\sigma_{TL} \right ]
\label{unpol}
\end{eqnarray}

When integrated against the electron azimuthal angle Eq.~\ref{unpol}
reduces to~:
\begin{equation}
\frac {d\sigma} {dt}=\frac{d}{dt}\left [\sigma_T
+\epsilon \sigma_L \right ]
\end{equation}

The Transverse and Longitudinal  cross-sections are related to the
matrix elements in the following way:
\begin{eqnarray}
\frac{d\sigma_T}{dt} &=& \frac{\alpha_{em}}{4s^2}\;
\frac{1}{2}\left[\sum\mid M_X\mid^2+\sum\mid M_Y\mid^2\right]
\nonumber \\
\frac{d\sigma_L}{dt} &=& \frac{\alpha_{em}}{4s^2}\;
\frac{Q^2}{\nu^2}\sum\mid M_Z\mid^2
\end{eqnarray}
where  the averaging over the magnetic quantum numbers of the target
proton and the summation over those of the outgoing proton are
implicitly understood, and where $M_X$, $M_Y$ and $M_Z$ are the
projections of the hadronic matrix element on the three cartesian axis.

The Transverse-Transverse and the Tranverse-Longitudinal interference cross sections are defined as:
\begin{eqnarray}
\frac{d\sigma_{TT}}{dt} &=& \frac{\alpha_{em}}{4s^2}\;
\frac{1}{2}\left[\sum\mid M_X\mid^2-\sum\mid M_Y\mid^2\right]
\nonumber \\
\frac{d\sigma_{LT}}{dt} &=& -\frac{\alpha_{em}}{4s^2}\;
\sqrt{\frac{Q^2}{4\nu^2}}\left[\sum M_Z^*M_X + \sum M_X^*M_Z\right]
\end{eqnarray}

At the real photon point, the photon asymmery is simply:
\begin{equation}
\Sigma= \frac{d\sigma_{TT}}{dt} / \frac{d\sigma_{T}}{dt}
\end{equation}

In these expressions the time component of the current have been expressed in term of its longitudinal component, since the gauge invariance of the electromagnetic currents relates them:
\begin{eqnarray}
q^{\mu}M_{\mu} &=& \nu M_0 - \mid \vec{k}\mid M_Z  = 0 \nonumber \\ 
q^{\mu}j_{\mu} &=& \nu j_0 - \mid \vec{k}\mid j_Z  = 0
\end{eqnarray}
Therefore:
\begin{eqnarray}
j^{\mu}M_{\mu} &=& j_0 M_0 -  j_ZM_Z -  j_XM_X-  j_YM_Y \nonumber \\
 j^{\mu}M_{\mu}&=& \frac{-q^2}{\nu^2}j_ZM_Z -  j_YM_X-  j_YM_X
\end{eqnarray}

So, the various electro-production cross sections rely only on the spatial part of the hadronic current.


\begin{thebibliography}{99}
\itemsep -2pt

\bibitem{gui97} M. Guidal, J.-M. Laget and M. Vanderhaeghen, \Journal{\NPA}{627} {645} {1997}
\bibitem{jml} J.-M. Laget, \Journal{\PLB} {489} {313} {2000}
\bibitem{bro} S.-J. Brodsky and G.-P. Lepage, \Journal{\PRD} {22} {2157} {1980}
\bibitem{lag04} J.-M. Laget, \Journal{\PRD} {70} {054023} {2004}
\bibitem{do89} A. Donnachie and P.-V. Landshoff, \Journal{\NPB} {311} {509} {1989}
\bibitem{me95} J.-M. Laget and R. Mendez-Galain \Journal{\NPA} {581} {397} {1995}
\bibitem{cano} F. Cano and J.-M. Laget, \Journal{\PRD} {65} {074022} {2002}
\bibitem{phiprl} E. Anciant {\it et al.}, \Journal{\PRL} {85} {4682} {2000}
\bibitem{bat01} M. Battaglieri {\em et al.}, \Journal{\PRL} {87} {172002} {2001}
\bibitem{bat03} M. Battaglieri {\em et al.}, \Journal{\PRL} {90} {022002} {2003}
\bibitem{physrep} J.-M. Laget, \Journal{\PREP} {69} {1} {1981}
\bibitem{lag07} J.-M. Laget, \Journal{\PRC} {76} {052201(R)} {2007}
\bibitem{lag10} J.-M. Laget, \Journal{\PLB} {685} {146} {2010}
\bibitem{lag11} J.-M. Laget, \Journal{\PLB} {695} {199} {2011}
\bibitem{mor05} L. Morand {\em et al.}, \Journal{\EPJA} {24} {445} {2005}
\bibitem{par13} K. Park {\em et al.}, \Journal{\EPJA} {49} {16} {2013}
\bibitem{air08} A. Airapetian {\it et al.}, \Journal{\PLB} {659} {486} {2008}
\bibitem{lei99} D.-B. Leinweber {\it et al.}, \Journal{\PRD} {60} {094507} {1999}
\bibitem{bo96} J. Bolz and P. Kroll, \Journal{\ZPA} {356} {327} {1996}
\bibitem{mcc} K. McCormick  {\it et al.}, \Journal{\PRC} {69} {032203(R)} {2004}
\bibitem{ser94} M.-N. Sergeenko, \Journal{\ZPC} {64} {315} {1994}
\bibitem{col84} P.D.B. Collins and P.J. Kearney, \Journal{\ZPC} {22} {277} {1984}

\bibitem{sch79} M.-A. Schupe {\it et al.}, \Journal{\PRD} {19} {192} {1979}
\bibitem{can03} F. Cano and J.-M. Laget, \Journal{\PLB} {551} {317} {2003}
\bibitem{can03E} F. Cano and J.-M. Laget, \Journal{\PLB} {571} {260} {2003}
\bibitem{and70} R.-L. Anderson {\it et al.},  \Journal{\PRL} {25} {1218} {1970}
\bibitem{bus70} G. Bushhorn {\it et al.}, \Journal{\PLB} {37} {207} {1970}

\bibitem{fan15} C. Fanelli {\it et al.}, \Journal{\PRL} {115} {152001} {2015}
\bibitem{kun18} M. Kunkel {\it et al.}, \Journal{\PRC} {98} {015207} {2018}
\bibitem{Des68} M. Braunschweig  {\it et al.}, \Journal{\PLB} {26} {405} {1968}
\bibitem{Sla76} R.-L. Anderson {\it et al.}, \Journal{\PRD} {14} {679} {1976}
\bibitem{schu} R. Bradford, {\em et al.}, \Journal{\PRC} {73} {035202} {2006}
\bibitem{Bre00} J. Breitweg {\it et al.}, \Journal{\EPJC} {14} {213} {2000}
\bibitem{Adl00} C. Adloff {\it et al.}, \Journal{\PLB} {483} {23} {2000}
\bibitem{Bli82} M. Blinkey {\it et al.}, \Journal{\PRL} {48} {73} {1982}
\bibitem{bro01} S.-J. Bodsky, E. Chudakov, P.Hoyer and J.-M. Laget, \Journal{\PLB} {498} {23} {2001}
\bibitem{Ali19} A.Ali {\it et al.}, \Journal{\PRL} {123} {072001} {2019}; arXiv:1905.10811 [nucl-ex] (2019)
\bibitem{Aub85} J.-J. Aubert {\it et al.}, \Journal{\PLB} {161} {203} {1985}; \Journal{\PLB} {133} {370} {1983}
\bibitem{san08} J.-P. Santoro {\it et al.}, \Journal{\PRC} {78} {025210} {2008}
\bibitem{mor09} S. Morrow {\em et al.}, \Journal{\EPJA} {39} {5} {2009}
\bibitem{kas11} M.-M. Kaskulov and U. Mosel, \Journal{\PRC} {81} {045202} {2010}
\bibitem{Bra78} F.-W. Brasse {\it et al.}, \Journal{\PLB} {58} {467} {1978}
\bibitem{Cam10} E. Fuchey {\it et al.}, \Journal{\PRC} {83} {025201} {2011}
\bibitem{bed14} I. Bedlinskiy {\it et al.}, \Journal{\PRC} {90} {025205} {2014}
\bibitem{Co97} J.-C. Collins, L.Frankfurt and M. Strikman, \Journal{\PRD} {56} {2982} {1997} 
\bibitem{Co99} J.-C. Collins and A. Freund, \Journal{\PRD} {59} {074009} {1999}
\bibitem{def14} M. Defurne, {\it et al.}, \Journal{\PRC} {92} {055202} {2015}; C. Munoz Camacho {\it et al}, \Journal{\PRL} {97} {262002} {2006}
\bibitem{Ai01} A. Airapetian {\it et al}, \Journal{\PRL} {87} {182001} {2001}
\bibitem{Ai07} A. Airapetian {\it et al}, \Journal{\PRD} {75} {011103(R)} {2007}
\bibitem{lag06} J.-M. Laget, \Journal{\PRC} {73} {044003} {2006}
\bibitem{reg59} T. Regge, \Journal{\NC} {14} {951} {1959}
\bibitem{reg60} T. Regge, \Journal{\NC} {18} {947} {1960}
\bibitem{jones80} L.-M. Jones, \Journal{\PLB} {72} {144} {1980}

\bibitem{cli77} R.-W. Clifft {\it et al.}, \Journal{\RMP} {52} {545} {1977}
\bibitem{bal73} J. Ballam {\it et al.}, \Journal{\PRD} {7} {3150} {1973}
\bibitem{jaf89} R.-L. Jaffe, \Journal{\PLB} {229} {275} {1989}

\bibitem{PDG} S. Eidelman {\it et al.}, \Journal{\PLB} {572} {1} {2004}
\bibitem{La72} T. Lasinsky {\it et al.}, \Journal{\NPB} {37} {1} {1972}
\bibitem{AnXY} R.-L. Anderson {\it et al.},  \Journal{\PRD} {14} {679} {1976}

\bibitem{Do02} S. Donnachie {\it et al.}, Pomeron Physics and QCD, Cambridge University Press, Cambridge, England (2002).
\bibitem{An70} R.-L. Anderson {\it et al.}, \Journal{\PRD} {1} {27} {1970}
\bibitem{An71} R.-L. Anderson {\it et al.}, \Journal{\PRD} {4} {1937} {1971}
\bibitem{lag05} J.-M. Laget, \Journal{\PRC} {72} {022202(R)} {2007}

\bibitem{PiroXX} B. Haber {\it et al.},  \Journal{\PRD} {10} {137} {1974}

\bibitem{ant18} G. Antchev {\it et al.},  arXiv:1712.06153 [hep-ex] (2018)
\bibitem{aba94} S. Abatzis {\it et al.},  \Journal{\PLB} {324} {509} {1994}

\bibitem {La90} P.-V. Landshoff, \Journal{\NPB} {18C} {211} {1990}
\bibitem{Do87} A. Donnachie and P.V. Landshoff, \Journal{\PLB} {185} {403} {1987} 
\bibitem{PP92} Review of Particles Properties, \Journal{\PRD} {45} {S1} {1992}
 \bibitem{Do88} A. Donnachie and P.V. Landshoff, \Journal{\NPB} {311} {509} {1989}
\bibitem{La87} P.V. Landshoff and O. Nachtmann, \Journal{\ZPC} {35} {405} {1987}
\bibitem{Cu90} J.R. Cudell, \Journal{\NPB} {336} {509} {1990}
\bibitem{Gu77} J.F. Gunion and D.E. Soper, \Journal{\PRD} {15} {2617} {1997}
\bibitem{Cu94} J.R. Cudell and B.U. Nguyen, \Journal{\NPB} {420} {669} {1994}
\bibitem{HuXY} J. Dolejsi and J. Hufner, \Journal{\ZPC} {54} {489} {1992}
\bibitem{colmart84} P.-D.-B. Collins and A.-D. Martin, Hadron Interactions, Adam Hilger LTD, Bristol (1984)

%\bibitem{collins84} P.-D.-B. Collins and P.-J. Kearney, \Journal{\ZPC} {22} {277} {1984}
%\bibitem{van98} M. Vanderhaeghen, M. Guidal and J.-M. Laget, \Journal{\PRC} {57} {1454} {1998}

\end{thebibliography}
\end{document}